\documentclass[12pt]{iopart}

\usepackage{color}
\usepackage{graphicx}
\usepackage{amsfonts,amsbsy}
\usepackage{Tomspeak}
\newcommand\hd[1]{{\color{blue}[\kern -2pt[\lower 1pt\hbox{$_{\scriptscriptstyle HD}$} #1 ]\kern -2pt]}}

\newcommand{\cch}{c^{\rm ch}} 
\newcommand{\mobs}{b^{\rm s}} 
\newcommand{\tenmobs}{\boldsymbol{\mathsf{b}^{\rm s}}}
\newcommand{\tenD}{D_{ij}} 
\newcommand{\teneta}{\eta_{ijkl}} 
\newcommand{\oseen}{{\cal G}}
\newcommand{\tenoseen}{\boldsymbol{\cal G}}
\newcommand{\tenPhi}{\boldsymbol{\Phi}}
\newcommand{\tenPsi}{\boldsymbol{\Psi}}
\newcommand{\tensigma}{\boldsymbol{\sigma}} 
\newcommand{\vecF}{\vec{F}}
\newcommand{\vecj}{\vec{j}}
\newcommand{\vecOmega}{\vec{\Omega}}
\newcommand{\vecq}{\vec{q}}
\newcommand{\vecr}{\vec{r}}
\newcommand{\vecR}{\vec{R}}
\newcommand{\vecV}{\vec{V}} 
\newcommand{\vecv}{\vec{v}} 
\newcommand{\vecomega}{\vec{\omega}} 
\newcommand{\xhat}{\hat{x}}
\newcommand{\yhat}{\hat{y}}
\newcommand{\zhat}{\hat{z}}
\newcommand{\RT}{\boldsymbol{\mathsf{T}}}
\newcommand{\Iop}{{\mathsf I}}
\newcommand{\Rop}{{\mathsf R}}
\newcommand{\IdentityMatrix}{{\mathsf 1}}
\newcommand{\xop}[1]{{\scriptstyle [\![} {#1}{\scriptstyle ]\!]}^\cross}
\newcommand{\cross}{\times}
\newcommand{\GG}{{\mathbf G}}

\renewcommand{\theequation}{\thesection.\arabic{equation}}
\renewcommand{\thefigure}{\thesection.\arabic{figure}}

\begin{document}

\title[Shaped particles in fluids. Copyright 2020 T. Witten and H. Diamant]{A review of shaped colloidal
  particles in fluids: Anisotropy and chirality}

\author{Thomas A.\ Witten$^1$ and Haim Diamant$^2$}

\address{$^1$ Department of Physics and James Franck Institute,
  University of Chicago, Chicago, Illinois 60637, USA}

\address{$^2$ Raymond and Beverly Sackler School of Chemistry, Tel Aviv University, Tel Aviv
  69978, Israel}

\ead{t-witten@uchicago.edu, hdiamant@tau.ac.il}

\vspace{10pt}
\begin{indented}
\item[] \today 
\end{indented}

\begin{abstract}
  

This review treats asymmetric colloidal particles moving through their host fluid under the action of some form of propulsion. The propulsion can come from an external body force or from external shear flow.  It may also come from  externally-induced stresses at the surface, arising from imposed chemical, thermal or electrical gradients.  The resulting motion arises jointly from the driven particle and the displaced fluid.   If the objects are asymmetric, every aspect of their motion and interaction depends on the orientation of the objects.  This orientation in turn changes in response to the driving.  The objects' shape can thus lead to a range of emergent anisotropic and chiral motion not possible with isotropic spherical particles.  
We first consider what aspects of a body's asymmetry can affect its drift through a fluid, especially chiral motion.  We next discuss driving by injecting external force or torque into the particles. Then we consider driving without injecting force or torque.  This includes driving by shear flow and  driving by surface stresses, such as electrophoresis.  We consider how time-dependent driving can induce collective orientational order and coherent motion.  We show how a given particle shape can be represented using an assembly of point forces called a Stokeslet object.  We next consider the interactions between anisotropic propelled particles, the symmetries governing the interactions, and the possibility of bound pairs of particles.  Finally we show how the collective hydrodynamics of a suspension can be qualitatively altered by the particles' shapes. The asymmetric responses discussed here are broadly relevant also for swimming propulsion of active micron-scale objects such as microorganisms.  

\end{abstract}

\vspace{2pc}
\par
\submitto{\RPP}
%
\maketitle
%

\tableofcontents
\title[Shaped particles in fluids. Copyright 2019 T. Witten and H. Diamant]{~}

\bigskip
\hrule
\bigskip

\section{Introduction}\label{sec:Introduction}

\begin{figure}[htbp]
\includegraphics[height=5in]{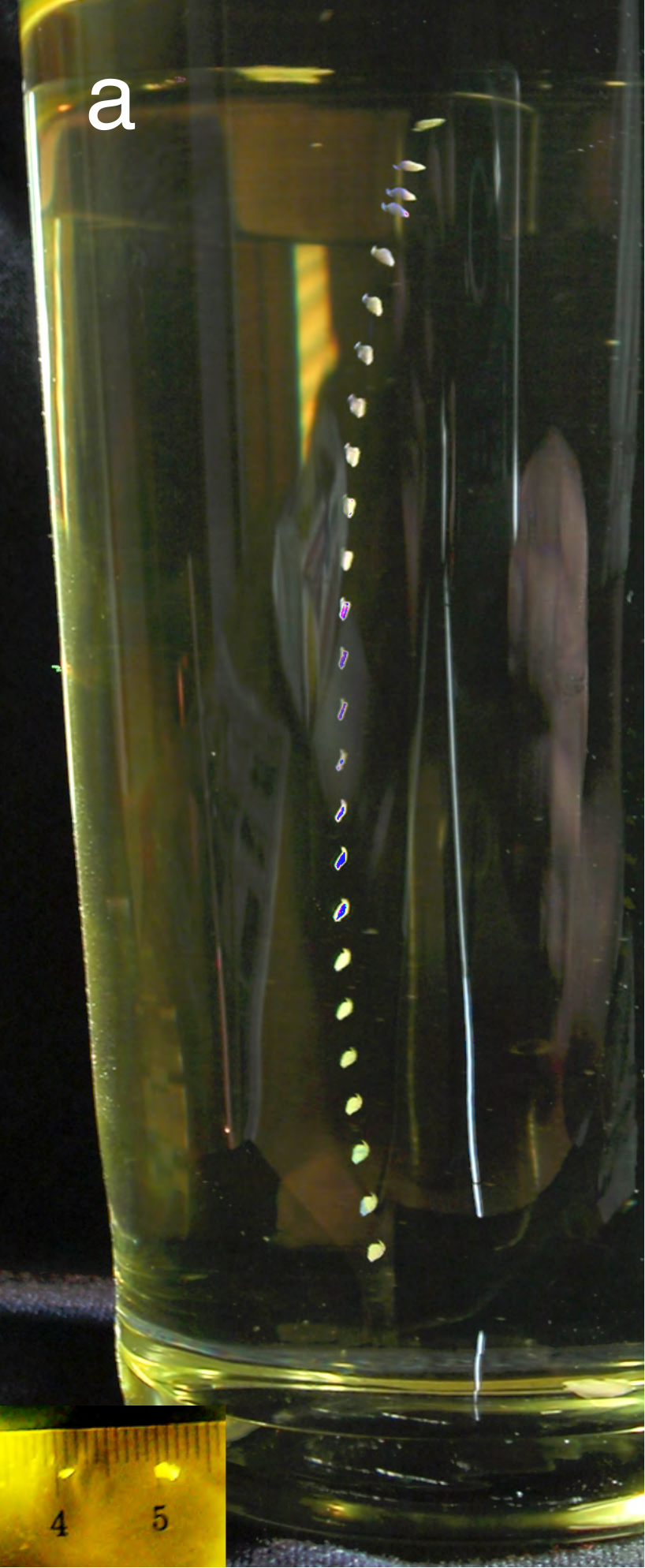}
\includegraphics[height=5in]{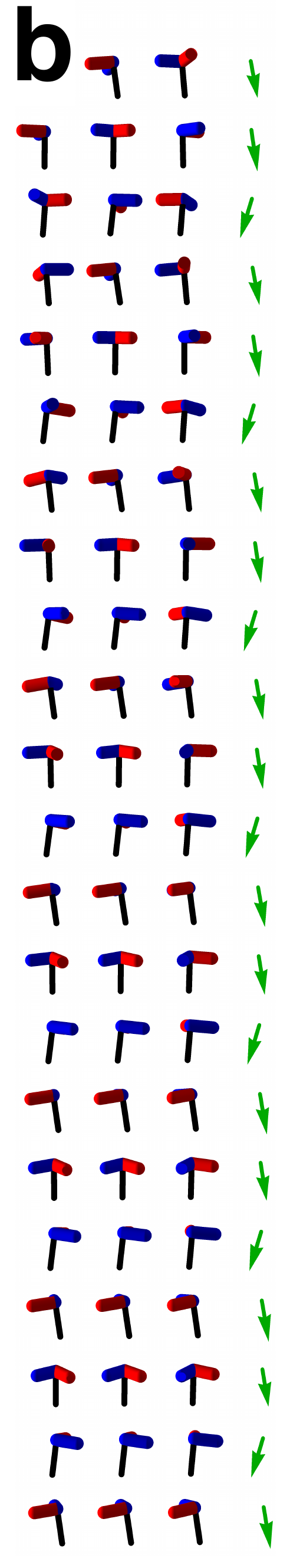}
\hfill
\includegraphics[width=3in]{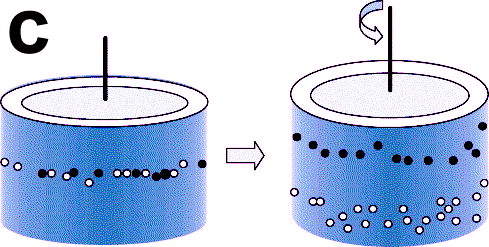}
\hfill
\caption{Chiral motion of asymmetric objects. a) Chiral sedimentation of a generically-shaped object\cite{Witten:2008cr}. Inset at lower left shows sedimenting objects, small pieces of plastic cut from a disposable spoon.  Main picture shows multiple exposure of the left object in the inset sedimenting in canola oil.  Height of oil was 9 cm. Exposures were taken every 10 seconds. b) Multiple-exposure view of three simulated identical chiral objects pulled by a rotating force (green arrow) \cite{Moths:2013mz}. Each object is represented by its three co-ordinate axes.  Initially the axes differ by a rotation. In the final exposure the axes are aligned. c) Sketch showing separation of chiral particles (white dots) from their mirror image particles (black dots) in a Couette flow, after \cite{Doi:2005yu}. 
}
\label{fig:plasticCrumb}
\end{figure}

Figure \ref{fig:plasticCrumb}a shows a crumb of broken plastic that was released into a still liquid.  The multiple exposure picture shows the crumb rotating as it gradually descends in a helical path.  If one releases it in a different orientation, it rapidly adopts the same motion after a brief transient.  The direction of twist is evidently an intrinsic property of the crumb.  The geometry of the motion is independent of extrinsic factors like the viscosity of the fluid.  Indeed, different viscosities alter the speed of descent and twisting but do not affect the pitch of the helical motion or the radius of the helix traced by a given point on the object.  This type of motion is a chiral signature of the object that must reflect some lack of inversion symmetry of the object.  This kind of chiral signature can be the basis for particle separation, \footnote{
Here and below we refer to objects or bodies as particles when we mean to focus on their collective behavior in a colloidal dispersion.
}
as sketched in Fig. \ref{fig:plasticCrumb}c.

               This chiral response clearly requires a surrounding fluid.  If the crumb had been released in a vacuum, it would not have rotated in this signature fashion.  Thus to understand how this motion arises, we must understand how the descending body perturbs the fluid.  As noted above, the spatial track of the motion depends only on geometric features of the bodies.  Mechanical quantities like mass, force and viscosity only affect the speed at which the motion follows this track, provided this speed is slow enough. The mathematics that predicts the motion from the shape is far simpler than in many fluid problems; all motion of a given object is expressible in terms of two ``mobility tensors" characteristic of the object.  We explain these in Sec. \ref{sec:Symmetry} below.     

                Figure \ref{fig:plasticCrumb}b  shows a further subtlety that arises from this same math.  The simulated picture shows the motion of three identical simulated objects, each descending under a downward force like the plastic crumb in Fig. \ref{fig:plasticCrumb}a.   As in Figure \ref{fig:plasticCrumb}a, the picture tracks their descent in multiple exposures.  The objects respond to the force independently without interacting, as they would if they were far apart.  What has changed is that the applied force on the bodies is no longer constant.  The force vector rotates at a speed similar to the objects' own rotation speed.  As we see, the changing force has an effect on the relative orientation of the objects.  Although the motion began with the objects in different orientations, it ended with them oriented in the same way.  The changing external force has dictated the phase as well as the direction of the responding motion.  Thus the three identical objects have ended up rotating in phase.  As we explain below, this phase-locking property is part of a general synchronization effect that happens under a wide class of time-dependent forcing.
                
\begin{figure}[htbp]
\includegraphics[width=3.5in]{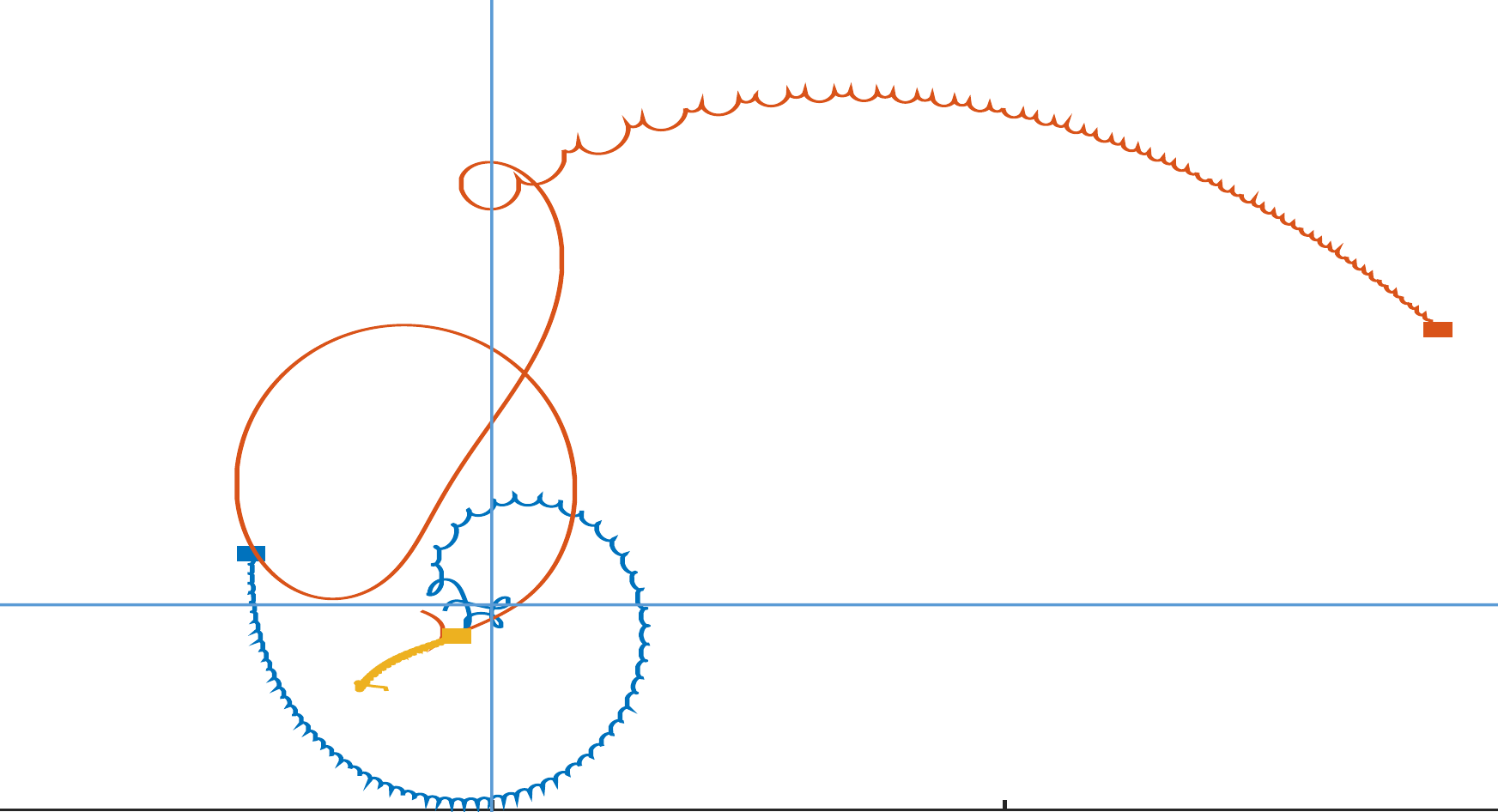}
\includegraphics[width=2.5in]{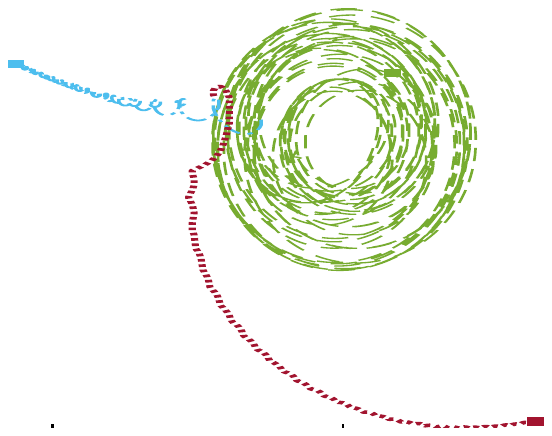}

\caption{Simulated sedimentation trajectories of a pair of identical, generic Stokeslet objects (\cf Sec.\ref{sec:StokesletObjects}), after \cite{Goldfriend:2015ek}. The horizontal projection of the vector displacement between the two objects over time is shown for three choices of initial condition, showing the effect of hydrodynamic interaction. Solid rectangles mark the final point of the simulation on each trajectory.  Both objects are subjected to a constant vertical force and a perturbing force rotating in the horizontal plane at roughly the natural rotation rate of each  body (as in Fig. \ref{fig:plasticCrumb}b). Left: particles with initially random orientations.  Fine straight lines mark the point where there is no horizontal displacement.  Right: particles with identical initial orientations showing a long-lived trapped state (circling (green dashed) curve on right.)}
\label{fig:TomerOrbits}
\end{figure}

                Figure \ref{fig:TomerOrbits} shows what can happen when two rotating objects feel the effects of the flow around each other.  This fluid flow is altered by the objects' anisotropy.  The fluid flow around the first object thus affects the second.  The anisotropy of the particles gives rise to new effects like the one shown here: the particles have become trapped in each other's flow.  Such trapping doesn't occur for isotropic particles.  We discuss In Sec. \ref{sec:HydrodynamicInteraction} the anisotropic tensors needed to describe this interaction.  
                
                In looking at Figures \ref{fig:plasticCrumb} and \ref{fig:TomerOrbits}, it's natural to ask what are the conditions for these behaviors and how strong they might be for a given object.  As noted above, these behaviors depend on how the body's asymmetric shape affects the fluid's motion. Thus we need a means to explore strongly asymmetric shapes.   To this end, we use ``Stokeslet objects"---assemblies of point sources of force on the fluid.  Such objects were invented by Kirkwood and Riseman\cite{Kirkwood:1948bf}  to describe tenuous objects like polymers and to determine how external flow penetrates into them. We will discuss Stokeslet objects in more detail in Sec. \ref{sec:StokesletObjects}. Stokeslet objects usefully approximate solid objects when local smoothness of the surface is not important; they complement the conventional boundary element methods \cite{Youngren:1975jx}.  One may determine the flow around \eg a pair of Stokeslet objects by simple linear algebra operations.  This in turn determines the motion and interactions of the objects.  Several results described below use these Stokeslet objects.  
 
 \begin{figure}[htbp]
 \begin{center}
\includegraphics[width=5in]{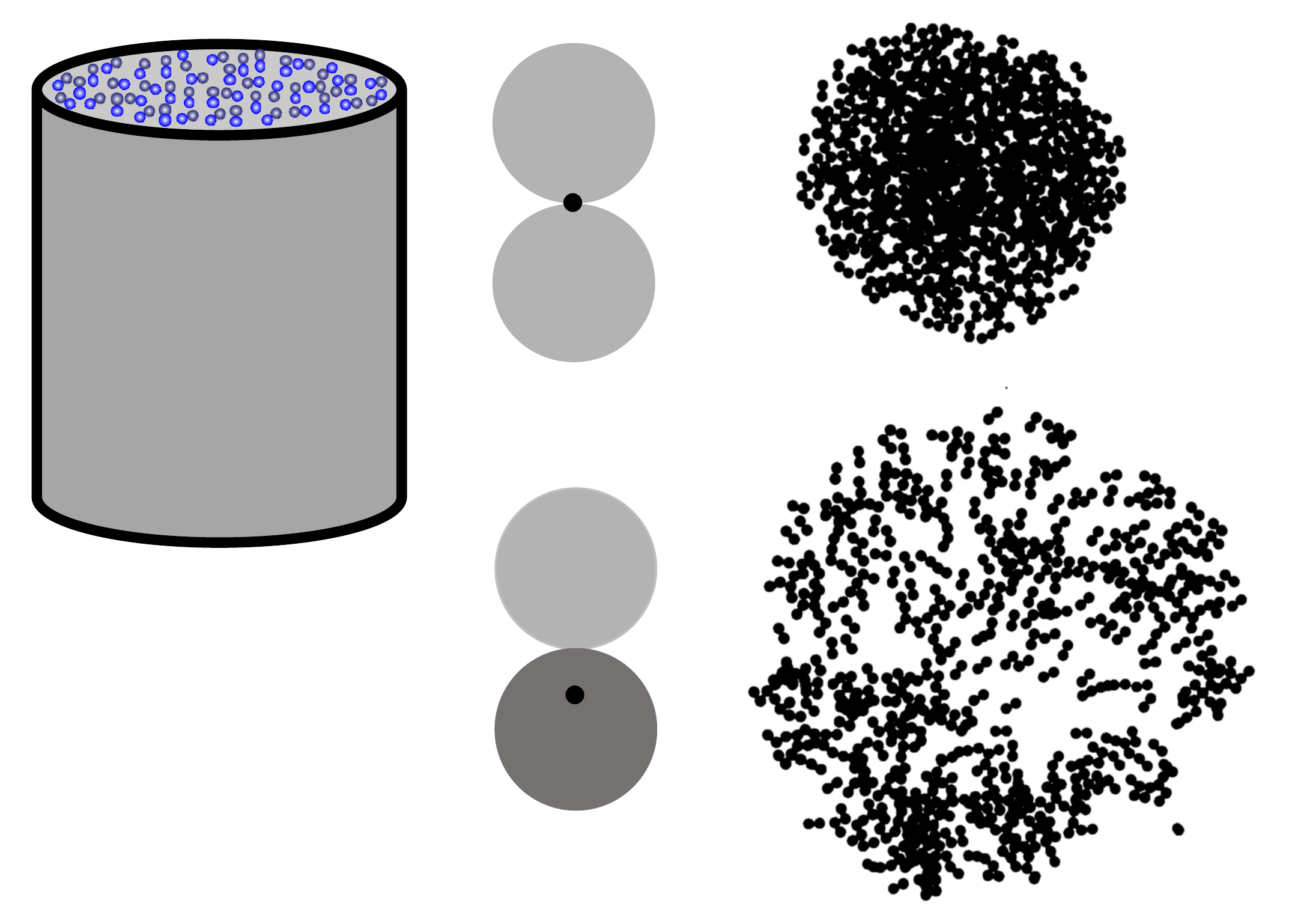}
\caption{Dependence of collective sedimentation properties on object shape, after \cite{Ma:2019zm}.  Left: collection of identical heavy dimers of millimeter-sized spheres at the top of a sealed cylindrical container filled with silicon oil.  The particles sediment to the bottom.  Middle column: shows a homodimer with center of mass in the middle and heterodimer with center of mass below the middle; this is its equilibrium orientation in sedimentation.  Right column: pattern of dimers after they have sedimented to the bottom of the container.  Heterodimers show mutual repulsion. Images courtesy of J. Burton and X. Ma.}
\label{fig:BurtonDimers}
\end{center}
\end{figure}
               
                The examples above show motions of individual bodies.  These shaped bodies also affect the collective motion of their embedding fluid.  Figure \ref{fig:BurtonDimers} illustrates collective effects of shaped particles on fluid behavior.  Fig. \ref{fig:sedimentation} shows how a dispersion of shaped sedimenting particles can suppress random density fluctuations.  This suppression makes a ``hyperuniform" state that is qualitatively more uniform than a random dispersion of particles\cite{Goldfriend:2017mz}, as discussed in Sec. \ref{sec:collective}.  

\begin{figure}[htbp]
\begin{center}
\includegraphics[width=1in]{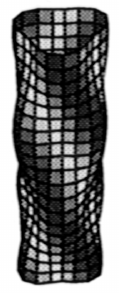}\hskip 1in
\includegraphics[width=2in]{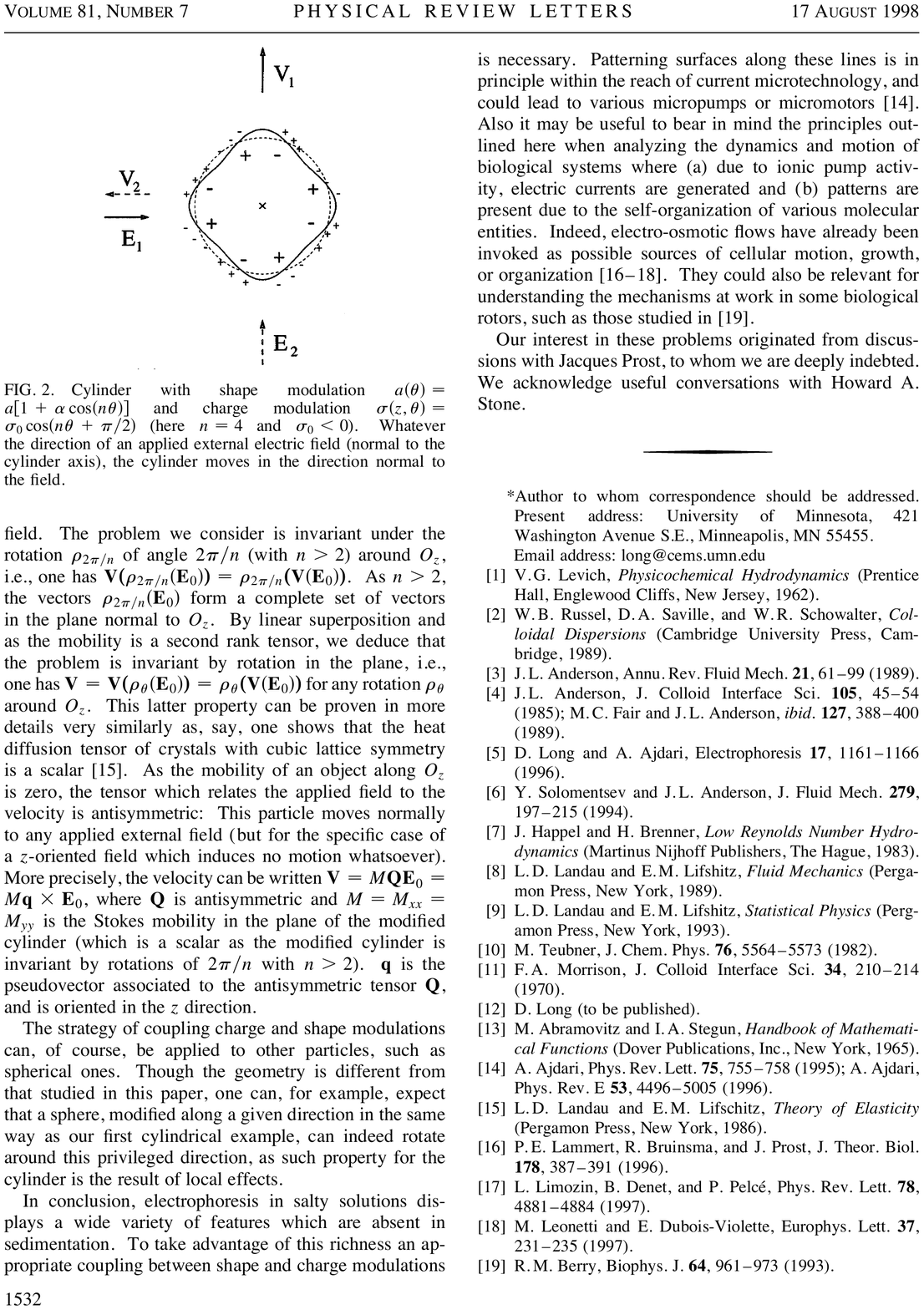}
\end{center}
\caption{Chiral electrophoretic motions of a patterned cylinder, after \cite{Long:1998fk}. 
left: cylinder with helical charge patterning (dark shading) and nonhelical shape distortion. In a vertical field, it moves with the helicity of a left-handed screw. right: cross-section of another cylinder with a four-fold pattern of charge and shape modulation. It migrates at right-angles to a transverse applied field as shown, with the velocity directed at 90 degrees to the left of the electric field.}
\label{fig:electrophoresis}
\end{figure}
                Figure \ref{fig:electrophoresis} shows the effect of a different kind of forcing. Here a charged body in a conducting solution is moving in a constant external electric field.  This kind of chiral response is qualitatively different from the cases treated above.  As with the gravitational forcing above, this motion is caused by an external vector.  But unlike gravity this electric field exerts no net force on the system, which is overall neutral.  Thus the connection between the object's geometry and its motion is quite different from the gravity case.  Normally the charge on a colloidal object is considered to be uniform.  Then even the most asymmetric or chiral shape can only move as though it were a sphere\cite{Morrison:1970ph}.   The authors of the figure showed that nonuniform charging can yield several anomalous motions\cite{Long:1998fk}.  Recently the Stokeslet-object representation has been used to quantify these anomalous tensorial responses in generic cases \cite{Braverman:2020oq}. 
                         
                The propulsion that occurs in electrophoresis comes from the Coulomb forces on the charged surface of the object and the opposing force on the adjacent screening charge in the fluid.  The stress induced by these equal and opposite forces causes the fluid to slip over the surface.  This slip occurs only where the surface is charged and its direction depends on the direction of the local electric field at that point.   Analogous slip flows can be caused by thermal or chemical gradients near the object.  These phenomena are known as phoresis\cite{Anderson:1989kx}.  
                
                In all these cases the motion is proportional to a vector driving influence.  As such they all share a common description: the velocity is a tensor times the driving vector.  In linear response, this tensor is independent of the driving vector; any anisotropy in it comes from anisotropy of the object itself.  The angular velocity of the object must be proportional to another such tensor.  The two tensors are analogous to the mobility tensors mentioned above for motion under gravity.  As with gravitational forcing, the rotational response changes the orientation, creating a complex rotational response.  One should thus expect any vectorial driving to create analogous rotational responses.
                
                Propulsion caused by an imposed slip velocity at the surface is more general than phoresis, where a simple external gradient is the cause of the slip.  The slip can be caused by a scalar influence such as a reactant in the solution that induces a nonuniform reaction on the object, as noted below.  It can occur in biological objects by means of mechanical swimming motions of structures on the surface.  Such forms of ``active particle" motion are of great current interest\cite{Marchetti:2013pi}.  The common property of imposed slip velocity sets constraints on the active motion that can occur.  The ideas in this review can help to identify these constraints.

The richness of dynamic behaviors exhibited by asymmetric objects in
viscous fluids can be used to develop ``steerable colloids"\,---\,\ie
colloids whose motions are controlled in intricate and programmable
ways. These capabilities may be used in biomedical applications, where
one aims, for example, to steer drug-carrying particles to specific
locations, or in the development of micro-robots. A large body of
research has been devoted to such systems. Here are a few examples.

Just as a forced chiral colloid would rotate, a torqued chiral colloid
would translate. Detailed experimental and theoretical studies have
investigated such propulsion of chiral magnetic objects under a
rotating magnetic field
\cite{GhoshNanolett2009,TottoriAdvMat2012,KeavenyNanolett2013,VachNanolett2015,MorozovNanoscale2014a,MorozovNanoscale2014b,WalkerNanolett2015,SachsPRE2018,MirzaeSciRobot2018}. The
chirality of these colloids is affected not only by their shape but
also by the intrinsic direction of their magnetic moment. The possible
usefulness of such colloids \cite{TottoriAdvMat2012} has motivated
attempts to optimize their shape for maximum propulsion efficiency
\cite{KeavenyNanolett2013,WalkerNanolett2015,VachNanolett2015}.

Another large class of systems involves particles with asymmetric
physical properties. 
One such property is chemical reactivity. An asymmetric distribution of reactants on the particle surface can cause asymmetric surface flows, leading to particle motion \cite{MoranAnnRevFluidMech2017}. Another property is dielectric anisotropy.  This enables particles to be
steered by an alternating electric field \cite{Lee:2019ye}. Finally, particles with asymmetric shapes or mass distributions may align
with a temperature gradient. Such particles can be steered by
thermophoresis \cite{TanSoftMatter2017,GardinPCCP2019,GittusEPJE2019}.

In this review we confine ourselves to a narrow sliver of driven motion phenomena in colloidal fluids where particle asymmetry plays a role.  We don't treat phenomena of active driving \cite{Marchetti:2013pi}, where the direction of motion is a built-in feature of the particle's structure.  We omit effects nonlinear in driving strength, such as rotational instability \cite{Quincke:1896fk} or the response of polarizable objects to an external field \cite{SachsPRE2018}. We omit chiral driving, and its generalized hydrodynamics\cite{Zuiden:2016gf}.  We omit phase separation phenomena caused by interaction \cite{Toner:2005rt,Vicsek:1995yq} or by circulating flow\cite{Cartwright:2010mz}.  We omit rheological and spatial organization such as shear thickening\cite{James:2019qr,Wyart:2014hl}, arising from packing constraints.  We say little about particular shapes, such as propellers\cite{Makino:2005ty,Andersen:2005ul}, or internal flexibility \cite{Shelley:2011fr} that give remarkable special motions.
We also pay little attention to advances in numerical methods that treat difficult-to-reach regimes such as closely-approaching surfaces \cite{Marenne:2017uo}. 
Instead we focus on the most basic, linear response phenomena that inevitably occur when particles are asymmetric.  We place extra emphasis on dispersions of identically-shaped particles, since these arise increasingly in practice and give remarkable collective behavior.  

                In what follows we first review the kinds of broken symmetry required of an object in order for it to show anisotropic or chiral response (Sec. \ref{sec:Symmetry}).  There we discuss what kinds of motion can occur when it is governed by mobility tensors.   Next in Sec. \ref{sec:ConstantForce} we specialize to motion of a single object induced by external forcing that adds momentum to the fluid---\eg sedimentation---first with constant forcing, then with time-dependent forcing.  In Sec. \ref{sec:NoForce} we consider the complementary situation where the object moves without exerting force or torque on the fluid.  This includes the case of motion induced by a shear flow and the case of phoretic driving induced by a slip layer on the surface, \eg electrophoresis. In Sec. \ref{sec:StokesletObjects} we show how real objects may be represented by the point assemblies called Stokeslet objects. In Sec. \ref{sec:HydrodynamicInteraction} we consider the hydrodynamic interaction between two driven particles and explore example motions numerically using Stokeslet objects. Then Sec. \ref{sec:collective} treats collective hydrodynamic interactions of an entire dispersion.  The Discussion to follow reviews open questions and future ways to use these distinctive driven motions.

\section{Symmetries and linear response}\label{sec:Symmetry} 
\setcounter{equation}{0}
\setcounter{figure}{0}

					
					This review is about how asymmetric objects interact with their surrounding fluid to produce asymmetric motions.  In the Introduction we emphasized a particular kind of asymmetry---chirality.  Objects, motions and flow fields can all have chiral asymmetry as well as other kinds of asymmetry.  In this section we review the conventional categories of asymmetry and how a given type of asymmetry in an object can lead to asymmetry of motion.

					All the fluid properties we will consider are in the regime of arbitrarily slow flow, or ``creeping flow".  Here the laws of hydrodynamics are at their simplest. The asymmetries we will encounter result directly from  the objects and the external forces rather than from the spontaneous asymmetries arising from the fluid itself.  
					
					If the forces causing flow are weak enough, one expects that the motion is proportional to the forces.  Since all motion dies out over time in the absence of forcing, one expects slowly varying forces to affect the motion only at the time $t$ when the force is applied.  The creeping flow regime can be defined as the regime where both of these statements are true.  
					
					Creeping flow is {\em quasistatic}.  That means that reducing all the forces by a common factor has the same effect as slowing the time scale by the same factor.  The sequence of configurations is the same independent of the speed.  Likewise, creeping flow is {\em time reversible}.  Reversing the forces simply reverses the motion.  Asymmetric particles in sedimentation and other forms of creeping flow are thoroughly treated in the monographs by Happel and Brenner \cite{Happel-Brenner} and by Kim and Karilla \cite{Kim:2005fr}.

					\subsection{Spatial symmetries of an object}  \label{sec:ObjectSymmetries}
					The shape of any object can be specified by giving the set of points in space that the object occupies.  Transformations such as translations and rotations move every point of the object into some other point.  These two transformations leave the distance between any pair of points unchanged---they ``preserve" distances. Thus they do not change the shape of the object.  Further, any translation or rotation can be performed continuously, in many tiny increments. Conversely, all continuous transformations that preserve distances are combinations of translations and rotations.  Now, other transformations also preserve distances, but cannot be performed continuously in three dimensions.  One example is a mirror transformation that takes every point $(x, y, z)$ into its mirror image $(-x, y, z)$. An object is said to have ``mirror symmetry" if all of its points transform into other points of the object. This mirror transformation can be made by first inverting space ($(x, y , z)\goesto (-x, -y, -z)$) and then rotating by 180 degrees about the $x$ axis, turning $(-y, -z)$ back to $(y, z)$.  In fact any distance-preserving transformation can be performed by translations, rotations and/or a spatial inversion.   An object is said to be {\em chiral} if the spatially inverted object cannot be continuously transformed into the original object, \ie by rotations and translations.  
					
					We'll be concerned with several familiar symmetries of objects.
\begin{itemize}
\item				A sphere: Any rotation about the center of the sphere leaves it invariant.  Spatial inversion about the center also leaves it invariant.  It is not chiral.

\item				A sphere with a marked point.  Any rotation about an axis through the center and the mark leaves this sphere invariant.  Inversion transfers the mark to the opposite side.  Many possible 180 degree rotations about the center can bring the mark to its original position.  It is not chiral.

\item				A sphere with two identical marked points.   One 180 degree rotation interchanges the two marked points.  Its axis contains the center and the midpoint between the marks.  This is the only rotational symmetry.  Inversion places the points on the opposite side of the sphere.  Either of two possible 180 degree rotations restore the points.  Thus this sphere is not chiral

\item				A sphere with two different marked points.  There is no rotational symmetry.  After inversion, one possible rotation can restore the positions of the two marks.  It is not chiral.

\item				A sphere with three different marked points.  There is no rotational symmetry.  After inversion there is one rotation that can restore any pair.  But no rotation restores all three unless the third point is midway between the other two.  Otherwise, this object is chiral.

\item 				An arrow or nail shape: it has complete rotational symmetry by any angle about its axis, and no other axis.  It necessarily has reflection symmetry for any mirror plane including this axis.  It has no inversion symmetry.  However a spatial inversion about a point on the axis followed by a 180 degree rotation at the same point perpendicular to the axis is a symmetry.	It is not chiral.
			
\item 				A rectangular prism or box: it has three reflection symmetries through the midpoint parallel to each face.  It also has inversion symmetry, since inversion amounts to a successive reflection in each of the three planes.  An ellipsoid has the same symmetry.  A symmetric matrix that maps one vector into another also transforms when the space is transformed.  It also has three orthogonal reflection planes that leave the matrix unchanged.  Such a matrix thus has the same symmetry as the box.  These objects are all not chiral.

\item				A box with a marked point: The situation is different from a sphere with a marked point.  The rotations used to bring the inverted marked sphere back to its uninverted state are not available.  Thus unless the mark is at a symmetrical point, the inverted box cannot be restored.  The marked box is in general chiral.  The same reasoning applies to a marked ellipsoid, which has the same symmetries as a marked box.
				
\item 				A vertical tennis racket: it has only two reflection planes: front to back and left to right.  Inversion also interchanges the top and bottom, which changes the set of points.   Still, one may restore the original top and bottom by a rotation of 180 degrees.  Thus, this object is not chiral.  
				
\item 				A curve in the shape of the letter ``h" in a plane: it has no rotational symmetry.  Also, there is no inversion point that allows it to be rotated back to its original position.  It is a chiral object.  This is true for a generic curve  confined to two dimensions.
				
\item 				A curve in the shape of the letter ``h" in space:  Now there is a mirror symmetry in the plane of the curve.  An inversion makes a backward ``h" as in the last example.  But rotation from front to back also makes this backward ``h".  Doing both operations in sequence restores the original ``h".  This is true for any planar curve in space.  The curve is not chiral in three dimensions.  

\item				A helical curve: It has 180-degree rotational symmetry about a line perpendicular to the axis of the helix and passing through the axis and a point of the helix.  However, it has no rotational symmetry about the helix axis.  Instead it has a more subtle symmetry.  For any small rotation about the axis there is a small translation along the axis that makes the helix invariant.  The amount of translation needed as well as its direction depends on the shape of the helix. The ratio of the translation to the rotation is called the helicity.  Many objects have this rotation-with-translation symmetry, and thus have a helicity.  By repeating these small transformations many times, one may make rotations of any size.  

					A helix is chiral.  If one inverts the space, the inverted helix has a helicity with opposite sign compared to the original helix.   No translation or rotation can change this sign.  The sign of the helicity thus distinguishes two classes of helix.  To decide whether two objects are in the same class, one aligns their axes, performs the same small rotation on each, and then compares the translations needed to leave each object invariant.  If the directions are the same, the two are in the same class.  The classes can be labeled by taking any common artifact that associates a rotation with a specific direction of translation along the same axis.  The convention is to use the human right hand as the artifact.  A right handed helix is one where the curling fingers and the extended thumb have the same relation as the helical rotation and translation.  Right-handed helices are said to have positive helicity.  
			
			\end{itemize}		
					Below we discuss more general objects that have a specific proportionality between translations (or some other vector influence) and rotations\cite{Efrati:2013fk}.  These need not have a definite helicity and need not be chiral.  
					
					\subsection{Symmetries of mathematical quantities}\label{sec:symmetryMath}
				Quantities describing the shape of objects or their motion, in general also change when the space is inverted.  The simplest of these is a vector, such as a displacement upward by one meter.  Rotation about a vertical axis does not change its magnitude or direction; thus it does not change the vector.  Any other rotation does change its direction and thus changes the vector.  In this way the vector is unlike an object with similar form, such as the nail mentioned above.   An upward pointing nail rotated to point down is the same object.  However, an upward pointing vector rotated to point down is not the same vector.  Likewise, a vector after spatial inversion is not the same vector.  Its relation to the original can now be stated quantitatively.  The inverted vector is the negative of the original one.  This is true of all vectors: they are ``odd under inversion".  
				
				The behavior studied below is based on rotation of objects in response to some external influence.  A rotation is a linear transformation from a given vector to its transformed vector. It may be represented by a matrix in a given co-ordinate system.  The rotation transformation itself is a tensor.  To see how a rotation changes under inversion, we consider a small rotation that turns the positive $x$ axis towards the positive $y$ axis, leaving the $z$ axis unchanged.  After inversion this rotation rotates the negative $x$ axis into the negative $y$ axis.  Evidently this inverted rotation turns the positive $x$ into positive $y$; like the original rotation.  The rotation operation is invariant or ``even" under inversion.   Below, we consider time-dependent rotations, represented by an angular velocity $\Omega$ pointing along the axis of rotation. Because rotations are unaffected by a spatial inversion, this $\Omega$ cannot be affected, either.  It is called an axial vector or pseudovector.  Axial vectors are no different from ordinary vectors under rotations; both rotate in the same direction by the same amount.  Thus we represent both axial vectors and ordinary ones by a vector sign: $\vector \Omega$ or $\vector F$.
				
				The motions considered in the Introduction are rotational responses to vector influences like forces and electric fields.   A rotational response is an angular velocity $\vector \Omega$ proportional to the external vector.  Thus each response can be represented by a tensor; in a given co-ordinate system it is a matrix of numbers.  The properties of these ``mobility tensors" for sedimentation as well as examples of specific objects are treated extensively in Refs.~\cite{Happel-Brenner,Kim:2005fr}.     The tensor is a property of the object; if the object rotates, the corresponding tensor is given by conventional transformation laws [\cite{Kelly:2013qe} Ch. 1.13], \cf Sec. \ref{sec:MotionLaws}.
				
				We now ask how a rotational response tensor should transform under a spatial inversion.  The external vector certainly reverses sign.  The $\vector \Omega$ does not, as described above.  Thus the proportionality between the external vector and the $\vector \Omega$ must reverse sign.  A rotational response tensor is necessarily {\em odd} under inversion.  Such tensors are called pseudotensors.  
				
				In our discussion of helices above, we saw that some chiral objects may be assigned a definite positive or negative chirality according to their resemblance to right-handed or left-handed helices.  We now ask whether rotational response tensors can be classified in the same way.  Such a tensor gives a proportionality between vectors and rotations, but it is not reducible to a single proportionality constant.  Arbitrary vectors give $\vector \Omega$'s that have components along the vector and hence a kind of helicity.  However, such helicities depend in general on the direction of the vector and thus need not provide a measure of chirality intrinsic to the response.  To find such intrinsic properties, we consider the rotationally invariant properties of the response tensor.  The one such invariant that is linear in the matrix elements is the trace---the sum of the diagonal elements.  If this trace is nonzero, it indicates an intrinsically chiral property corresponding to right or left handed helix.  Averaged over orientations, the object rotates in response to the external vector in a direction like one of the two types of helix.  			
				
				How is the chirality of an object related to the helicity of its response tensor? Evidently a non-chiral object must have a non-chiral response tensor, and thus its response can have no intrinsic helicity.  A chiral object may have a chiral response tensor, but this tensor need not have a trace.  Thus it need not have an intrinsic helicity.  Yet any small generic modification to such an object in general has a nonzero effect on any element of any tensor property of that object, including the trace of the response matrix.  Only exceptional chiral objects would happen to have a traceless response tensor.  Thus a general chiral object would have an intrinsic positive or negative helicity.
						
				\subsection{Response tensors of a rigid body} \label{sec:Response}
				
				Here we review the mathematics governing the motion of a rigid body under the action of some perturbing vector, denoted $\vector X$. If the response depends continuously and smoothly on $\vector X$, then in general the linear velocity $\vector V$ and angular velocity $\vector \Omega$ are proportional to $\vector X$ (\cf Sec. \ref{sec:Introduction}).  Any such response is a tensor, and its specific matrix in a given basis depends on the orientation of the object relative to that basis.  As we will encounter many such tensors in what follows, we adopt a general definition for them.   We shall denote the tensor giving the response $\vector Y$ to a vector $\vector X$ by $\RT_{Y X}$.  We denote the result of operating on $\vector X$ with tensor $\RT$ by $\RT \cdot X$. Thus
\begin{equation}
\vector Y = \RT_{Y X} \cdot\vector X
\end{equation}

The most basic response is the motion induced by an external force $\vector F$ or torque $\vector \tau$.  These are the responses illustrated in Fig. \ref{fig:plasticCrumb}.  The complete linear response to an external force $\vector F$ and torque $\vector \tau$ on an object is given by 
\begin{eqnarray}
\vector V = \RT_{V F} \cdot \vector F  ~+~ \RT_{V \tau} \cdot \vector \tau \nonumber\\
\vector \Omega = \RT_{\Omega F} \cdot \vector F ~+~ \RT_{\Omega \tau} \cdot \vector \tau
\end{eqnarray}
or in matrix form
\begin{equation}\label{eq:bigmateqn}
\left [ \matrix{\vector V \cr \vector \Omega} \right ] = \left [\matrix{\RT_{V F}& \RT_{V \tau} \cr
\RT_{\Omega F} &\RT_{\Omega \tau}} \right ] \left [ \matrix{ \vector F \cr \vector \tau }\right ]
\end{equation}
Happel and Brenner\cite{Happel-Brenner} consider the tensors for the inverse relations giving $\vector F$ and $\vector \tau$ in terms of $\vector V$ and $\vector \Omega$.  Their $\boldsymbol{\mathsf{K}}$ is $\RT_{F V}$ in our notation.  Their $\boldsymbol{\Omega}$ is our $\RT_{\tau \Omega}$ and their $\boldsymbol{\mathsf{C}}$ is our $\RT_{\tau V}$. The relation between these alternative sets of mobility vs.\ resistance tensors is further discussed in Ref.~\cite{Kim:2005fr}, Chapter 5. Since the dissipated power $\vector F \cdot \vector V + \vector \tau \cdot \vector \Omega$ must be non-negative for any $\vector F$ and $\vector \tau$, the diagonal tensors $\RT_{V F}$ and $\RT_{\Omega \tau}$ must each be symmetric, while $\RT_{\Omega F}$ and $\RT_{\tau V}$ (mixed Latin-Greek) must be transposes of each other.  (Further conditions are needed as well in order to assure this positivity\cite{Happel-Brenner, Kim:2005fr}).  

The diagonal tensors $\RT_{V F}$ and $\RT_{\Omega \tau}$ are invariant under spatial inversion. For example both $\vector V$ and $\vector F$ change signs, so their proportionality does not.  Thus they are insensitive to any chirality in the object.  Only the off-diagonal pseudotensors $\RT_{\Omega F}$ and its transpose $\RT_{V \tau}$ change sign under spatial inversion and can give a chiral response.  

These response tensors evidently provide the proportionalities relating $(\vector F, \vector \tau)$ to $(\vector V, \vector \Omega)$.  Thus given $(\vector F, \vector \tau)$ one can determine $(\vector V,  \vector \Omega)$.  More generally, any two of $(\vector F,\vector \tau, \vector V, \vector \Omega)$ determine the other two.  Thus, \eg if $\vector F$ and $\vector \Omega$ are known, the corresponding $\vector \tau$ and $\vector V$ may be found in terms of the same $\RT$ matrices.
The needed relation has the same form for any pair of quantities $(Q, P)$  linearly related to another pair $(q, p)$, via a proportionality matrix
\begin{equation}
\left [\matrix{Q \cr  P
} \right ] = 
\left [\matrix{a & b \cr c & d}
 \right ]
\left [\matrix{q \cr  p} \right ] 
, \end{equation}
Our example requires interchanging the two second quantities, $p$ and $P$.  Simple algebra gives the corresponding matrix expressing $(Q, p)$ in terms of $(q, P)$:
\begin{equation}
\left [ \matrix{Q \cr p}
 \right ]= 
 \left [\matrix{a - b ~d^{-1}~ c & & b~ d^{-1} \cr -d^{-1}~ c~  & & d^{-1}}
 \right ] 
\left [\matrix{q \cr  P}
 \right ]
\end{equation}
The same expression holds if the quantities $q, p, Q, P$ are vectors of the same dimension.  Each of the coefficients $a, b, c, d$ is then a square matrix.  

Applying this algebra to our case where the quantities $(q, p, Q, P)$ are respectively $(\vector V, \vector \tau, \vector V, \vector \Omega)$ yields
\begin{equation}
\vector V= (a - b ~d^{-1}~ c) \vector F + (b~ d^{-1}) \vector \Omega
; \quad
\vector \tau = (-d^{-1}~ c) \vector F + d^{-1} \vector \Omega
. \end{equation}
Then, substituting $(\RT_{V F}, \RT_{V \tau}, \RT_{\Omega F}, \RT_{\Omega \tau})$ for  $(a, b, c, d)$ we obtain 
\begin{eqnarray}
\label{eq:VofOmega}
\vector V= (\RT_{V F} - \RT_{V \tau} ~\RT_{\Omega \tau}{}\kern-3pt^{-1}~ \RT_{\Omega F}) \vector F + (\RT_{V \tau}~ \RT_{\Omega \tau}{}\kern-3pt^{-1}) \vector \Omega \\
\vector \tau = (-\RT_{\Omega \tau}{}\kern-3pt^{-1}~ \RT_{\Omega F}) \vector F + \RT_{\Omega \tau}{}\kern-3pt^{-1} \vector \Omega
\end{eqnarray}
A similar procedure gives analogous formulae for any two of the four vectors in terms of the other two. 

These formulas all involve taking the inverse of a $\RT$ matrix.  This raises the question whether the formulae are well defined. Happily, the only inverses that appear in these formulae are the ``diagonal blocks" $\RT_{V F}$ and $\RT_{\Omega \tau}$.  As noted above, these are necessarily positive-definite and thus invertible.

The case shown above is important in a number of current experiments, in which ferromagnetic particles are forced with a rotating magnetic field \cite{SachsPRE2018, Soni:2019ty}.  Here the forcing is strong enough that the magnetic moment of each particle must align with the external field.  Thus the angular velocity $\vector \Omega$ of the particle is obliged to match that of the field.  When the magnetic moment of the particle is not along a symmetry axis, it is in general a chiral object, as noted in the previous section.  Accordingly, it in general has a chiral response, translating along a helical path whose helicity is dictated by the particle.  When there is no external force $\vector F$, Eq.(\ref{eq:VofOmega}) gives the helicity $V/\Omega$: $V/\Omega =\RT_{V \tau} \RT_{\Omega \tau}{}\kern-3pt^{-1}$ \cite{SachsPRE2018}.

				\subsection{Motion induced by response tensors}\label{sec:MotionLaws}
				
				Though the motion is governed by fixed response tensors, the motion under constant force and torque need not be constant. This is because the result of \eg $\vector V = \RT_{V X}\cdot \vector X$ changes if the orientation of the object changes.  Thus $\vector V$ can change even if $\vector X$ does not.  If $\vector \Omega$ is nonzero, then the orientation is changing with time.  In a short time $dt$ the orientation has changed by a small rotation denoted $\Rop(dt)$.  This rotation transforms any response matrix $\RT_{ }$ according to
\begin{equation}\label{eq:MopRop}
\RT(t+dt) = \Rop(dt) \cdot \RT(t)\cdot  \Rop^T(dt) 
\end{equation}
In the limit of small time interval $dt$, $\Rop$ approaches the identity matrix $\IdentityMatrix$:
\begin{eqnarray}\label{eq:Rop}
\Rop(dt) =& ~\IdentityMatrix + dt~ \xop{\Omega}~ ; \nonumber\\
\Rop^T(dt) =& \Rop(-dt) .
\end{eqnarray}
Here $\xop{\Omega}$ is the antisymmetric matrix corresponding to the vector $\vector \Omega$, defined by 
\begin{equation}\label{eq:Omegaop}
\xop{\Omega} \cdot \vector A \definedas \vector \Omega \cross \vector A
\end{equation} 
for any vector  $\vector A$.  
Using Eqs. (\ref{eq:MopRop}) and (\ref{eq:Rop}), we find the time derivative $\dot{\RT}$:
\begin{equation} 
\dot{\RT}(t) =  \xop{\Omega}\cdot \RT(t) ~ - ~ \RT(t)\cdot \xop{\Omega} 
\end{equation}
or, in commutator notation
\begin{equation}\label{eq:RTdot} 
\dot{\RT} = [\xop{\Omega}, \RT(t)] 
\end{equation}

This allows us to infer any $\RT(t)$ whenever $\vector \Omega(t)$ and hence $\xop{\Omega}(t)$ is known.  It is sufficient to consider a single driving vector $\vector X$.  Then to determine $\vector \Omega(t) = \RT_{\Omega X}(t)\cdot \vector X$ we use the closed pair of equations
\begin{eqnarray} \label{eq:RTOmegaXdot}
\dot{\RT}_{\Omega X} = [\xop{\Omega}, \RT_{\Omega X}(t)]  \nonumber\\ 
\hbox{where} \nonumber\\
\vector \Omega(t) = \RT_{\Omega X}(t)\cdot \vector X 
\end{eqnarray}
Thus $\dot{\RT}_{\Omega X}$ depends quadratically, not linearly, on $\RT_{\Omega X}$ itself.

\begin{figure}[htbp]
\includegraphics[width=\textwidth]{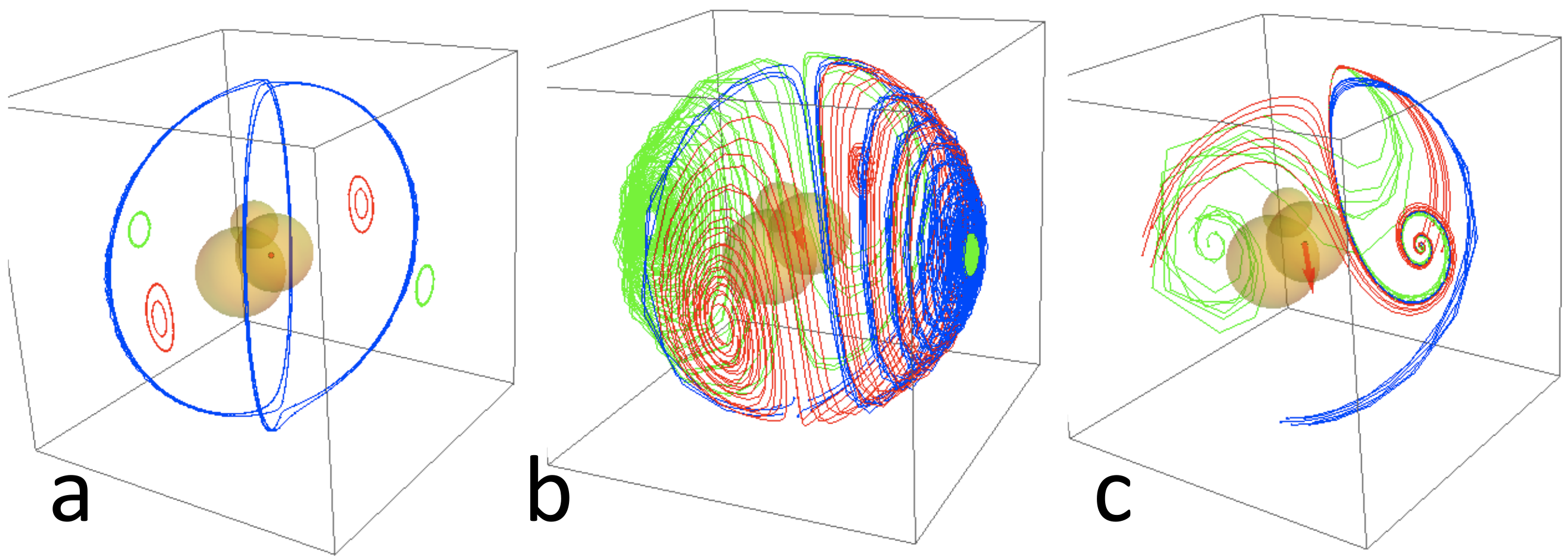}
\caption{\label{fig:Orbits} 
Example motions induced by $\RT_{\Omega X}$ under constant driving vector $\vector X$ showing the effect of the antisymmetric part of $\RT_{\Omega X}$, after \cite{Braverman:2020oq}. 
Cluster of shaded spheres in the middle of each cube represents the object.  Colored lines are orbits traced by $\vector X$ as viewed in the frame of the object, beginning near each fixed point. Color of orbit indicates the eigen-axis from which it started. 
a) no antisymmetric part.  All orbits are closed. Two pairs of orbits (on the side faces of the cube) remain localized near their starting points. Orbits starting near the bottom and top fixed point form a single connected orbit that oscillates between the two starting points.  
b) small antisymmetric part.  Orbits starting near the three unstable fixed points spiral away from their starting point.  All converge to the stable fixed point on the right.  Orbits near the right and rear stable fixed points converge to the local stable fixed point. Since the motion of any point on an orbit depends only on its location on the sphere, no two orbits may cross.  
c) large antisymmetric part.  All starting points converge to the stable fixed point on the right. }
\end{figure}

Even when the driving $\vector X$ is constant, the  resulting motion is rich.  This is easier to see by choosing a frame of reference fixed in the body, so that the response tensors are constant.  In this frame, the $\vector X$ rotates in time according to 
\begin{equation}\label{eq:Xdot} 
\dot{\vector X} = -\vector \Omega \cross \vector X = -\left (\RT_{\Omega X}\cdot \vector X \right ) \cross \vector X.
\end{equation}  
The motion does not change $|X|$; $\vector X$ thus moves on a sphere in time.  We note that if $\vector X$ happens to be in an eigendirection of $\RT_{\Omega X}$, \ie $\vector \Omega \parallel \vector X$, then $\vector X$ remains fixed.  Conversely, if $\vector X$ and $\RT_{\Omega X}\cdot \vector X$ are both nonzero, then any fixed-point $\vector X$ with $\dot{\vector X} = 0$ must be an eigenvector of $\RT_{\Omega X}$.  

This same equation describes the familiar case of an object of arbitrary shape rotating in free space\cite{Gonzalez:2004la}.  Here the angular momentum vector $\vector L$ plays the role of the constant driving vector $\vector X$.  Here too the rotational velocity $\vector \Omega$ is proportional to the angular momentum via the inverse inertial tensor $\Iop^{-1}$: $\vector \Omega  = \Iop^{-1}\cdot \vector L$.  This is a special case of Eq. (\ref{eq:Xdot}) in which the tensor is symmetric.  Fig. \ref{fig:Orbits}a shows the motion that results in this case.  Here there is a constant of the motion, namely, the kinetic energy ($ = \half \vector L \cdot \vector \Omega$ or $\half \vector L \cdot \Iop^{-1} \cdot \vector L$).  This constraint of constant energy forces the motion for any starting point to remain on a closed orbit on the sphere.  There are three pairs of fixed points where $\vector X$ remains constant.  These fixed points are in the (mutually orthogonal) eigen-directions of $\Iop^{-1}$.  The two points of a pair are equivalent, reflecting the mirror symmetry of $\Iop^{-1}$ about its eigen directions.  In general the three eigenvalues are distinct.  For the axes of largest and smallest energy any $\vector L$ near the fixed point remains near it: these are neutrally stable.  Points near the intermediate fixed axis remain in the vicinity of a large orbit that bridges the two fixed points.    No orbits can cross, since $\dot{\vector L}$ is a deterministic function of $\vector L$.  

For an asymmetric object in a fluid, the tensor $\RT_{\Omega X}$ can also be symmetric as discussed in Sec. \ref{sec:ShapeMass}.  Then the motion follows neutrally stable orbits, like the inertial case above.  But now the eigenvalues can have either sign.  That is, when the body is rotating about an eigen-axis,  the proportionality between the driving vector and the rotational response may have either sign.  As noted above, the eigenvalue is a helical signature, the same for both fixed points on the eigen-axis.  The object may have a net helicity if the rotational average of $\RT_{\Omega X}$ (\ie its trace) is nonzero.  For a non-chiral object, this average evidently must be zero  \cite{Efrati:2013fk}, \cf Sec. \ref{sec:symmetryMath}.

Now, $\RT_{\Omega X}$ need not be symmetric; there is in general an antisymmetric part.  Just as the vector $\vector \Omega$ can be expressed as an antisymmetric matrix, the antisymmetric part of $\RT_{\Omega X}$can be expressed as a vector, denoted $\vector d$.  This $\vector d$ points in a specific direction in the object.  With symmetric and antisymmetric parts present, the symmetry of $\RT_{\Omega X}$ is like that of a box with a distinguished mark on it, \cf Sec. \ref{sec:ObjectSymmetries}.  It is necessarily chiral apart from exceptional, symmetric choices of $\vector d$. The two fixed points on an eigen-axis are no longer equivalent.  When this antisymmetric part is nonzero, the orbits are no longer closed\cite{Moths:2013qf, Gonzalez:2004la}, as shown in Fig. \ref{fig:Orbits}b, c.  

The generic behavior for small antisymmetric part is shown in Fig. \ref{fig:Orbits}b.  As in the symmetric case of Fig. \ref{fig:Orbits}a, the response tensor $\RT_{\Omega X}$ has three eigen-axes---no longer orthogonal---for which $\vector \Omega \parallel \vector X$. They lie close to the eigen-axes of the symmetric part of $\RT_{\Omega X}$.  These eigen-axes remain as fixed points of the $\vector X$ motion, at which $\vector X$ is constant. \cf Eq. (\ref{eq:Xdot}).  The body simply rotates about the $\vector X$ axis at a rate $\Omega$ equal to the eigenvalue of $\RT_{\Omega X}$.  The ``pseudo energy" $\GG \definedas \half \vector X \cdot \RT_{\Omega X} \cdot \vector X$ is no longer constant but can slowly vary during the motion.   The two fixed points near the maximal-$\GG$ directions acquire opposite stability\cite{Gonzalez:2004la}.  That is, if $\vector X$ is near one of these fixed points, it converges to one of them and moves away from the other.  The $\vector d$ of the object determines which is the stable fixed point.  The same is true for the minimal-$\GG$ pair.  

These stable and unstable fixed points do not account for the qualitative behavior completely.  In some cases there is a stable fixed {\em orbit} [\cite{Happel-Brenner} Chap. 5-1] in addition to the two stable fixed points.  Finally, any small perturbation of the symmetric $\RT_{\Omega F}$ leads to {\em convergence} from a generic initial orientation of $\vector X$ to no more than three final states--- the two fixed points and possibly a fixed orbit.  In such a final orbit, $\vector X$ is not fixed but moves periodically around the orbit at a rate that varies with its position.  The final states depend only on the body, while the choice among the final states depends on the initial orientation.  We know of no general explanation of these final states or the general feature of convergence to discrete final states.  Section \ref{sec:SummaryForced} discusses this behavior in further detail.

As the antisymmetric part grows larger and larger, the mutual influence among the different fixed points also grows.  When the antisymmetric part grows arbitrarily large relative to the symmetric part, one finds \cite{Gonzalez:2004la} that there is only one stable fixed point and one real eigenvalue.  Its magnitude and sign are controlled by the symmetric part of $\RT_{\Omega X}$.  The final motion of this fixed point has a definite helicity, which may be nonzero even when the symmetric part has no net helicity. 
Section \ref{sec:ShapeMass} discusses how the antisymmetric part arises physically for a sedimenting object.  
		
\subsection{Translation and helicity}\label{sec:translation}

The preceding section showed how an object can adopt a continual rotation in response to a fixed driving vector $\vector X$. In general this same rotation creates helical motion of the object through space. We discuss this translating motion. In the body frame, the driving vector $\vector X$ has reached a fixed direction, denoted $\vector X^*$ and the angular velocity $\vector \Omega$ has reached a fixed direction $\vector \Omega^* \parallel \vector X^*$, as explained 
above. In this body frame, the velocity adopts a fixed value $\vector V^* = \RT_{V X}\cdot \vector X^*$.

In the lab frame, $\vector \Omega^*$ and $\vector X^*$ remain as in the body frame, but the velocity $\vector V$ is not in general constant. It rotates around $\vector \Omega^*$ (and $\vector X^*$) according to $\dot{\vector V} = \vector \Omega^* \cross \vector V^*$. The component of $\vector V$ along $\vector \Omega^*$, denoted $\vector V_\parallel$, remains fixed, while the component perpendicular to $\vector \Omega^*$, denoted $\vector V_\perp$, moves in a circle. Thus the object's origin migrates along a helical path, always moving at an angle $\alpha$ to the helix axis given by $\tan \alpha = |V_\perp|/V_\parallel$. This
is the motion seen in Fig. \ref{fig:plasticCrumb}a. The sign of the helicity is the sign of $\vector \Omega^* \cdot \vector V^*$. It can happen that the terminal rotation rate $\vector \Omega$ is zero. Nevertheless there is a nonzero terminal translation velocity $\vector V^*$ that is not in general parallel to $\vector X^*$. Thus even without helicity, the rotational dynamics enforces a particular ``glide angle" $\alpha$.

\section{Object under external force}\label{sec:ConstantForce} 
\setcounter{equation}{0}
\setcounter{figure}{0}

															
					Using the background from the last section, we now turn to the chiral behavior seen in Fig. \ref{fig:plasticCrumb}.  The elements needed to explain this behavior are long known and well understood\cite{Happel-Brenner, Doi:2005qv}.  Here we emphasize the generality of this chiral motion for asymmetric objects.    Such asymmetric objects are the rule rather than the exception in the world of colloidal particles.   Thus the chiral behavior is there to be exploited for controlling and probing these particles.  In this section we consider chiral motion induced by applying external force to noninteracting, asymmetric particles dispersed in a fluid.  The main practical case is sedimentation under gravity.  We ask what an observed colloidal motion can tell us about the particle's structure, how can the motion be exploited to manipulate the objects, and how these properties survive under experimental conditions. We begin with this latter question.

					\subsection{Orders of magnitude} \label{sec:Magnitudes}
				Chiral sedimentation occurs in principle for any size of object, from the molecular to the millimeter scale and above.  We focus on the colloidal length scale of a few microns because the potential utility is greatest at this scale\cite{Moths:2013mz,Moths:2013qf}.  Macroscopic objects like Fig. \ref{fig:plasticCrumb} can readily be probed and manipulated individually, so there is less benefit from inferring their structure via chiral motion.  Accordingly we concentrate on particles the size of a small cell or bacterium, about 1 micron in radius, in earth's gravity with a ten percent larger density than water. Such a particle requires several seconds to descend its own radius in water. 
As noted below, helicities are typically 0.1 revolution per radius or less.  Thus to make a rotation requires tens of seconds.  Meanwhile rotational diffusion is randomizing the orientation and obscuring the rotation.  The orientation becomes random in a few seconds for this sphere in water.  To attain the mutual alignment pictured in Fig. \ref{fig:plasticCrumb}b requires tens of rotations, hence, many seconds.  If the buoyancy is closer to neutral than the 10 percent assumed here, time for the needed rotations increases even further. Thus unassisted gravity is insufficient to take advantage of any chiral effects by a few orders of magnitude  for particles of this size.  Reducing the size $a$ of the particles only worsens the randomizing effects of diffusion.  One needs larger particles and/or stronger gravity, \eg via centrifugation.  Thus, the practical applicability of chiral response via sedimentation is limited.  Still this case serves as a simple context to explain the type of chiral responses that can be anticipated.  Sections to follow will present more practical realizations.
				
					\subsection{Effects of shape and of mass distribution}\label{sec:ShapeMass}
								
								
				The object's motion depends on its mass distribution as well as its shape. For a given shape and total mass, the total force $F$ on the object is fixed.  It acts on the body as if concentrated at a single point $\vector C$ in the object, called the center of force.  Redistributing this mass moves the center of force by a displacement denoted $\vector U$ .  This changes the torque by an amount $\Delta \vector\tau = \vector U \cross \vector F$.  This added torque changes the linear and the angular velocity. In particular the angular velocity $\vector \Omega$ changes by an amount $\RT_{\Omega \tau} \cdot [\vector U \cross \vector F]$. This change is controlled by the symmetric $\RT_{\Omega \tau}$ matrix.  We may rewrite the cross product using the matrix form of $\vector U$, \viz $\xop{U}$ from Eq. (\ref{eq:Omegaop}).  Then the change in $\vector \Omega$ becomes $\RT_{\Omega \tau} \cdot \xop{U} \cdot \vector F$. 

This shift in $\vector C$ alters the overall response of $\vector \Omega$ to $\vector F$. We denote the original responses as $(\RT_{\Omega F})_{\vector C}$ and $(\RT_{\Omega \tau})_{\vector C} $.  Then the shift from $\vector C$ to $\vector C + \vector U$ gives 
\begin{equation}\label{eq:shiftThm}
\vector \Omega = \left [ (\RT_{\Omega F})_{\vector C} + (\RT_{\Omega \tau})_{\vector C} \cdot \xop{U}  \right ]\cdot \vector F ~~\definedas~~ (\RT_{\Omega F})_{\vector C +\vector U} \cdot \vector F
\end{equation}
The two terms in $[...]$ in general have antisymmetric parts.  By solving a linear system one may in general find a $\vector U$ such that the antisymmetric part of $[...]$ vanishes.  When the force is applied at this point, the $\RT_{\Omega F}$ is symmetric and the motion is as shown in Fig. \ref{fig:Orbits}a.  This point is called the ``center of twist" \cite{Krapf:2009qp}.  We now take this point to be our origin, so that $\vector U$ is measured from the center of twist.  

As in Section \ref{sec:MotionLaws} some general features of the motion emerge independent of shape when the center of force is very close to the center of twist or very far from it.   When $\vector U$ is small, $(\RT_{\Omega F})_{\vector U}$ is nearly symmetric, and we expect three stable fixed points close to eigendirections of the symmetric $(\RT_{\Omega F})_0$, as described in Sec. \ref{sec:MotionLaws}. 

 Conversely, when $\vector U$ is large, there is only one stable fixed point, and it is arbitrarily close to the $\vector U$ direction, as we now discuss.  As we observed at Eq. (\ref{eq:Xdot}), a fixed-point $\vector F$ in the object's body frame must be an eigendirection of $(\RT_{\Omega F})_{\vector U}$.  It's convenient to denote these directions as unit vectors, \eg $\hat F$.  Thus we ask whether an eigendirection of $(\RT_{\Omega F})_{\vector U}$ approaches $\hat U$ when $U$ becomes large.  We may verify this by examining Eq. (\ref{eq:shiftThm}).  Using $\xop{U} \cdot \vector F = \vector U \cross \vector F$, this may be written
\begin{equation}\label{eq:Omega}
\vector \Omega =  (\RT_{\Omega F})_{0}\cdot \vector F + (\RT_{\Omega \tau})_{0} \cdot \vector U \cross \vector F
\end{equation}
For large $U$ the second term dominates.  However, this term evidently vanishes when $\vector F \parallel \hat U$.  Thus $\hat U$ is an eigen vector of the second term with eigenvalue 0.  The small first term can alter the eigendirection slightly to some direction denoted $\hat E_1$ near $\hat U$; this first term can also make the eigenvalue slightly nonzero.  Thus when $\vector F \parallel \hat E_1$,  the body rotates slowly around the $\hat E_1$ direction and $\vector F$ remains fixed.    

It remains to show that the $\hat E_1$ direction is the only stable fixed point for large $U$  \cite{Weinberger:1972kx}.   The situation is nearly the same as in Sec. \ref{sec:MotionLaws} above, where only the antisymmetric part of $\RT_{\Omega F}$ was taken to infinity.  Here we show this by observing that the $\vector F(t)\cdot \hat E_1$ must increase with time for every $\vector F$ not parallel to $\plusorminus \hat E_1$.  

Since the motion (Eq. \ref{eq:Xdot}) does not change the magnitude of $\vector F$, it suffices to consider the motion of the unit vector $\hat F$. Then whenever $\hat E_1 \cdot \dot{\vector F}(t) > 0$, $ \hat E_1 \cdot \hat F$ is obliged to increase until it reaches a maximum value at $\hat F = \hat E_1$ \cite{Weinberger:1972kx}. 
In terms of $\vector \Omega$ the needed $ \hat E_1\cdot \dot{\hat F}$ is given by 
\begin{equation}
\hat E_1 \cdot \dot{\hat F}  = -\hat E_1 \cdot (\vector \Omega \cross \hat F) 
\end{equation}
This triple product vanishes whenever its three vectors are coplanar.  Thus the only part of $\vector \Omega$ that contributes is the component in the $\hat E_1 \cross \hat F$ direction.  Accordingly, without altering the right hand side, we may replace $\vector \Omega$ by $v \hat E_1 \cross \hat F$, for some coefficient $v$:
\begin{equation}
\hat E_1 \cdot \dot{\hat F}  = - v~ \hat E_1 \cdot( \hat E_1 \cross \hat F)  \cross \hat F)  
\end{equation}
Simplifying the double cross product yields
\begin{equation}\label{eq:vForm}
\hat E_1 \cdot \dot{\hat F}  = v~ (1 - (\hat E_1 \cdot \hat F)^2)
\end{equation} 

We can now show that the coefficient $v$ is positive for large enough $U$, so that the right side of Eq. (\ref{eq:vForm}) cannot be negative. The component $v$ is simply $(\hat E_1 \cross \hat F) \cdot \vector \Omega$. 
For large $U$, we may isolate the dominant part of $\vector \Omega$ in terms of $\hat E_1$.  In Eq. (\ref{eq:Omega}) we write $\vector U$ as $U ~\hat U$ or $U \hat E_1 - U (\hat U - \hat E_1)$.  Since $\hat E_1\goesto \hat U$ for large $U$, this second remainder term is arbitrarily small.  Thus $\vector \Omega$ may be written
\begin{equation}\label{eq:OmegaLimit}
\vector \Omega =  U F(\RT_{\Omega \tau})_{0} \cdot (\hat E_1 \cross \hat F) + U F(\RT_{\Omega \tau})_{0} \cdot (\hat U - \hat E_1) \cross \hat F  + F(\RT_{\Omega F})_{0}\cdot \hat F  
\end{equation}
The dominant first term in this $\vector \Omega$ contributes a part $v_1$ to $v$:
\begin{equation}
v_1 = U F (\hat E_1 \cross \hat F) \cdot (\RT_{\Omega \tau})_{0} \cdot (\hat E_1 \cross \hat F)
\end{equation} 
Now, $(\RT_{\Omega \tau})_0$ is a positive definite matrix, as noted in Sec. \ref{sec:Response}.  That is, the dot product $\vector V \cdot(\RT_{\Omega \tau})_0 \cdot\vector V$ is necessarily positive for any vector $\vector V$, including $\hat E_1 \cross \hat F$.  Thus the coefficient $v_1$ is positive.  
The remaining parts of $v$, though not positive, are arbitrarily smaller than $v_1$.  Thus for large enough $U$ this $v$ remains positive.  In view of Eq. (\ref{eq:vForm}) this means that $\hat E_1 \cdot \dot{\hat F}$ is positive except when $\hat U = \plusorminus \hat F$.  A general $\vector F$ must change until alignment with $\hat U$ is reached.  All motion leads to this aligned state.  It can be shown that this aligning property holds as long as the $\RT_{\Omega F}$ has only one real eigenvalue \cite{Gonzalez:2004la}.

We now have a general picture of how an asymmetric body moves under any constant external force $\vector F$, such as gravity.  In general there is an initial rotation.  And typically  this rotation leads to an orientation where $\vector F$ is aligned with an eigendirection of the rotational response tensor $\RT_{\Omega F}$.    Then in the lab frame the body's eigendirection aligns with $\vector F$ and the body typically rotates around this axis. If the center of force is close to the center of twist, the motion resembles that of a rotating rigid body in free space, as discussed in Sec. \ref{sec:Response}. There are two possible final orientations in the two eigendirections having maximum and minimum eigenvalues of $\RT_{\Omega F}$.  There may also be a closed orbit (in the body frame) corresponding to the intermediate eigenvalue \cite{Moths:2013qf}. (For special shapes with higher symmetry, such as that of Fig. \ref{fig:PaddleTwist}, more elaborate final states are possible \cite{Doi:2005yu}.) The initial orientation of the body determines which of these final states is reached.  

If one changes the mass distribution to increase the displacement $\vector U$ of the center of force further and further from the center of twist, two of the three eigenvalues approach, merge, and become a complex conjugate pair.  The remaining real stable eigendirection becomes a direction of global stability \cite{Gonzalez:2004la}.  These objects are termed ``axially aligning".  The stable state has a fixed proportionality between the force vector and the angular velocity vector.  Its motion thus has a fixed positive or negative helicity determined by the object itself.  This chiral response does not require a chiral shape, but the center of force must be displaced from the center of twist.  Since there exist possible force centers that preserve the reflection symmetries of the shape, there are exceptional cases that depart from the general scheme leading to helical motion.  

In the next section we consider ways of organizing the motion of colloidal bodies by varying the forcing in time.

				\subsection{Motion under programmed forcing}\label{sec:ProgrammedForcing}

				The motion described above was qualitatively altered by the antisymmetric part of $\RT_{\Omega F}$.  Without this antisymmetric part, the motion is obliged to retain its initial value of the pseudo energy $\GG[\vector F] \definedas \half \vector F \cdot \RT_{\Omega F} \cdot \vector F$ of Sec. \ref{sec:MotionLaws}.  If the bodies in the colloidal sample are initially randomly oriented, they remain so.  Thus the forcing has no organizing effect on the orientations.  The addition of an antisymmetric part changes that.  The motion evolves into a small subset of orientations.  These are fixed rotation axes or orbits occupying a vanishingly small part of the set of initial possible orientations.  The orientations are strongly ordered.  Yet they are not completely ordered.  The states are distributed between the two fixed axes and distributed over the stable orbit if present.  Moreover, even the objects with a common axial alignment are disordered, since the objects may have any orientation (\ie phase of rotation) about this fixed axis. 
				
				This disorder is detrimental for studying or manipulating the colloidal objects.  Their random orientations make them respond to the same force with different motions.  On the other hand the different response of different orientations offers a way to manipulate these orientations to reduce the disorder.  When all the objects are oriented identically, their common motion can be manipulated in ways not possible with isotropic objects.  Below we describe ways of producing dispersions with a common orientation.  
				
				The potential utility of elementary objects that have an internal orientation is well illustrated by the case of magnetic moments in atoms.  Time-dependent magnetic fields can rotate a population of atoms so that the atoms in a given region all point in the same direction as they rotate together\footnote{More accurately, a small subset of the atoms is precisely aligned} .  It is this ordered rotation that makes the flexible spatial probing of magnetic resonance imaging\cite{Callaghan:1993fj} possible.
				
				Here we focus on the simple case of axially aligning objects with a globally stable fixed direction $\vector E_1$ pointing along $\vector F$.  The objects then differ only by a rotation about this axis.  We ask whether there is a means of applying a time-dependent force to make these angles the same for all the objects in the dispersion.   We review two methods,  rotating force, and random tilts in the forcing direction.  
			
				There is a simplistic reason to expect that mild variations in the forcing should lead to increased order.  In all the cases we have encountered, the forcing concentrates the initial orientations into a smaller range of orientations.  As the system approaches its final axially aligned state, this concentration process stops.  That is, the concentration only occurs during the transient leading to the final state.  Thus, further concentration requires further transients, \eg by tilting the force.  The object must then readjust to the new force by undergoing a new transient.  This deliberate changing of the driving force is what we term ``programmed forcing".
					\subsubsection{Random alternation}\label{sec:RandomAlternation} \quad
			Here we follow the above intuition that anything that induces  transients should increase order.  We imagine the simplest possible change in the force: a tilt in the direction of the driving force $\vector F$ by an angle $\theta$.  We imagine a sequence of such tilts, back and forth with alternating $\theta$ and $-\theta$ tilts, applied to the whole sample.  We avoid issues related to one tilt influencing the next one by waiting after each tilt for a new final orientation to be reached.   This primitive programmed forcing was explored in Refs. \cite{Eaton:2016ee} and \cite{Moths:2013qf}.  The authors found that this scheme indeed creates ordering within well-defined limits, as we now describe.
			
			\begin{figure}[htbp]
\includegraphics[width=.8\textwidth]{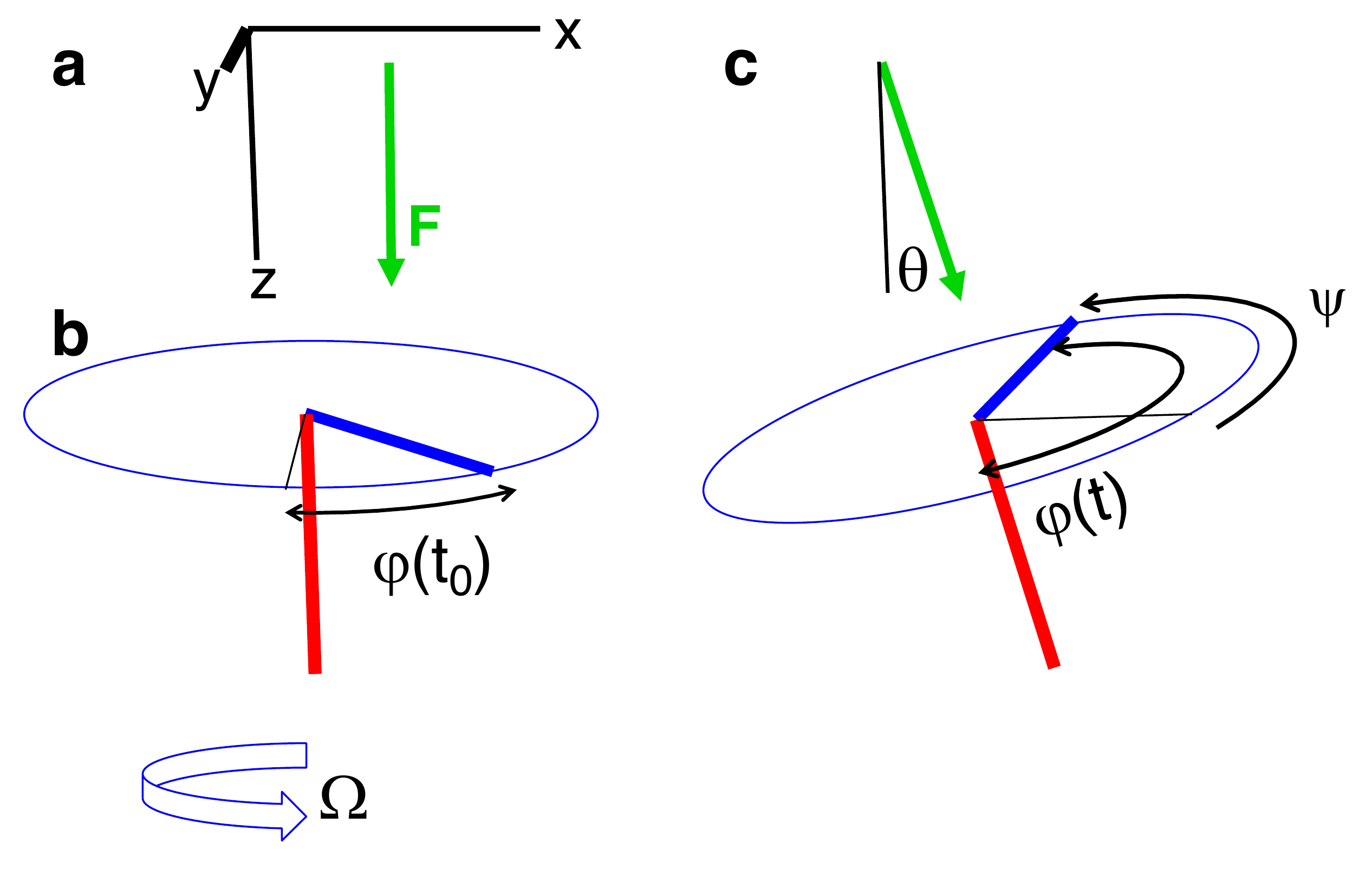}
\caption{\label{fig:Tiltphipsi} 
Definitions for phase angles between identical objects rotating in response to a common force $\vector F$.  a) Arrow indicates initial force $\vector F$ (green) and basis in lab frame.  b) object rotating in steady-state response to force $\vector F$ at the time of a tilt $t_0$.  All the objects are represented by two basis vectors.  Downward (red) basis vector is the aligning direction in the object, parallel to $\vector F$.  A second orthogonal basis vector is shown in dark color (blue).  The (blue) axis of a second ``fiducial" object, which points along the $y$ at time $t_0$, is marked with a thin line.  The phase angle between the  two blue axes is denoted $\phi(t_0)$.   c) the object at a time $t$ long after the tilt.  Green arrow indicates the force tilted by angle $\theta$.  The angle of the blue axis and the $y$ axis $\phi(t)$ increases with time.  The corresponding axis of the fiducial object (thin line) increases at the same rate.  Thus the relative angle $\psi$ between the two blue axes is constant in time. }
\end{figure}

			A simple function suffices to deduce the outcome of this procedure, regardless of the complexities of the transient response. The only aspect of an object that can affect its response to a tilt is its orientation angle $\phi(t)$ about the $\vector F$ axis.  For definiteness, we suppose that the tilts in $\vector F$ occur in the $x$--$z$ plane.  Then we can consider an arbitrary point on the object that rotates around the alignment axis. At any moment $t$ this point makes a well-defined angle with the $y$ axis.  We define this angle as an explicit measure of the orientation angle $\phi(t)$, as shown in Fig. \ref{fig:Tiltphipsi}.  Under a constant force this $\phi$ increases at a constant rate equal to the aligning eigenvalue $\Omega$.  
			\begin{figure}[htbp]
		\begin{center}
\includegraphics[width=.3\textwidth]{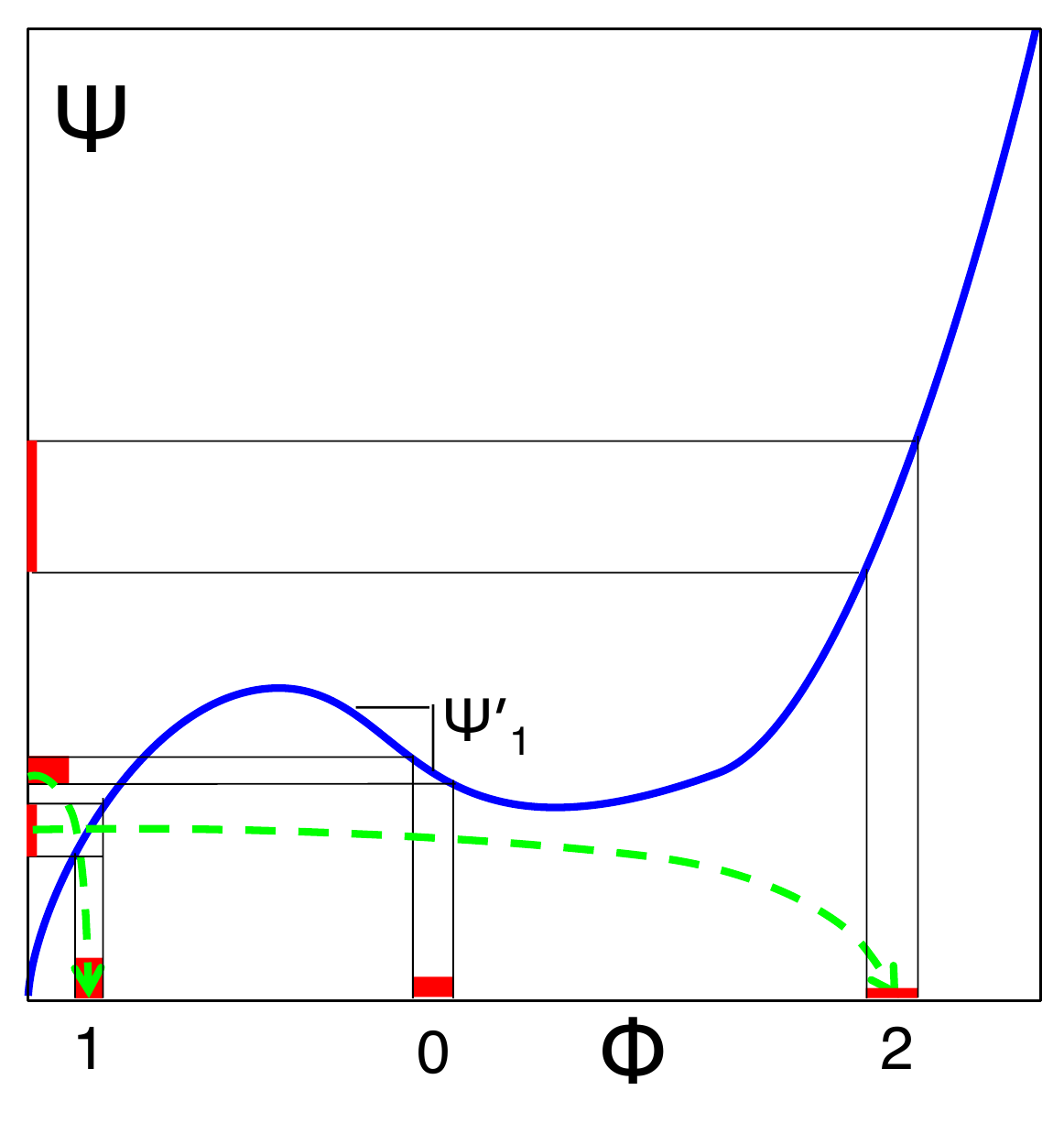}
		\end{center}
\caption{\label{fig:IteratedPhaseMap} Random iteration of a phase map.  Graph shows a phase map $\psi(\phi)$. Colored segment labeled $0$ is a typical small interval $\Delta \phi_0$.  The vertical and horizontal lines indicate the corresponding interval $\Delta \psi$ resulting from one tilt. The interval is widened by a factor $|\psi'(\phi_0)|$ relative to the initial $\Delta \phi_0$.  The dotted arrow indicates the next $\Delta \phi_1$ at the next tilt, \ie the initial $\Delta \psi$ plus a random phase $\rho$. A second typical iteration, with a different random $\rho$, leading to $\Delta \phi_2$, is shown.
}\end{figure}

\begin{figure}[htbp]
			\begin{center}
\includegraphics[width=.9\textwidth]{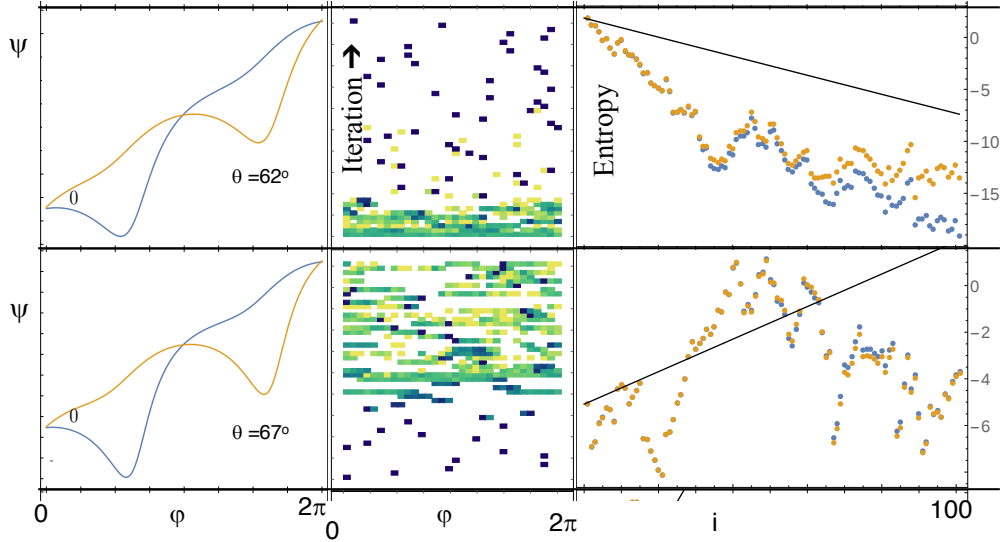}
\end{center}
\caption{\label{fig:Jonahbig} Ordering effect of alternating tilts for a particular object, after \cite{Eaton:2016ee}.  Top row shows results for tilt angle $\theta = \plusorminus 62^o$.  Left panel shows the two phase maps for forward and backward tilts.  Center panel shows kymograph of the distribution $p(\phi)$. Shading at a given $\phi$ indicates $p(\phi)$. Bottom row shows initial, uniform distribution; top row shows final distribution, concentrated in a single block.   Right panel shows the entropy $H$ of $p(\phi)$ distribution over 100 iterations for a particular sequence of phases $\rho_i$ induced by random waiting times. Light and dark colored dots show two different methods of calculation. Solid line shows the upper bound on the statistical average stated in the text.  Bottom row shows the same quantities for a slightly larger $\theta = \plusorminus 67^o$, with Lyapunov exponent $\Lambda>0$.  Here the initial state had $p(\phi)$ confined to a single narrow interval.  It became random over time.  In the right panel the entropy tends to increase, in rough consistency with the statistical bound shown by the solid line.
}
\end{figure}

			When the force tilts, a body with orientation $\phi$ undergoes some transient gyration that ultimately subsides to a new steady rotation. The effect of the tilt depends on the orientation of the body at the moment of the tilt; \ie it depends on $\phi$.  Long after the tilt the bodies in the sample are again rotating at the same rate, so that the phase difference between any two bodies remains constant.  Only the phase differences among the objects are of interest.  Accordingly, we shall measure each object's phase relative to that of an object whose initial $\phi$ at the moment of tilt was 0.  We denote this relative phase as $\psi(\phi)$. Evidently $\psi(0) = 0$.  This difference $\psi(\phi)$  depends on the size $\theta$ of the tilt
			---it goes to zero as the tilt angle $\theta$ goes to zero.  Every object with a given $\phi$ before the tilt is in the same dynamical state at the moment of the tilt and follows the same path to the new state with the same final $\psi$.  The complete effect of a tilt is thus given by $\psi(\phi)$.  This  so-called ``phase map" is the only property of the system needed to determine the fate of the original population \cf Fig. \ref{fig:IteratedPhaseMap}.  Typical phase maps\cite{Eaton:2016ee} using actual $\RT_{\Omega F}$'s and tilt angles $\theta$ are shown in Fig. \ref{fig:Jonahbig}.	

			With this knowledge of the effect of a single tilt, we now investigate the effect of many tilts.  The tilts must be separated by a sufficient time to allow all initial states to converge to the asymptotic $\psi(\phi)$.  To determine the effect of the second tilt, it is necessary to know the absolute phase $\phi(t_2)$ from the preceding tilt at the moment $t_2$ of the new tilt.  This $\phi(t_2)$ depends on the particular object. It is not sufficient to know $\psi$.  Still, we may determine the effect of $\psi(\phi)$ by performing the next tilt after  a {\em random} waiting time $t_2$.  
This random time adds an arbitrary common phase $\rho$ to the $\phi$ of all the objects.  Thus a phase that was $\phi_2$ just before the second tilt has become $\psi(\phi_2 + \rho)$ after that tilt.  Evidently, $\phi_2 + \rho$ is completely random, like  $\rho$ itself. Thus the statistical behavior of the $i$th iteration is now independent of  $\phi_i$.  Fig. \ref{fig:IteratedPhaseMap} depicts the process of successive iterations of an initial $\phi_0$.

			Increasing order in this process amounts to the concentration or bunching of the random distribution of phase angles into a few narrow intervals.  We may see this bunching effect by examining the fate of a small interval of $\phi$'s, denoted $\Delta \phi$, during a tilt, as shown in Fig. \ref{fig:IteratedPhaseMap}.  A typical $\Delta \phi$ is marked on the horizontal axis.  The new width after one tilt is marked on the vertical axis.  Evidently the change of width depends only on the absolute slope of the $\psi(\phi)$ function \ie $|\psi'(\phi)|$.  If this slope is smaller than unity, the interval shrinks, and bunching of the points occurs; conversely if it is larger than unity, the interval expands, and spreading of the points occurs.  This bunching factor depends on $\phi$ but not on how many tilts have occurred.  The particular fate of a given $\Delta \phi$ depends on where the previous random phase $\rho_i$ has placed it on the $\phi$ axis.  We may find the average effect after one tilt by averaging over $\rho_i$.  
The net amount of spreading $f_n$ after $n$ tilts is evidently $|\psi'_1| |\psi'_2|... |\psi'_n|$.  Thus $\log f_n$ is a simple sum of the $\log|\psi'_i|$: 
\begin{equation}
\log f_n =  \sum_i \log |\psi'_i|
\end{equation}

Being the sum of independent variables in a well-behaved range, this sum obeys the law of large numbers: $\expectation{\log f_n} = n \expectation{\log |\psi'(\phi)|}$.  Thus the expected amount of spreading $f_{\rm eff} = \exp (n \expectation{\log f(\phi)})$.  It grows or shrinks exponentially according to the sign of $\expectation{\log f(\phi)}$.  Many dynamical systems show bunching and spreading of small intervals in this way, governed by an analogous factor, called the Lyapunov exponent\cite{Ott:2002fu} and denoted $\Lambda$.			

			Using this Lyapunov statistic, we may characterize the growth of order quantitatively.  An initially random system with a uniform distribution $p_0(\phi)$ evolves into a $p_i(\phi)$ that is larger for some $\phi$ than for others.  As in statistical physics, we may measure the degree of order of  $p(\phi)$ by the entropy\cite{Ott:2002fu}, defined as $-\expectation{\log p}$ or $-\integral d\phi ~p(\phi) \log p(\phi)$.  For definiteness we measure entropy relative to a uniform distribution.   This entropy, denoted $H$, can be written $-\expectation{\log(p(\phi)/p_0)}$.   Thus if a uniform probability in a certain interval becomes uniformly bunched into an interval of half its  length, $p(\phi)/p_0$ doubles and the entropy decreases by an amount $\log 2$.  

			By carefully analyzing the flow of probability implied by the phase map $\psi(\phi)$, Ref. \cite{Eaton:2016ee} showed that the average entropy change during one tilt must be more negative than the Lyapunov exponent $\Lambda$ defined in the last paragraph. 
			Thus whenever $\Lambda$ is negative, the entropy must go to $-\infinity$ relative to the fully random state. These authors investigated the changes of entropy through actual sequences of tilts, using a realistic $\RT_{\Omega F}$ with various tilt angles $\theta$, as shown in Fig. \ref{fig:Jonahbig}.  The actual change of entropy fluctuates from tilt to tilt.  But over many tilts, the entropy decreases linearly as dictated by the averages.  Thus the statistical bound has predictive value for individual cases.
		
			The ability of strong iterated tilting to produce order is limited.  The phase map function $\psi(\phi)$ develops strong maxima and minima as the tilt angle $\theta$ grows.  Larger $\psi(\phi)$ produces larger slopes $|\psi'(\phi)|$ and ultimately the Lyapunov exponent becomes positive.  Then the tilting process erodes order instead of enhancing it.  
			However, the Lyapunov exponent does not fully capture the tendency towards ordering.  It accounts for the contraction of contiguous intervals of $\phi$.  But ordering processes also occur beyond the contraction of contiguous intervals.  Two disjoint intervals may also merge to increase the ordering.  This disjoint merging can compensate for a small positive Lyapunov exponent to produce a steady state of nonzero entropy \cite{Eaton:2016ee}, \cf \cite{Nakao:2007vn}.
			
			Not surprisingly, this phenomenon of reducing disorder in a population of similar systems by random but equal perturbations in all of them occurs beyond our domain of chiral colloids.  It is known as ``noise-induced synchronization" \cite{Nagai:2009ys,Eaton:2016ee}.   Many forms of noise beyond the simple tilts treated here are known to produce order.  This behavior supports the primitive notion that inducing transients promotes order.   
			
			Notably, the orientation induced by the random tilt method is not predictable.  Though the objects in the sample are oriented in the same direction, the ordering process gives no information about what this direction is.  The periodic forcing described below {\em does} give information about the ordering direction.
			
				\subsubsection{Periodic forcing}\label{sec:PeriodicForcing} \quad
				A more straightforward form of programmed forcing is the case pictured in Fig. \ref{fig:plasticCrumb}b, where a constant force is perturbed by a small perpendicular rotating vector, so that its magnitude remains constant.  The equation of motion in Eq. (\ref{eq:RTOmegaXdot}) depends only on the instantaneous $\vector X$ and hence has the same form whether $\vector X$ is varying or not.  Ref.\cite{Moths:2013mz,Moths:2013qf} explored a range of objects and magnitudes of the rotating force.  By ramping the rotation rate from slow to fast, one could entrain the rotation of objects with a range of intrinsic rotation rates.  The mechanism for this entrainment is most easily seen when the magnitude of the rotating force is small and its rotation period is close to the natural period of the object.  If the variable forcing is weak, any mutual alignment effect must require many cycles of rotation.

				One may account for this entrainment by viewing the rotating object under constant force as a general dynamic system of period $S$.  This periodic motion need not be sinusoidal; it may move quite irregularly as it traverses its cycle.  Every variable within this system takes on the same value at a given point in the cycle.  Thus to know the system's internal state, it suffices to know the time $t$ relative to the cycle time, \ie the phase $\phi \definedas 2 \pi t/S$.   Now we introduce an external periodic system called the ``pacer",  with a period $T \aboutequal S$.  We monitor the phase $\phi$ at every cycle $i$ of this pacer system.  Since $S \notequal T$ these phases $\phi_i$ change gradually on each successive cycle.  By tuning the external period $T$ to be close to the system period $S$, we may make this motion of the successive $\phi_i$ as slow as we like.  
				
				We now allow the pacer to perturb the system very weakly.  The perturbation is so weak that the system's state is still completely determined by its phase $\phi$.  But now this $\phi$ is influenced by the pacer.  We follow the system with initial phase $\phi_i$ through to cycle $i+1$. The phase $\phi_{i+1}$ will be altered by the perturbation.  During this cycle the (strictly periodic) pacer exerts a set sequence of forces on the system.  Thus, whenever the initial phase is $\phi_i$ the resulting phase $\phi_{i+1}$ must be the same.  That is, $\phi_{i+1}$ is a deterministic function of $\phi_i$. This function, denoted $f(\phi)$, is called a return map of the system under the given driving by the pacer\cite{Ott:2002fu}.  The return map plays the same role as the phase map $\psi(\phi)$ of Sec. \ref{sec:ProgrammedForcing}.     
				
				For some ranges of $\phi$ the perturbation advances the phase, for other ranges of $\phi$ the perturbation retards it.  If $|S-T|$ is small enough, there is some $\phi$ denoted $\phi^*$ for which $\phi_{i+1} = \phi_i \definedas \phi^*$, \ie $f(\phi^*) = \phi^*$.  Thus $\phi^*$ is a fixed point of the dynamics.  If $|f(\phi) - \phi^*| < |\phi - \phi^*|$ then successive $\phi_i$'s approach $\phi^*$ and the fixed point is locally stable.  The continuity of $f(\phi)$ assures that if there is a fixed point, there is at least one {\em stable} fixed point\cite{Ott:2002fu}.  At this fixed point the $\phi_i$ of the system is the same at every cycle of the pacer; the system is ``phase locked" to the pacer.  For any given weak coupling to the pacer, this phase locking occurs for some range of mismatch $S-T$; stronger couplings can lock the phase over broader ranges.  
				
				Applied to the rotating forces, this means that the eigenvector of the object (downward-pointing basis vector in Fig. \ref{fig:plasticCrumb}b) rotates with the force (green arrow in Fig. \ref{fig:plasticCrumb}b) and maintains a fixed directional difference from it.  The fixed angle is necessarily the same for all identical objects.  They are now oriented identically, not merely along a common axis.  Analogous phase locking occurs whenever $S$ is ``commensurate" with $T$, \eg $S = 2T$ or $3S = 4T$.

				This phase locking is somewhat delicate.  If the mismatch of the pacer period with the system period is too strong, chaos ensues:  there is no predictable, steady state relation between the pacer's motion and the system's. The progression from a predictable phase-locked state to a chaotic state is governed by the ``KAM" (Kolmogorov Arn'old Moser) route to chaos\cite{Ott:2002fu}.  This KAM scenario has not been demonstrated (even numerically) in the context of chiral sedimentation.  For the purpose of inducing ordering, the KAM picture shows that delicacy is required; one can easily overdo the strength of the programmed forcing.  	

	   			\subsubsection{Non-axially-aligning objects}\label{sec:non-self-aligning} \quad
				
			The ordering by programmed forces discussed above presupposed that all the objects could be pre-aligned about a common axis, so that the only remaining disorder was the orientation about this axis.  There remains the large category of non-self-aligning objects, \eg where $\RT_{\Omega F}$ has only a small antisymmetric part.  Then the sample after transients have died out contains up to three populations, two for the two stable fixed points and a further one if there is a stable orbit.  The programmed forcing can order the orientation within either of the fixed-point subpopulations, by the logic above.  To unite the three subpopulations, obtaining a common orientation for all the objects, requires something further.  It is possible to add steps in the forcing that transfer objects from one subpopulation to another.  It is not clear how to design such forces without knowing about the existing orientations.  This is an open subject.
			
				\subsubsection{Implementing programmed forcing}\label{sec:Implementing} \quad

		 The organizing influences illustrated above are not straightforward to implement for real sedimenting objects.  To benefit from these influences, one needs a strong enough forcing to give many rotations of the object within the disordering time set by rotational diffusion.  Within this time it must be possible to alter the direction of the force repeatedly.  Further, the apparatus must be tall enough to contain the linear motion accompanying this rotation.  Ideally one would wish a large helicity, with little translation for each rotation.  But hydrodynamics sets strong limits on this helicity as noted in Sec. \ref{sec:LimitsOfChirality}.  

		 
		 In the domain of turbulent flow, many of these requirements are fortuitously satisfied.  Small particles in a rapidly accelerating fluid feel Stokes drag proportional to their buoyant mass.  Their motion relative to the fluid can be small despite the strong inertial effects in the fluid as a whole.  Further, circulating vortex motion found in turbulence provides the type of cyclic time dependent forcing that gives mutual alignment\cite{MitchellPrivate2014}

			\subsection{Limits of chirality}\label{sec:LimitsOfChirality}
Many of the phenomena treated above depend on the strength of the chiral response to driving. 
This strength can be characterized by the pitch angle $\alpha$ of the helical path traced out by some representative point on the object (\cf Fig.\ref{fig:plasticCrumb}a) This is the angle defined in Sec. \ref{sec:translation}.  For a nonrotating object, the helix becomes an oblique line and $\alpha$ becomes the glide angle of the object.  

For a nongliding forced object that moves in the force direction, the helix axis passes through the object.  Points at distance  $r_m$ furthest from this axis have the largest $\alpha$.   Then in terms of the speed $V$ and angular speed $\Omega$ the maximal pitch angle is given by $\tan\alpha =\Omega ~r_m/V$.  An object with large $\alpha$ is one with maximal $\Omega/V$ for a given lateral size $r_m$.  The $\Omega$ and $V$ in turn are determined by the response tensors $\RT_{\Omega ~F}$ and $\RT_{V~F}$, as explained in Sec. \ref{sec:translation}. 
				
To see what limits this pitch angle, we can consider the two-paddle object of Fig. \ref{fig:PaddleTwist}a forced to the right.  Since the paddles are oriented obliquely to the force, the upper paddle will glide into the page as it moves towards the force.  The two paddles glide in opposite directions to create a helical motion.  The helix angle $\alpha$ is then the glide angle.  If the paddles are small enough, the glide angle of each is unaffected by the other.  Thus the maximum $\alpha$ is the maximum glide angle of a paddle.  Now, glide angles of sedimenting objects are limited by the anisotropy of their stokes drag.  The ratio $A$ of maximum to minimum drag is smaller than 1.5 for a disk and smaller than 2 for a rod [\cite{Happel-Brenner}, Eqs. 5-11.19--5-11.25].  The corresponding maximum glide angles $\alpha_{\rm max}(A)$  are 0.20 and 0.34 radians [\cite{Happel-Brenner} Sec. 5-11]. 

Analogous reasoning gives similar limits on chirality when the object is driven by torque. Such a torque necessarily produces a terminal angular speed $\Omega$ for any object.  A chiral object may also have a terminal speed $V$.  Without this translation, the torque drives the object point in a circle and $\alpha = \pi/2$, with no characteristic chirality.  A finite helicity requires nonzero $V$: $\tan (\pi/2 - \alpha) = 1/\tan \alpha = |V|/(\Omega~ r_m)$.  The paddles of Fig. \ref{fig:PaddleTwist} are now pushed into and out of the page by torque.  Their glide now makes them translate to the left or right with speed $V$.  Just as the forced paddles are limited to small $\alpha$, the torque-driven paddles are limited to small $\pi/2 - \alpha$. 

The strengths found in various experimental and theoretical studies are also small. Optimization over the shapes of helical objects \cite{KeavenyNanolett2013,WalkerNanolett2015} and arbitrarily-shaped ones \cite{VachNanolett2015,MirzaeSciRobot2018} have not yielded values of $\tan \alpha$ or $\tan(\pi/2 - \alpha)$ larger than about $0.2$, consistent with the glide-based reasoning above.  By this measure the chiral response to force or torque is limited and modest.  It is in the collective phenomena discussed in Sec. \ref{sec:collective} that the effects become dramatic.

				\subsection{Summary}\label{sec:SummaryForced}
			The discussion above shows that the orientational response of an asymmetric object to forcing is in general self-organizing.   An initially random set of orientations evolves towards a less random final state. To some degree this is to be expected.  Asymmetry means not all orientations are equivalent; thus it is natural that a process that depends on orientation should favor some orientations over others.  The self organization shown above goes beyond a mere tendency however.  The initially random state becomes concentrated into an {\em indefinitely} small fraction of the initial states.  Though the specific outcome depends on the particular tensor, the thoroughness of the self-organization is generic.  In the cases we examined, this self-organization required an antisymmetric part in the response tensor arising from some asymmetry in the object.  
			
			Analogous concentration of many states into few occurs in energy-minimizing systems.  But it is not clear how the forced colloids above act to minimize an effective energy.  This kind of self organization is by no means generic among simple dynamical systems as a class.  Instead one often finds that narrowly-defined initial states evolve to fill their configurational space ergodically\cite{Ott:2002fu}.  This is true for both conservative Hamiltonian systems and for dissipative, driven systems such as ours.  Even our forced colloids become random when the time-dependent driving is too strong, as seen in Sec. \ref{sec:ProgrammedForcing}.
			
			We now turn to motions driven not by external forces on the object but from driving of the fluid.  These are all governed by response tensors analogous to the $\RT_{\Omega F}$ studied above.  In some cases the existence of strong self organization is an open question.

\section{Motion without external force}\label{sec:NoForce}
\setcounter{equation}{0}
\setcounter{figure}{0}

				In the preceding section we introduced the formalism for describing the response of a rigid object to a force or torque.  In a quiescent fluid, motion requires force or torque.  The same is not true in a driven fluid in nonuniform motion, such as shear.  In this section we describe the linear response to a weak imposed fluid motion.  A major example is simple shear in a fluid.  Another important example is a driven flow at the surface of the object.  The flow can be driven by an applied electric field in a conducting medium or by active sources of stress on the surface. Both of these give new forms of chiral response.  
				
				Finding the motion resulting from a specified flow requires untangling the imposed flow from that caused by the motion.  Some of this complexity can be sidestepped by considering an adjoint problem.  The adjoint system has the same object and the same imposed flow as the desired system, except that the object is held fixed by constraining forces.   By determining the flow around the adjoint system, one may readily determine the force and torque needed to constrain it.  If the object is released, its motion causes an additional drag force and torque equal  and opposite to those of the fixed system \cite{Burelbach:2019sz}\footnote{
We first consider an immobilized object with the imposed fluid velocities on its boundary as $\vector u_{F \tau}(\vector r)$.  Here $\vector F$ and $\vector \tau$ are the external force and torque required to immobilize it.  This force and torque are thus transmitted to the fluid.  Next we denote the flow field $\vector v_{F \tau}(\vector r)$ as that of the object in a quiescent fluid, without imposed fluid velocities, under the same $\vector F$ and torque $\vector \tau$. In these quiescent conditions the object would move with $\vector V$ and $\vector \Omega$ given by the mobility tensors of Eq. (\ref{eq:bigmateqn}).   We now consider the flow field  $\vector u(\vector r) \definedas \vector u_{F \tau}(\vector r) - \vector v_{F \tau}(\vector r)$.  The addition of $-\vector v$ adds a rigid body motion $-\vector V$, $-\vector \Omega$ to the imposed surface velocity.  It does not change the surface velocity relative to that of the body.   This added motion produces a drag (force, torque) equal and opposite to the original constraint (force, torque) $(\vector F, \vector \tau)$.  The field $\vector u(\vector r)$ then satisfies both the boundary condition on the surface velocity and the condition of free motion with no external force or torque.  Thus it is the desired field. The same reasoning applies also to the motion of a body induced by a pre-defined external flow such as a simple shear.  For spherical bodies, this adjoint property follows from the Fax\'en theorem \cite{Happel-Brenner, Kim:2005fr}.  
}.
The relation between this motion of the real system and the constraining force and torque of the adjoint system is precisely the response treated in  Sec. \ref{sec:ConstantForce} above.  
				
				\subsection{Chiral effects in shear}\label{sec:ChiralShear} 
\begin{figure}[htbp]
\hfill\includegraphics[width=.8 \textwidth]{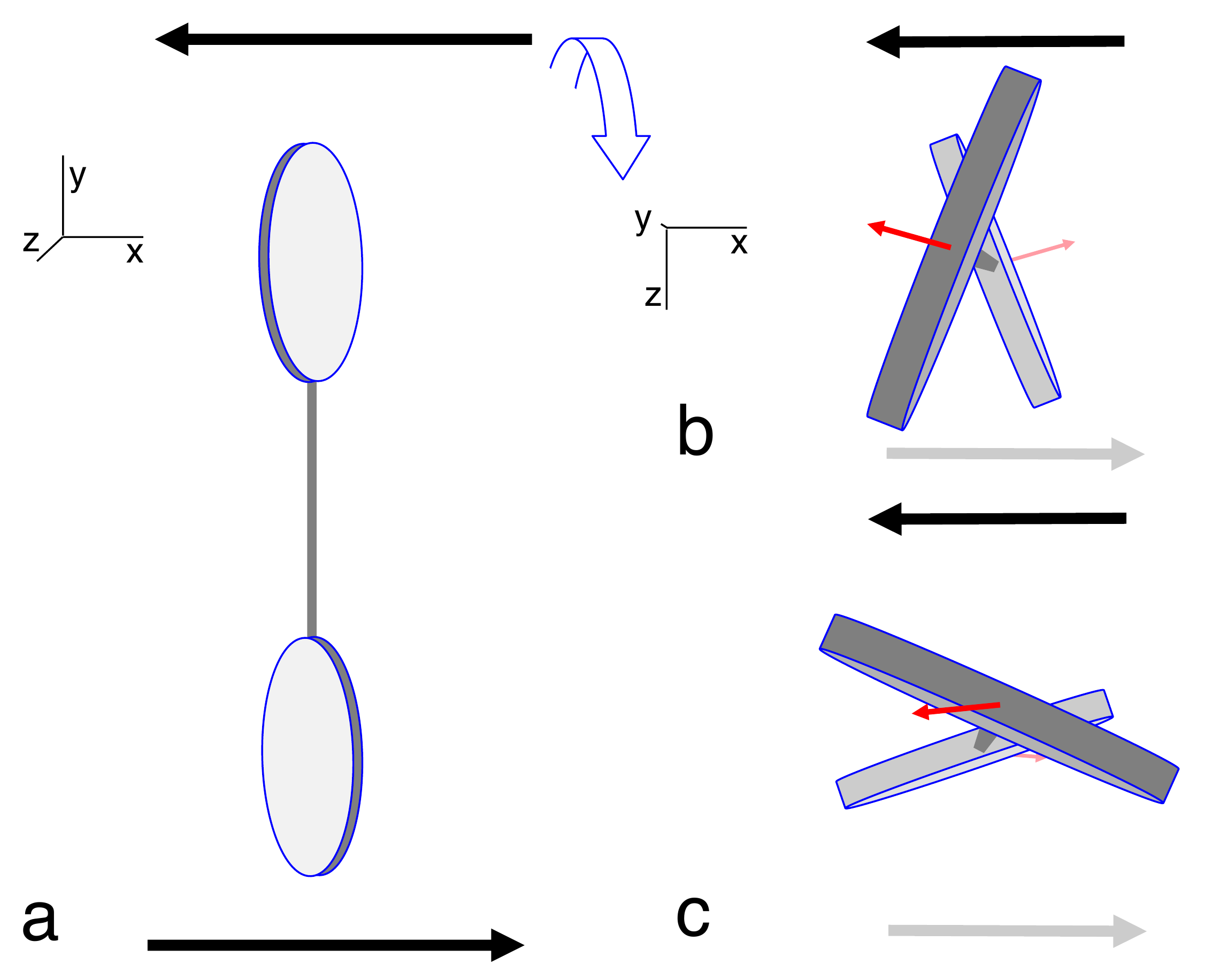}\quad
\caption{Chiral migration of a twisted double-paddle object in a shear flow, after \cite{Doi:2005yu}.  a) Overview of flow and paddle as it would appear in a top view of Fig. \ref{fig:plasticCrumb}c.  Flow is in horizontal $\plusorminus x$ direction, increasing vertically with increasing $y$.  Paddle is aligned vertically.  b) Downward view along paddle axis, front view in Fig. \ref{fig:plasticCrumb}c, showing closer paddle as larger.  Imposed fluid velocity at closer paddle is to left as shown by dark arrow.  Imposed fluid velocity at further paddle is to right as shown by light arrow.  If the paddles are held fixed, the closer paddle receives a force canted upward from the velocity because of the tilt angle of the panel.  The further paddle receives a force also canted upward from its rightward velocity, because of its opposite tilt.  If the paddles are released, they respond to this force by moving upward; if the paddles are spatially inverted, the direction of this common force and its motion also reverse.   c) Same as b with paddles rotated 90 degrees about their axis. The net force also reverses sign but has weaker magnitude than in b) because the paddles meet the flow at a lower angle.  The combined effect of orientations b) and c) is a net chiral motion.  The chirality of the object, and also this motion vanish when the paddles are at right angles to each other.
}
\label{fig:PaddleTwist}
\end{figure}
			
Suspended objects rotate, translate and perturb their surrounding fluid when put into a flowing liquid with a nonuniform velocity field $\vector v(\vector r)$.  Here we consider objects with a chiral shape in a shear flow and ask what distinctive chiral motions result.  This type of motion has been extensively investigated for specific shapes by Makino and Doi\cite{Doi:2005yu,Makino_2004,Makino:2003jk,Doi:2005qv,Makino:2005ty,Makino:2008fj} and more recently by Powers \etal and Speer \etal \cite{Marcos:2009fk,Speer:2010rw}.  They proposed and later demonstrated that chiral objects placed in shear flow have motions that change sign when the objects are replaced by their spatial inversions.  Fig. \ref{fig:plasticCrumb}c shows a proposed device exploiting this phenomenon.  The twisted double-paddle shape used by Doi (Fig. \ref{fig:PaddleTwist}) illustrates why such motion is expected, using the adjoint argument above:  if the object is held fixed, there should be a force in the vorticity direction (perpendicular to the velocity and its derivative).

This example suggests that the object's orientation as well as its shape  determine its force response.  Further, shear flow has a strong effect on this orientation.  Thus we expect some noteworthy linear response dynamics analogous to Eq. (\ref{eq:RTdot}).  In these phenomena, the driving quantity is not a vector, but the velocity gradient tensor $\vector \del \vector v$.  As discussed above, the resulting motion can be found via determining the force  $\vector F$ and torque $\vector \tau$ on the object, necessarily proportional to the velocity-gradient tensor $\vector \del \vector v$.  The proportionality  between these quantities is expressed by a tensor property of the body defined by
\begin{equation}
\vector F = \RT_{F ~\del v}\cdot (\vector\del \vector v)     ~;\quad       \vector \tau  = \RT_{\tau~ \del v}\cdot (\vector \del \vector v)
\end{equation}
From these tensors one can get the motion vectors $\vector V$ and $\vector \Omega$ using the four response tensors of Eq. (\ref{eq:bigmateqn}) for the object:
\begin{eqnarray} \label{eq:TDelv}
\vector V =  \left [-\RT_{V F} \cdot \RT_{F, \del v} - \RT_{V \tau} \cdot \RT_{\tau, \del v} \right ] (\vector \del \vector v) \nonumber \\
\vector \Omega = \left [-\RT_{\Omega F} \cdot \RT_{F, \del v} - \RT_{\Omega \tau} \cdot \RT_{\tau, \del v} \right ] (\vector \del \vector v)
\end{eqnarray}
We may term the combined tensors in $[...]$ as $\RT_{V ~\del v}$ and $\RT_{\Omega ~\del v}$. 
By dimensional analysis, $\RT_{V ~\del v}$ must be proportional to the particle size $a$.

Any migration effect ultimately arises from the $\RT_{V ~\del V}$ response tensor.  Thus we first ask whether this tensor produces a net chiral migration of an object as in  Fig. \ref{fig:PaddleTwist} when averaged over orientations.  In fact, there can be no such response \cite{Makino:2005ty}.  This is a subtle consequence of the rotational transformation properties of third-rank tensors like $\RT_{V ~\del v}$.  Any isotropic average of such a third-rank tensor must be a rotational scalar.  However, any such rotational average $A$ must arise from a tensorial form proportional to the completely antisymmetric tensor $\epsilon_{ijk}$ \cite{Jerphagnon:1970jk}.  Thus, if $A$ is nonzero, the equation $\vector V^i =  (\RT_{V ~\del v})^{ijk}~ \grad^j v^k$ reduces to $\vector V = A ~\grad \cross \vector v$.  Now, even a fluid in uniform rigid rotation has a nonzero, constant $\grad \cross v$.  Yet such a fluid cannot produce a relative motion such as $\vector V$.  Thus the constant $A$ must vanish \cite{Makino:2005ty,De-Groot:2013mz}.  Conversely, the effect seen in Fig. \ref{fig:PaddleTwist} requires some anisotropic distribution of orientations.  

Although this chiral shear-induced migration requires biased orientations, the shear-induced rotation, governed by $\RT_{\Omega ~\del v}$, requires no such bias.   Shear flow produces a rich rotational response akin to the force response treated in Sec. \ref{sec:ConstantForce}.  As in Sec. \ref{sec:MotionLaws} the rotation of the object gives a basis for ordering effects. That is, $\vector \Omega$ once calculated influences $\RT_{\Omega ~\del v}$ and this in turn influences $\vector \Omega$.  In analogy to Sec. \ref{sec:MotionLaws}, we may view the motion in the body frame, where $\vector \del \vector v$ rotates in time.  To be sure, this dynamics is more complicated than in Sec. \ref{sec:MotionLaws}: here the role of the vector $\vector X$ is played by the tensor $\grad \vector v$.  The possible orientations of $\vector \del \vector v$ cover a three-parameter compact set of orientations \vs the two-dimensional spherical surface covered by $\vector X$.  In Sec. \ref{sec:SummaryForced} we found that under forcing the arbitrary initial orientations became concentrated into an indefinitely small set of final orientations.  Analogous asymptotic states induced by shear are known as Jeffery orbits \cite{Jeffery:1922fd}.  For example, the Jeffery orbit of a thin rod is a sequence of half rotations about the vorticity axis.  Like the response to a force, the Jeffery orbits give an anisotropic distribution of orientations\cite{Hinch:1979ix}.  Thus \eg thin rods tend to point in the flow direction.  This anisotropy is sufficient to allow the shear-induced migration studied in Ref. \cite{Makino:2005ty}.   Still, the biased orientation arising from Jeffrey orbits is weaker than we observed in Sec. \ref{sec:ConstantForce};  it does not produce concentration into an arbitrary set of orientations (Sec. \ref{sec:SummaryForced}).

We'll discuss hydrodynamic consequences of this chiral migration in Sec. \ref{sec:collective}.
				\subsection{Phoresis}\label{sec:Phoresis}
In the preceding section, the motion was induced by an imposed flow far from the body.  Motion can also be caused by an imposed flow at the surface of the body itself.  The classic example is electrophoresis.  Electrophoresis occurs for bodies that become charged when immersed in a conducting fluid.  Such a body moves when an electric field $\vector E$  is applied to the dispersion.  The motion $(\vector V, \vector \Omega)$ is proportional to the electric field via tensors denoted $\RT_{V ~E}$ and $\RT_{\Omega~ E}$.  As above, it is helpful to consider the adjoint problem, in which the object is held fixed and the required force and torque are proportional to $\vector E$ via tensors $\RT_{F~E}$ and $\RT_{\tau ~E}$.  Electrophoresis gives a useful example to show the subtleties of motion driven by surface flow.  Henceforth we will denote this externally imposed field as $\vector E_0$.  

The most familiar uses of electrophoresis are to characterize biological molecules such as DNA\cite{Delgado:2007qy,Lee:2018uo}.  Evidently the magnitude of the charge on the body is important.  This charge is generally expressed in terms of the electrostatic potential between the surface of the body and the bulk of the fluid.  This potential, denoted the ``zeta potential" $\zeta$, vanishes in uncharged regions and has the sign of the surface charge.  Naturally, the charge on the surface attracts oppositely charged ions in the fluid. This opposite charge leads to an exponential decrease of the body's electric field away from the surface.  In practice the decay length of the exponential is much smaller than the micron-scale size of the bodies considered here.  Accordingly, we consider the limit in which this screening length is indefinitely smaller than the body.  In this limit the motion is governed by the characteristic velocity scale called the Smoluchowsky velocity $v^s$ \cite{Smoluchowski:1903fv}.  It is proportional to the external electric field $\vector E_0$, to the dielectric constant $\epsilon_r \epsilon_0$ and to the zeta potential; it is inversely proportional to the viscosity $\eta$ of the fluid \cite{Delgado:2007qy}:
\begin{equation}\label{eq:vsi}
\vector v^s \definedas \zeta {\epsilon_r \epsilon_0\over \eta} ~\vector E_0  ,
\end{equation}
Specifically, $v^s$ is the velocity of the fluid relative to a uniformly charged wall in the presence of a transverse field $\vector E_0$.

In most situations the bodies may be considered as uniformly charged and electrically insulating.  Thus neither fluid nor electric current exist in the interior.  In this case the fluid flow around the moving body is a potential flow whose potential is proportional to the electrostatic potential caused by the external field $\vector E_0$, as  shown by Morrison \cite{Morrison:1970ph}.  Then any shape asymmetry in the body has no effect; instead all bodies move in the direction of $\vector E_0$ at a speed $v^s$.  Thus any tensorial effects like those treated above can arise only when the charge (and thus $\zeta$) are nonuniform over the surface.

When the charge is nonuniform, a fully tensorial response is expected.  Indeed, calculated examples \cite{Long:1998fk,Long:1996yk} show motion perpendicular to $\vector E_0$ regardless of the body's orientation.  Other possible motions are chiral rotation from a nonchiral body shape, or rotation without translation.   Here the laws of motion are formally identical to those of sedimentation: \viz, rotation of the body at an angular velocity proportional to a non-symmetric $\RT$ tensor times the driving vector $\vector E_0$.  On the other hand, the basis for this motion is quite different from sedimentation.  There is no external force on the fluid, and thus the flow around the moving body is entirely different from the sedimentation case.   

Before discussing electrophoretic motion, it is useful to contrast the two sources of electric field near the body.  Even with no charge on the body, the external field $\vector E_0$ produces a field near the body.  This field causes an electric current $\vector J$ in the fluid, but this current cannot enter the body.  Thus any charge driven to the surface causes a ``depolarization charge" that stops current from entering.  At any point $p$ on the surface the current $\vector J(p)$ must have no normal component.  Since $\vector J$ is proportional to the local field $\vector E$, the field $\vector E_s(p)$ at the surface must also have no normal component.  This field is the external $\vector E_0$ plus the field induced by the depolarization charge;  this is just sufficient to cancel the normal component of $\vector E_0$ at every point $p$ on the surface.  Below we will consider the motion caused by a nonzero $\vector E_0$ in linear response.  To determine this response, we may consider the strength of $\vector E_0$ to be as small as we wish.  In particular, we may consider $\vector E_0$ to be too small to cause appreciable disturbance to pre-existing charges in the body or the fluid. 

Even with no external field, there is a second field in the fluid due to the charge residing on the body.  As discussed above, this charge attracts an opposite charge from ions in the fluid; this ``screening charge" cancels the field except immediately adjacent to the surface.  Thus the charge density in the fluid vanishes and the potential is uniform except immediately adjacent to the body.  Any nonuniformity is caused by nonuniform charge on the body.  The potential difference between point $p$ and the bulk fluid beyond is the zeta potential caused by the charge density at $p$.  The screening charge around this body charge is in equilibrium; its electric field creates no organized motion and no dissipation.  

We now consider the effect of adding the external field $\vector E_0$.  The two fields exist independently.  Each has its independent sources.  The external field exerts forces on the body charge and on the screening charge in the fluid.  As in Sec. \ref{sec:ChiralShear}, we hold the body fixed.  Thus the only motion resulting from $\vector E_0$ is that of the screening charge.  All the forces exerted on the screening charge are transmitted to the fluid, resulting in flow of the screening layer.  Since the electric field $\vector E_s$ there is tangential, so is the force and so is the flow.  Thus the force on the screening charge produces a transverse velocity field $\vector u_s(p)$ just outside the screening layer.  It is proportional to the screening charge per unit area and thus proportional to $\zeta(p)$.  The electric field thus induces a sheath of fluid to slip over the body.  A careful discussion of the origin of this slip layer is given by Anderson et al. \cite{Anderson:1985fk,Anderson:1989kx}.

This slip velocity field is dictated by a) the charge distribution over the body and b) the surface electric field $\vector E_s(p)$, which itself is proportional to the external $\vector E_0$.  It is this imposed velocity that produces the force and torque needed to hold the body fixed.  When the body is released, it is this slip velocity that generates electrophoretic motion, given by the sedimentation $(\vector V, \vector \Omega)$ corresponding to the constraining $(\vector F, \vector \tau)$ (Sec. \ref{sec:NoForce}).

\subsubsection{Determining force and torque}\label{sec:Teubner}
Finding the constraining (force, torque) required by a given $\vector u_s(p)$ can be done in several ways.  We first note that the relation between $\vector u_s(p)$ and the resulting force and torque is a linear one: the force is the sum of independent contributions from each surface point $p$.   This is because the force and torque depend linearly on the velocity field $\vector v(\vector r)$ in the fluid, which is in turn linearly dependent on the surface velocity profile $\vector u_s(p)$.  

Teubner \cite{Teubner:1982kq} recognized a fundamental law relating the $\vector u_s(p)$ profile and the resulting force and torque.  This relationship is an example of the Lorentz reciprocal theorem of hydrodynamics[\cite{Happel-Brenner} Sec. 3-5]. This theorem allows one to decouple the effect of shape from that of the $\vector u_s$ profile.  For example we suppose that we know $\vector u_s(p)$ for a given object and wish to know the $x$ component of the force needed to hold it fixed, denoted $\vector F^{(x)}$.  For this, it suffices to know the viscous drag stress caused by moving the uncharged object at speed $V$ in the $x$ direction without rotation. We denote the local stress profile as $\sigma_{Vx}(p)$.  Then $\vector F^{(x)}$ for the {\em charged} body in terms of the known $\vector u_s$ is given by $\vector F^{(x)} = \integral_p ~\sigma_{Vx}(p)\cdot \vector u_s(p)/V$. Analogously, knowing the stress fields for a basis set of $\vector V$ and $\vector \Omega$ for this object allows one to find the full $\vector F$ and $\vector \tau$ needed to keep it fixed resulting from an {\em arbitrary} flow $\vector u_s(p)$ over it.  As noted above, these six drag stresses are all one needs in order to determine the motion of the object with surface flow $\vector u_s(p)$ when released.  This prescription makes clear that there are many possible surface flows that produce identical motion of the object.  However, these do not produce the same flow around the object.  To determine this local flow field one needs the full $\vector u_s(p)$.  

\subsubsection{Flow at large distance}\label{sec:large_distance}
The flow generated by a driven body extends to infinity.  But the character of the flow at large distances depends strongly on the nature of the driving.  It is useful to characterize the distant flow by considering flows made from assemblies of point forces.  A single point force creates a ``force monopole" or Oseen field, discussed in Sec. \ref{sec:StokesletObjects}.  It falls off inversely with distance from the source.  Since the electrophoretic body has no external force on it, this force monopole must vanish.  A pair of equal and opposite forces separated by a vector $\vector a$ is called a force dipole.  If the forces are transverse to $\vector a$, generating such forces requires a torque.  Since the electrophoretic body moves without external torque,  this transverse dipole field must also vanish.  There is no such restriction on longitudinal dipoles, whose forces are parallel to $\vector a$.  Such dipole fields at infinity are found by taking the gradient of the monopole field for the given $\vector F$ and multiplying by $\vector a$.  It falls off inversely with the {\em square} of distance from the source.  Such a force dipole field is produced, \eg by a charged colloid tethered to a neutral colloid by a tether of length $\vector a$.  Naturally sources made by pairs of dipoles, which fall off faster than inverse distance squared, can also be generated by electrophoretic bodies.  

One such flow must be present for any body moving at velocity $\vector V$.  Any solid body with a nonzero volume $W$ must displace fluid as it moves.  The net displacement of fluid in a time $\Delta t$ is the same as if the body had been removed from its initial position, and then re-inserted at the new position $\Delta t$ later. The removal results in an inward displacement $-W \vector r/ r^3$ at large distance $r$, like the electric field from a point charge.  The reinsertion creates a similar outward displacement about the new position, displaced by $\vector V \Delta t$.  The net fluid displacement during time $\Delta t$ is thus $W(-\vector r /r^3 + (\vector V \Delta t)/|\vector r + \vector V \Delta t|^3)$ for large distances $r$.  This gives a fluid velocity $\vector v(\vector r)= W((3 \vector V \cdot \hat r) \hat r - \vector V)/r^3$.  This flow is called a ``mass dipole" .  

\subsubsection{Non-electrical surface driving}\label{sec:nonelectrical} 
Other forms of driving act via production of surface flows.  The electrophoresis described above is induced by an electrical potential.  Analogous flows may be induced by thermal or chemical gradients.   The term ``phoresis" denotes this class of motions. The review by Anderson\cite{Anderson:1989kx} gives a unified description of these phoretic phenomena.  Their common element is the generation of a velocity field $\vector u_s(p)$ analogous to the electrophoretic $\vector u_s(p)$ discussed above.  

Other, internal, driving mechanisms can also create a surface velocity $\vector u_s(p)$.  Many implementations of active particles are of this form. Here each body's direction is set internally without being proportional to an external vector. We call such objects ``swimmers" \cite{Marchetti:2013pi}. For example one part of the object may be the site of a chemical reaction that drives flow toward that site \cite{MoranAnnRevFluidMech2017}.  Alternatively the object may be a living organism with cilia or other propulsive structures \cite{Marchetti:2013pi}. 

Whenever the active velocity takes the form of an imposed slip velocity at the surface, methods described above are useful.  In particular the Lorentz Reciprocal relations allow one to infer how the force on the immobilized object depends on its shape.  Once the force is known, the mobility tensors of Eq. (\ref{eq:bigmateqn}) allow one to determine the swimming motion.

\section{Stokeslet Objects}\label{sec:StokesletObjects}
\setcounter{equation}{0}
\setcounter{figure}{0}

				The sections above have surveyed a range of possible motions of asymmetric objects under different forms of driving.  Up to now we have been silent about how these response tensors depend on the geometric shape of the object. There is a standard way to address this question.  First, one determines the flow around the object given the boundary conditions at its surface.  Next one integrates the surface stress to find the force and torque on it.  Finally, one relates the velocity and angular velocity to this force and torque.  \cite{delaTorre:1981hp, Youngren:1975jx}.  
				
				If the shape is compatible with a standard symmetric co-ordinate system, such as an ellipsoid, then the flow field and forces can be determined by symmetry-dependent multipole expansions\cite{Fair:1989kx}.  Various discrete boundary element methods\cite{Read:1997fk, Youngren:1975jx,Yan:2015ek} are available to find the flow and the response tensors.  Another method is to represent the hydrodynamic field by random walkers \cite{Hubbard:1993ye}.  An alternative approach\cite{delaTorre:1981hp,Ortega:2011rb,Cichocki:1994xr,Ekiel-Jezewska:2009fe} approximates the desired object as an assembly of standard building blocks (\eg spheres). In this section we introduce a simplified and concrete way of determining response tensors for general, asymmetric objects.  

\begin{figure}[htbp]
\begin{center}
\includegraphics[width=.15\textwidth]{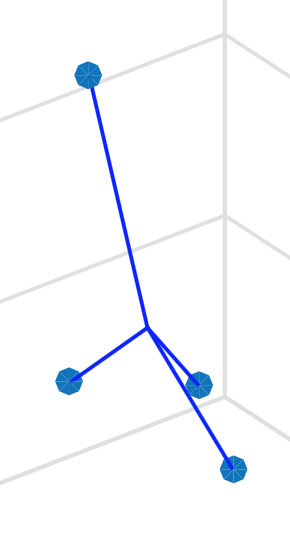}\quad\quad\quad
\includegraphics[width=.3\textwidth]{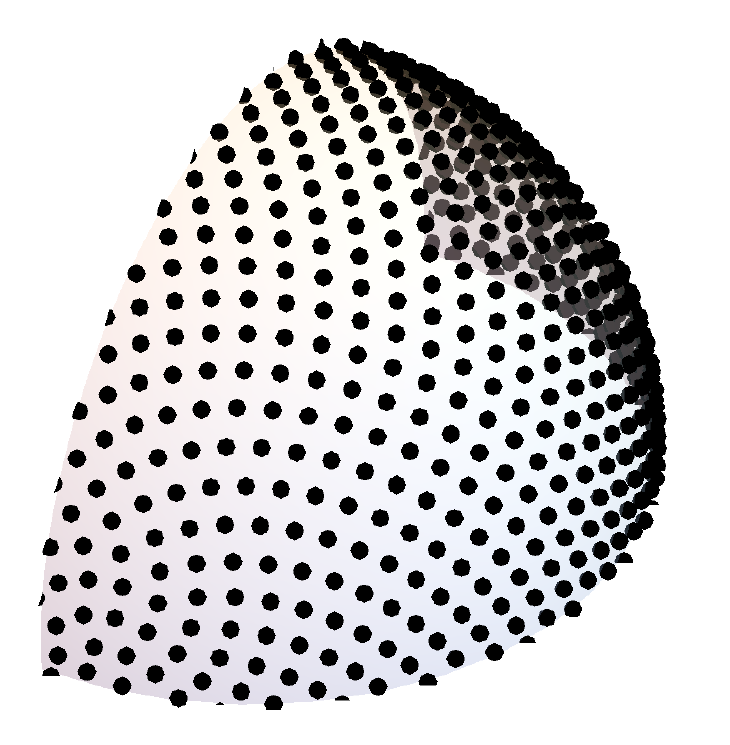}
\end{center}
\caption{\label{fig:stokesletExamples}
Examples of Stokeslet objects. left: a four-Stokeslet object used to show interactions described in Sec. \ref{sec:HydrodynamicInteraction}.  right: 1999 Stokeslets spread over an imaginary quarter sphere.  The distance between the spheres is much larger than their radius.
}
\end{figure}

				To show the method clearly, we shift our focus from rigid solid objects to a conceptually simpler object: an assembly of arbitrarily small spheres of radius $a$ held at fixed displacements from each other by hydrodynamically invisible rods.  Such spheres are known as Stokeslets, and so we call the assembly a ``Stokeslet object."  Two examples are pictured in Fig. \ref{fig:stokesletExamples}.  Using Stokeslet objects to represent physical bodies was introduced by Kirkwood and Riseman \cite{Kirkwood:1948bf} to describe polymers in a fluid.  

				If a Stokeslet object is caused to move, the Stokeslets exert a force on the fluid and cause it to flow.  The velocity field at position $\vector r\,'$ in the presence of an isolated Stokeslet at $\vector r$ exerting a force $\vector f$ on an otherwise quiescent fluid is given by the well-known Oseen law \cite{Happel-Brenner,Kim:2005fr}
				\begin{equation}\label{eq:Oseen}
\vector v(\vector r\,') = {1 \over 8 \pi \eta |\rho|}~ (\vector f + (\vector f \cdot \hat\rho)~ \hat \rho )
\definedas \tenoseen(\vector \rho) \cdot \vector f
\end{equation}
where $\vector\rho \definedas \vector r\,' - \vector r$, $\hat \rho$ is the unit vector along $\vector \rho$, and $\eta$ is the viscosity. That is, $\vector v$ is proportional to $\vector f$ via the ``Oseen tensor" $\tenoseen$.  In a Stokeslet object, this velocity caused by a given Stokeslet is felt by the other Stokeslets.

A Stokeslet moving at velocity $\vector u$ in a fluid which is itself moving at velocity $\vector v$ gives a force $\vector f$ to the fluid.  This $\vector f$ is proportional to $\vector u - \vector v$ via the well-known Stokes drag coefficient $\gamma = 6\pi \eta a$: 
\begin{equation}\label{eq:StokesDrag}
\vector f = \gamma~ (\vector u - \vector v)
\end{equation}
This same force produces the Oseen flow generated by the Stokeslet.  When other Stokeslets are present, the $\vector v_\alpha$ at Stokeslet $\alpha$ comes from the Oseen flow from the other Stokeslets.  We denote the velocity at Stokeslet $\alpha$ due to Stokeslet $\beta$ by $\vector v_{\alpha\beta} = \tenoseen(\vector r_\alpha - \vector r_\beta) \cdot \vector f_\beta$.  The total force $\vector f_\alpha$ is then given by 
\begin{equation}\label{eq:ficonsistency}
\vector f_\alpha = \gamma~ (\vector u_\alpha - \vector v_\alpha) = 
\gamma \vector u_\alpha - \gamma \sum_\beta \vector v_{\alpha\beta} = \gamma~\vector u_\alpha - \gamma \sum_\beta \tenoseen_{\alpha\beta} \cdot \vector f_\beta
\end{equation}
Here $\vector u_\alpha$ is the velocity of Stokeslet $\alpha$ and $\tenoseen_{\alpha\beta} \definedas \tenoseen(\vector r_\alpha - \vector r_\beta)$.
From these facts we can readily find the force on a Stokeslet object like the ones in Fig. \ref{fig:stokesletExamples}.  This equation amounts to a linear relationship between the $\{\vector f_\alpha\}$ and the $\{\vector u_\alpha\}$ \cite{Kirkwood:1948bf}
\begin{equation}\label{eq:fi2ui}
\vector f_\alpha + \gamma \sum_\beta \tenoseen_{\alpha\beta} \cdot \vector f_\beta = \gamma ~\vector u_\alpha  
\end{equation}
If there are $N$ Stokeslets, this amounts to $3N$ equations for the $3N$ force components $\{\vector f_\alpha\}$.  The equations determine the $\{\vector f_\alpha\}$.  Further, since the Stokeslet object is a rigid object, the $\vector u_\alpha$ are determined by the object's velocity $\vector V$ and angular velocity (about the origin) $\vector \Omega$.  
\begin{equation}\label{eq:VOmega2ui}
\vector u_\alpha = \vector V + \vector \Omega \cross \vector r_\alpha
\end{equation}
Thus Eqs. (\ref{eq:VOmega2ui}) and (\ref{eq:fi2ui}) determine the $\vector f_\alpha$ via a system of linear equations.  Finally, the total force $\vector F$ and torque $\vector \tau$ given to the fluid are simply 
\begin{equation}\label{eq:fi2F}
\vector F = \sum_\alpha \vector f_\alpha \quad \hbox{and} \quad \vector \tau = \sum_\alpha \vector r_\alpha \cross \vector f_\alpha
\end{equation}
This prescription determines the proportionality between motion and (force, torque), and thus specifies the response tensors such as $\RT_{V F}$ and $\RT_{\Omega F}$.  Thus for Stokeslet objects, finding these response functions is a matter of straightforward algebraic operations.  

Qualitatively, the effect of the inter-Stokeslet interactions is to weaken the individual drag forces.  A lone Stokeslet moving at velocity $\vector V$ experiences the full stokes drag force $\vector f$.  However there are many other Stokeslets moving at the same $\vector V$, their Oseen flows lie in the same general direction, thus reducing the drag force on any one Stokeslet.  Indeed, $N$ Stokeslets within a radius $R$ create a total Oseen flow of order $f N/R \goesas V N (a/R)$.   Thus even a number of Stokeslets $N \greaterthanorabout R/a$ suffices to reduce the flow by a factor of order unity.   (This ``screening threshold" $N$ is much smaller than that of a fixed density $\rho$ , namely $\rho R^3$.)  For $N$ much larger than the screening threshold, the Stokeslets exert forces much smaller than $f$, and their relative velocities $\vector u_\alpha - \vector v_\alpha$ are much smaller than $\vector V$.  Nearly all of $\vector V$ is shielded or screened from the Stokeslets.  Thus even a dilute set of Stokeslets with separation much greater than $a$ suffices to produce strong hydrodynamic screening \cite{Zimm:1956xu}. 

\subsection{Solid bodies}\label{sec:solidBodies} 

\begin{figure}[htbp]
\begin{center}
\includegraphics[width=.6\textwidth]{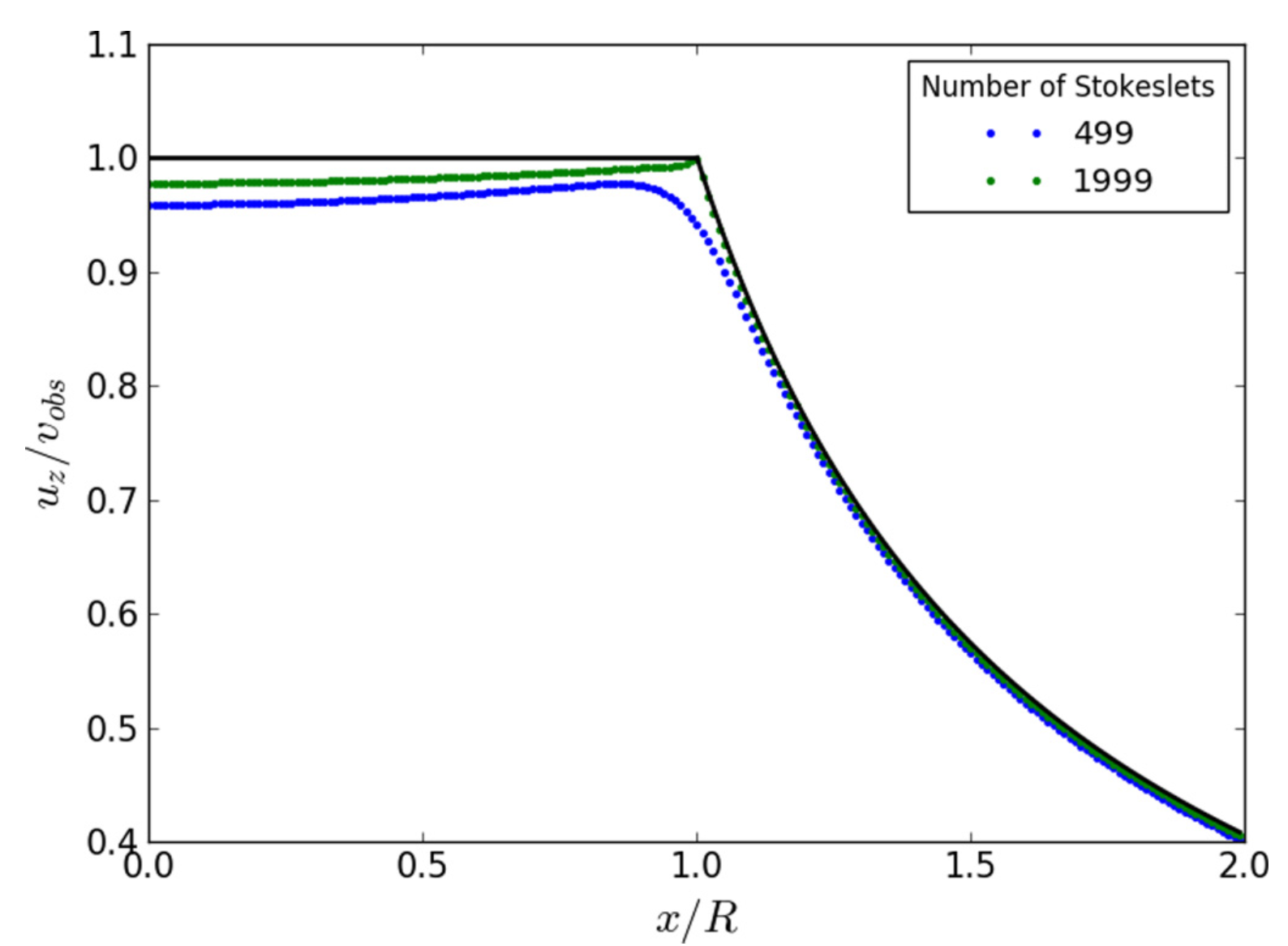}
\caption{Flow around a stokeslet sphere under unit force. Plot shows component of fluid velocity in the direction of the force vs transverse distance from the center, after \cite{Mowitz:2017kx}. Velocities are normalized to Stokes velocity.  Black solid line, solid sphere.  Upper dotted line: 1999-Stokeslet sphere pictured in Fig. \ref{fig:stokesletExamples}. Stokeslet cross-sections cover 11 percent of the surface area calculated using Eqs. (\ref{eq:VOmega2ui}) and (\ref{eq:fi2F}). Velocity at sphere surface is 0.9 percent lower than the stokes velocity for the applied force.  Lower dotted line: 499-Stokeslet sphere with the same area fraction.  To good accuracy, fluid within the sphere moves with the Stokeslets.  
}
\label{fig:ScreenedSphere}
\end{center}
\end{figure}

This strong hydrodynamic screening enables an economical way to represent a solid body as a stokeslet object \cite{Mowitz:2017kx,Braverman:2020oq}.  Fig. \ref{fig:ScreenedSphere} shows the results for a sphere represented by Stokeslets as shown in Fig. \ref{fig:stokesletExamples}.  Similar results have been demonstrated for other shapes \cite{Braverman:2020oq}.

		  Though Stokeslet objects do not implement exactly the no-slip boundary condition of a solid surface, they do exclude flow from the interior, and thus they give an accurate representation of the exterior flow and thence the response tensors.  Evidently, this method is not restricted to a special class of shapes or conformity with particular co-ordinate systems.  Thus they are suitable for studying unrestricted asymmetric shapes that are not specified {\em a priori}. It is also suitable for objects that admit some penetration of the external flow, such as colloids with a polymeric corona.  The only requirement for accurate representation is that the distance between Stokeslets be smaller than the local radius of curvature.   

In Sec. \ref{sec:Phoresis}, we noted that the distinctive dynamics of driven asymmetric objects is preserved in types of driving without external forces.  Here too one must determine the appropriate response tensors, \eg $\RT_{F E}$. For this it is necessary to determine the force on the immobilized object with the phoretic velocity field $\vector u_s$ acting over its surface.  Here, the Stokeslet-object representation also appears promising \cite{Braverman:2020oq}.

Likewise, the spacing should be small compared to the separation of the object from external sources.  With this condition met, one may use Stokeslet objects to examine hydrodynamic interactions between asymmetric bodies.  The next section explores these interactions.

\section{Two objects under external force: Hydrodynamic interaction}
\label{sec:HydrodynamicInteraction}
\setcounter{equation}{0}
\setcounter{figure}{0}

Two forced colloids interact hydrodynamically over large distances. As
discussed in Sec.~\ref{sec:StokesletObjects}, a forced colloid of size
$a$ creates a long-range flow whose velocity falls off with distance
$r$, to leading order in $a/r$, as $1/r$. This is the Stokeslet flow
introduced in Eq.~(\ref{eq:Oseen}). A second colloid located a
distance $r$ away has its velocity changed, to the same leading order,
by that flow velocity. Similarly, the change in the angular velocity
of the second object is equal to the flow vorticity, which decays as
$1/r^2$. Thus, the leading-order (monopole-monopole) translational and
rotational hydrodynamic interactions are long-ranged and independent
of object shape; the two objects might as well be replaced by
featureless points. Higher-multipole interactions do depend on
object shape.  Thus they are of particular interest to us. Although
these interactions fall off more quickly than the dominant $1/r$ term,
they still have slow power-law decays.

This discussion implies that the forced motion of a single object, as
described in Sec.~\ref{sec:ConstantForce}, may change significantly in
the presence of another object, even if the two are
well-separated. The change should be sensitive to the objects'
shapes. This raises basic questions, for example, whether the objects move
together, or apart, or keep a fixed distance, or whether they are trapped in
an oscillating bound state.  A further question is the effect of the interaction on
the alignment of objects with the force as described in
Sec.~\ref{sec:ProgrammedForcing}.  Besides the interest in the pair
interaction itself, this alignment has implications for the collective dynamics of
a suspension of many particles, as will be shown in
Sec.~\ref{sec:collective}.

Let us first briefly summarize the relevant discussion in
Sec.~\ref{sec:ConstantForce} regarding the forced motion of a single
asymmetric colloid. The motion of a single object already forms a
complicated dynamical system. The orientation with respect to the
force determines the rotational response, which in turn determines the
change of orientation and position in the next instant. The resulting
orientational dynamics may show various features such as stable and
unstable fixed points, closed orbits, etc.. The object's response is
accounted for by four $3\times 3$ response tensors, $\RT_{VF}$,
$\RT_{\Omega\tau}$, $\RT_{V\tau}$, and $\RT_{\Omega F}$. The
orientational behavior under force is encoded in the off-diagonal
$3\times 3$ block $\RT_{\Omega F}$.

It is clear that a similar characterization for a pair of forced
asymmetric colloids would be more complicated[\cite{Kim:2005fr}, Chapter 7]. Apart from the
variables required to define the orientations of the two objects,
there are the additional components of the vector $\vecR$ connecting
their forcing points. We will restrict most of the discussion,
therefore, to a simple limit\,---\,a pair of identical, axially
aligning objects under an identical, constant force and no
torque\,---\,and focus in particular on the case where the
inter-particle distance is large. We begin with a general
consideration of the instantaneous response of two arbitrarily shaped
objects, \ie the structure and symmetries of the required response
tensors. We will proceed to the more detailed properties of the
hydrodynamic interaction between the two objects in the limit of large
separation, $R\gg a$. Finally, we describe various types of
trajectories which can be observed for pairs of asymmetric, axially
aligning objects.

\subsection{General response of a particle pair}
\label{sec:pairtensors}

To characterize the linear response of two objects or particles, we need 16 response
tensors like the ones defined in Sec.~\ref{sec:Response}. These are
$\RT_{Y_\alpha X_\beta}$, where $Y=(V,\Omega)$, $X=(F,\tau)$, and the
indices $(\alpha,\beta)=(1,2)$ label the two particles. For example,
the tensor $\RT_{\Omega_2F_1}$ relates the force acting on particle 1
with the angular velocity of particle 2, $\vecOmega_2
= \RT_{\Omega_2F_1}\cdot \vecF_1$. All the tensors depend on the
configuration of the pair. The self ($\alpha=\beta$) tensors, similar
to the single-particle ones presented in Sec.~\ref{sec:Response}, give
the response of one of the objects to the force and torque acting on
itself. They depend, nevertheless, on the entire configuration of the
pair. The off-diagonal $\alpha\neq\beta$ tensors give the hydrodynamic
interaction between the two objects, which is the focus of this
section.

For the same reasons presented in Sec.~\ref{sec:Response}, the whole
response matrix must be positive-definite (and therefore also
symmetric). That is, for any set of forces and torques $\vec{X}_\beta$
the dissipation rate $\sum_\beta\vec{X}_\beta\cdot\vec{Y}_\beta
= \sum_{\alpha,\beta} \vec{X}_\beta\cdot\RT_{Y_\beta
X_\alpha}\cdot\vec{X}_\alpha$ must be positive.  For the self tensors
the symmetry implies
\begin{equation}
  \alpha=\beta:\ \ \ \ \RT_{V_\alpha\tau_\alpha}=(\RT_{\Omega_\alpha F_\alpha})^T,
\end{equation}
as in the case of a single object(even though these tensors are
perturbed by the second object). For the interaction tensors, we have
\begin{equation}
  \alpha\neq\beta:\ \ \ \ \RT_{Y_\alpha X_\beta}=(\RT_{Y_\beta X_\alpha})^{\rm T}.
\end{equation}
This eliminates six independent tensors out of the sixteen.

In the special case where the two objects are in the exact same
orientation, \ie fully aligned, they are indistinguishable and satisfy
another symmetry\,---\,the simultaneous exchange of object labels and
inversion of their separation vector,
\begin{equation}
  \mbox{aligned pair, $\alpha\neq\beta$:}\ \ \ \ \RT_{Y_\alpha
  X_\beta}(\vecR)=\RT_{Y_\beta X_\alpha}(-\vecR).
\label{exchange}
\end{equation}
The way to achieve these {\it a priori} improbable aligned states has
been the subject of Sec.~\ref{sec:ProgrammedForcing}. The additional
symmetry, therefore, should help us clarify the effect of the
hydrodynamic interaction on alignment.

Each tensor can be separated into even ($+$) and odd ($-$)
contributions according to whether the contribution keeps or changes
its sign under inversion of the separation vector $\vecR\rightarrow
-\vecR$. We call this symmetry/antisymmetry $\vec{R}$-inversion
symmetry/antisymmetry (as opposed to full inversion). We consider the
corresponding responses $\vec{Y}_\alpha^\pm$ (velocity or angular
velocity) to identical drives $\vec{X}_\alpha=\vec{X}_\beta=\vec{X}$
(force or torque) applied to the two objects. Due to the exchange
symmetry (\ref{exchange}) we have
\begin{equation}
  \mbox{aligned pair, $\alpha\neq\beta$:}\ \ \ \ \RT^\pm_{Y_\alpha X_\beta}(\vecR)
  = \pm \RT^\pm_{Y_\alpha X_\beta}(-\vecR) = \pm \RT^\pm_{Y_\beta X_\alpha}(\vecR).
\end{equation}
Hence,
\begin{equation}
  \mbox{aligned pair, $\alpha\neq\beta$:}\ \ \ \ \vec{Y}^\pm_\alpha
  = \RT^\pm_{Y_\alpha X_\beta} \cdot \vec{X} = \pm \RT^\pm_{Y_\beta
  X_\alpha} \cdot \vec{X} = \pm \vec{Y}^\pm_\beta.
\end{equation}
We conclude that, generally, for an aligned pair under the same drive,
even contributions to the interaction tensors cause the two objects to
move together, and odd contributions, to move oppositely to one
another. Take two sedimenting spheres, for example. By the
$\vecR$-inversion symmetry their response to forces is even, while their response to torques is odd. As a result, they do not develop relative velocity but do
cause one another to rotate in opposite senses \cite{Happel-Brenner}. A particularly
important conclusion is that {\em odd contributions to
$\RT_{\Omega_\alpha F_\beta}$ would cause two aligned objects to
develop relative angular velocity and degrade their alignment} \cite{Goldfriend:2015ek}.

\subsection{Hydrodynamic interaction at large separation}
\label{sec:multipole}

When the separation between the two particles is much larger than
their size, their instantaneous hydrodynamic interaction can be
conveniently decomposed using a multipole expansion in $a/R$, similar to
electrostatics. We write
\begin{equation}
  \RT_{Y_\alpha X_\beta}(\vecR) = \sum_{n=0}^\infty \RT^{(n)}_{Y_\alpha X_\beta}(\vecR),
\end{equation}
where the $n$th term is proportional to $1/R^n$. We restrict the analysis to the response under force,
with no torque, $X_\beta=F_\beta$. For a more complete discussion see
Ref.~\cite{Goldfriend:2015ek}.

The $R^0$ term in this expansion does not include interactions. The self
($\alpha=\beta$) tensors are the single-object response tensors of
Sec.~\ref{sec:Response}, and the interaction ($\alpha\neq\beta$)
tensors vanish.
\begin{equation}
  n=0:\ \ \ \RT^{(0)}_{Y_\alpha X_\beta} = \RT_{YX} \delta_{\alpha\beta}.
\end{equation}

In the $1/R$ term the self tensors remain the same. In fact, the
single-object self tensors get their first correction only at
$n=4$ \cite{Goldfriend:2015ek}. The translational interaction is as if
the two particles were points. It is given by the Oseen tensor,
$\tenoseen \sim 1/R$, introduced in Sec.~\ref{sec:StokesletObjects},
Eq.~(\ref{eq:Oseen}). The rotational interaction is zero because its
leading term is proportional to the vorticity
$\nabla\times\tenoseen \sim 1/R^2$.
\begin{equation}
  n=1:\ \ \ \RT^{(1)}_{V_\alpha F_\beta}(\vecR) = \tenoseen(\vecR)
  (1-\delta_{\alpha\beta}),\ \ \ \
  \RT^{(1)}_{\Omega_\alpha F_\beta}(\vecR) = 0.
\end{equation}
It follows from Eq.~(\ref{eq:Oseen}) that the $n=1$ interaction
tensors are even under $\vecR$-inversion.

The $1/R^2$ term is where the particles' shape first enters. Its self
tensors vanish as mentioned above. The interaction is as if one of the
objects were a point and the other were not. The translational
interaction has two contributions. The first arises from the response
of object $\alpha$ to the flow-velocity gradient $\grad\vecv$ around
it, created by the force acting on point-like particle $\beta$. The
velocity response to a flow gradient is a shape-dependent property of
the object. We characterize it by a dimensionless tensor $\tenPhi$,
such that, in general, $V_i=a\Phi_{i,jk}\partial_k v_j$. (This response
includes the shear response depicted in Fig.~\ref{fig:PaddleTwist}.)
The rank-3 tensor $\Phi_{i,jk}$ transforms a rank-2 tensor
($\partial_k v_j$) into a vector ($V_i$). In the present case $V_i$ is
the correction to particle $\alpha$'s velocity and $v_j
= \oseen_{jl}(\vecR)F_{\beta,l}$ is the flow generated by point-like
particle $\beta$. The second contribution to the translational
interaction originates from the force dipole induced in particle
$\beta$ by the force $\vecF_\beta$ acting on it. The induced force
dipole $S_{ij}$ due to an applied force $\vecF$ is another
shape-dependent property. We characterize it by another tensor,
$\boldsymbol{\tilde{\Phi}}$, such that, in general,
$S_{ij}=a\tilde{\Phi}_{ij,k}F_k$. The rank-3 tensor
$\tilde{\Phi}_{ij,k}$ transforms a vector $F_k$ into a rank-2 tensor
$S_{ij}$. One can prove that the two response tensors are simply
related as
$\boldsymbol{\tilde{\Phi}}=\tenPhi^T$ \cite{Goldfriend:2015ek}. The
induced force dipole is projected by
$\nabla\tenoseen(-\vecR)=-\nabla\tenoseen(\vecR)$ to create a flow
velocity at the position of point-like particle $\alpha$ and advect
it. To sum up this part,
\begin{equation}
  n=2:\ \ \ [\RT^{(2)}_{V_\alpha F_\beta}(\vecR)]_{il} = a\Phi_{i,jk}\partial_k
  \oseen_{jl}(\vecR) - a\Phi^T_{jk,l}\partial_j \oseen_{ki}(\vecR).
\end{equation}
The $n=2$ rotational interaction arises from the vorticity
$\nabla\times\vecv$ of the flow $\vecv=\tenoseen\cdot\vecF_\beta$
generated at the position of particle $\alpha$ by the force acting on
the point-like particle $\beta$. Thus, it does not depend on particle
shape,
\begin{equation}
  n=2:\ \ \ [\RT^{(2)}_{\Omega_\alpha F_\beta}(\vecR)]_{il} = \epsilon_{ijk}
  \partial_j \oseen_{kl}(\vecR).
\end{equation}
The $n=2$ interaction tensors are proportional to $\nabla\tenoseen$;
hence, they are odd under $\vecR$-inversion.

Since each subsequent multipole involves another derivative of
$\tenoseen$, the parity of the interaction tensors alternates as
$(-)^{n+1}$ \cite{Goldfriend:2015ek}.

Finally, if the configuration of the pair is symmetric under
$\vecR$-inversion, all even-$n$ translational ($V_\alpha F_\beta$)
multipoles and all odd-$n$ rotational ($\Omega_\alpha F_\beta$)
multipoles of the interaction must vanish.

We now turn to the consequences of this analysis for the dynamics of a
particle pair.

\subsubsection{Relative translation.}

Combining the results of Secs.~\ref{sec:pairtensors} and
\ref{sec:multipole}, we reach the following
conclusions concerning the instantaneous translation of two forced
objects:
\begin{enumerate}

\item Two arbitrarily shaped forced objects generally develop
relative translation.  To leading order in large separation the
relative velocity decays as $a/R^2$. Whether the objects move closer
or further apart under this leading interaction depends on their shape through
the (instantaneous) tensor $\tenPhi$. If the objects are aligned, only
even powers of $1/R$ contribute to their relative translation.
\label{item:relativetrans}

\item Two forced objects whose configuration is symmetric under
$\vecR$-inversion move together without relative translation
velocity. The interaction causes their common velocity to change only
through odd powers of $1/R$.
\label{item:inversion}

\end{enumerate}

Conclusion~(\ref{item:relativetrans}) has consequences for the
collective dynamics of suspensions of asymmetric particles, as will be
discussed in Sec.~\ref{sec:collective}.

Conclusion~(\ref{item:inversion}) clarifies experimental and
theoretical results concerning relative motion of forced spheres in
systems where $\vecR$-inversion symmetry is
broken \cite{SquiresJFM2001}. For example, two spherical particles
repelled from a planar surface move toward one another parallel to the
surface \cite{SquiresPRL2000}. Two spherical particles pushed along a
curved path move radially in opposite
directions \cite{SokolovPRL2011}.

Concerning configurations with $\vecR$-inversion symmetry
conclusion~(\ref{item:inversion}) has limited use. While the symmetry
turns off relative translation it turns on relative rotation; see the
next sub-section. Thus, unless the objects are spherical, the
configuration will lose its inversion symmetry in the next instant.

\subsubsection{Relative rotation.}
\label{sec:relativerotation}

The parallel conclusions concerning the instantaneous rotation of two
forced objects are as follows.
\begin{enumerate}

\item Two arbitrarily shaped forced objects generally develop relative
rotation. To leading order in large separation the relative angular
velocity decays as $1/R^2$ and is insensitive to particle size and
shape. If the objects are aligned, only even powers of $1/R$
contribute to their relative rotation.
\label{item:relativerot}

\item Two forced objects whose configuration is symmetric under
$\vecR$-inversion have relative rotation.

\end{enumerate}

An important consequence of conclusion (\ref{item:relativerot}) is
that the hydrodynamic interaction causes relative rotation of two
aligned colloids. Thus, it should linearly degrade alignment. At large
separation the degradation effect falls off as $1/R^2$. If the objects
translate away from one another, the degradation will weaken and
alignment with the force will be eventually restored. In the presence
of many particles, however, the slowly decaying degradation effect
sets a strong upper bound on the volume fraction which would allow
large-scale alignment \cite{Goldfriend:2015ek}.

\subsection{Trajectories of particle pairs}

This section has dealt so far with the instantaneous (\ie
configuration-dependent) response of particle pairs to the force. We
turn now to the resulting time-integrated trajectories. Calculations of
such trajectories for the simple examples of axially aligning
spheroids (spheroids whose forcing point is shifted from their
geometrical center) and Stokeslet objects (see
Sec.~\ref{sec:StokesletObjects}) have been presented in
Refs.~\cite{Goldfriend:2015ek,Goldfriend:2016rr}. As expected of this
multi-variable dynamical system, the spatial and temporal behaviors
are diverse. We will briefly present here two exemplary
behaviors. Many more can be found in Ref.~\cite{Goldfriend:2015ek}.

\subsubsection{Repulsion of two axially aligning objects.}

An important effect noted in Sec.~\ref{sec:relativerotation} is the
degradation of alignment by hydrodynamic interaction and its recovery
with increasing separation. To examine this scenario consider two
identical axially aligning objects. At $t=0$ they are perfectly
aligned, positioned at the same height, and separated by $R(0)$
perpendicular to the force. By symmetry the trajectories of the two
particles are identical up to reflection. We denote by $\theta(t)$
their tilt angle away from alignment.

The equations of motion for the separation $R(t)$ and tilt angle
$\theta(t)$ are fully determined by the response tensors
$\RT_{Y_\alpha F_\beta}(\vecR)$ introduced in
Sec.~\ref{sec:pairtensors}. For small tilt ($\theta\ll 1$) and large
separations ($R\gg a$) these equations take the simple
form \cite{Goldfriend:2016rr},
\begin{eqnarray}
  \dot{R} &=& c_1\theta + c_2/R^2,
  \label{Rdot}\\
  \dot{\theta} &=& -c_3\theta + c_4/R^2,
  \label{thetadot}
\end{eqnarray}
where the coefficients $c_i$ are derivable from the response
tensors. The physical origins of the four terms are as follows. The
first term corresponds to the glide of the individual objects when
they are tilted. If we define $\theta>0$ for objects that glide away
from each other, then $c_1>0$. The second term is the large-separation
translational ($V_\alpha F_\beta$) interaction (repulsion for $c_2>0$
and attraction for $c_2<0$). The third term describes the decay of
tilt due to alignment with the force, where $c_3>0$ characterizes the
strength of alignment. The fourth term is the large-separation
rotational ($\Omega_\alpha F_\beta$) interaction ($c_4>0$
corresponding to an interaction that tilts the objects further away
from each other). Even within the simple limit described by
Eqs.~(\ref{Rdot}) and (\ref{thetadot}), the competition between these
four effects leads to rich behavior.

\begin{figure}
\centerline{\includegraphics[width=0.45\textwidth]{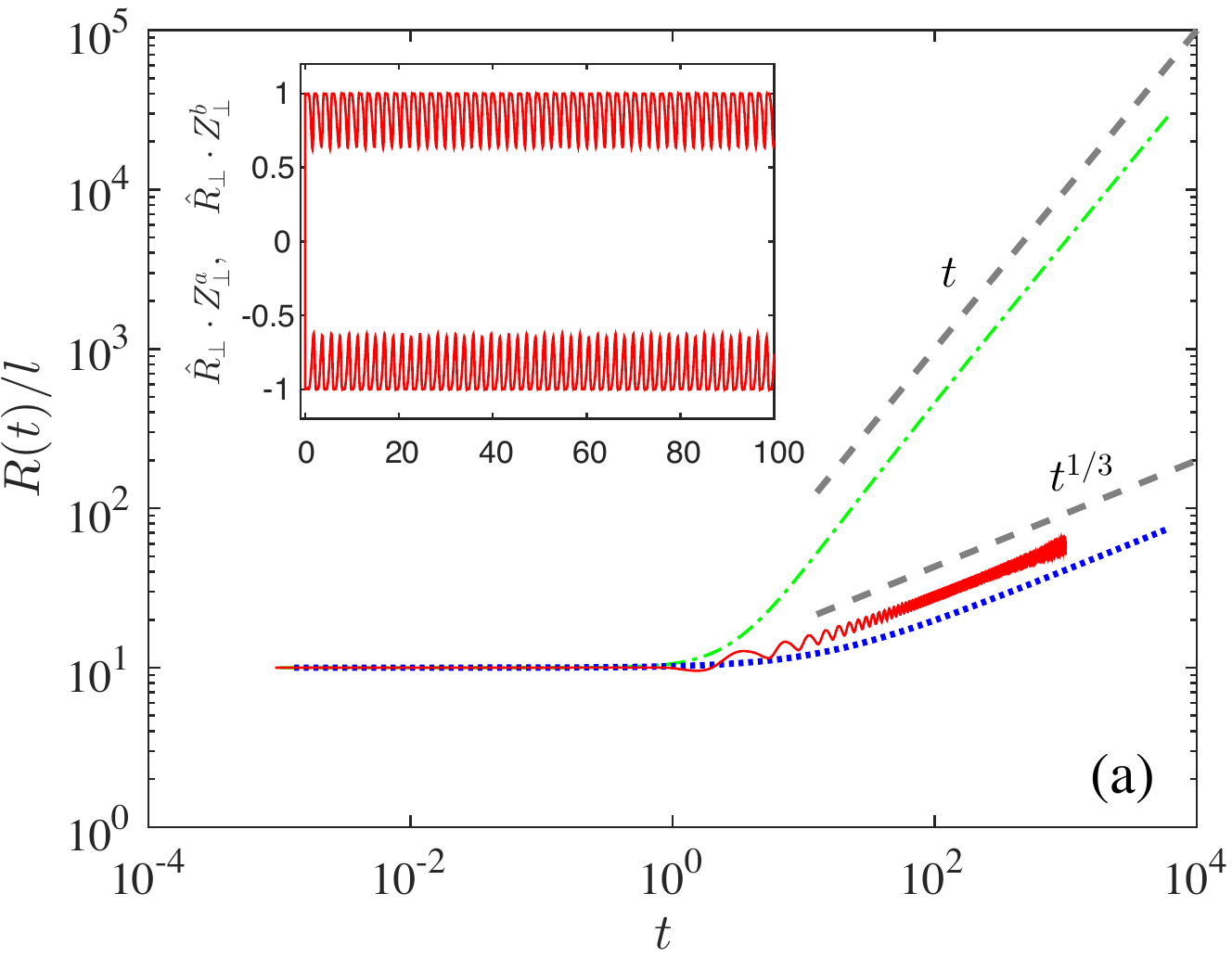}
\hspace{0.3cm} \includegraphics[width=0.45\textwidth]{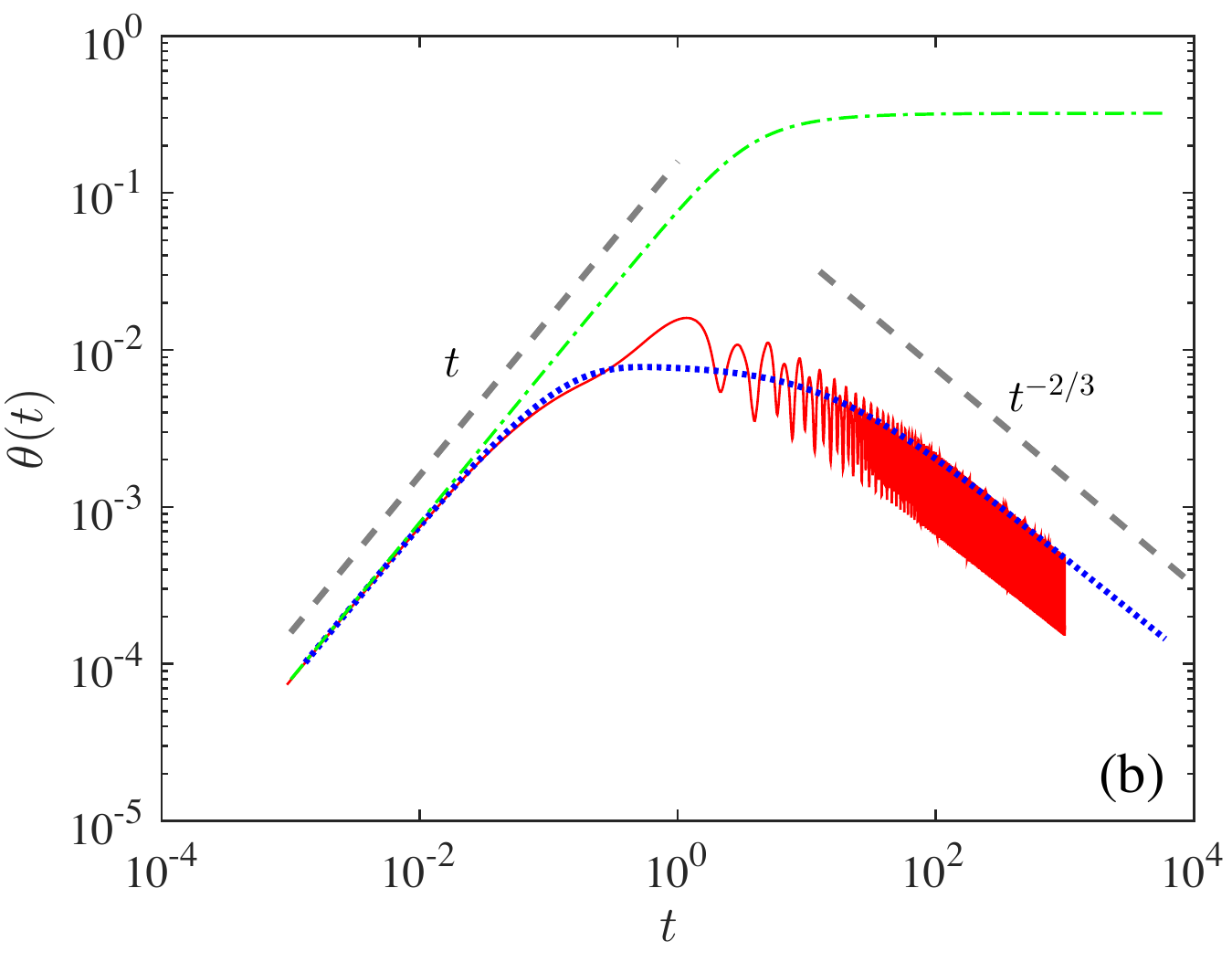}}
\caption{(a) Separation and (b) tilt angle as a function of time for
a pair of axially aligning 4-Stokeslet objects (solid red curves),
axially aligning spheroids (dotted blue curves), and symmetric
spheroids (dash-dotted green curves). After
Ref.~\cite{Goldfriend:2016rr}.}
\label{fig:repulsion}
\end{figure}

If the overall interaction is repulsive (for example, when both $c_2$
and $c_4$ are positive), the separation between the objects grows
indefinitely, and Eqs.~(\ref{Rdot}) and (\ref{thetadot}) become
increasingly accurate. Deriving the asymptotic motion from these
equations is straightforward \cite{Goldfriend:2016rr}. After a linear
accumulation of tilt due to the rotational interaction, the tilt angle
decays back to zero because of alignment. At the same time the
separation increases. Asymptotically we obtain
\begin{equation}
  R(t)\sim t^{1/3},\ \ \ \ \theta(t) \sim t^{-2/3}.
\label{repulsion_asymptotic}
\end{equation}
Figure~\ref{fig:repulsion} demonstrates this repulsion-realignment
behavior. The results were obtained from numerical integration of the
full equations of motion for two axially aligning spheroids and two
irregular 4-Stokeslet objects. The asymptotic laws of
Eq.~(\ref{repulsion_asymptotic}) are confirmed. The figure shows also
the very different dynamics of two symmetric spheroids. The symmetric
particles reach a constant tilt and continue to glide away from each
other at constant velocity (see also Ref.~\cite{Claeys:1993zr}).

\subsubsection{Bound state of two axially aligning objects.}

The translational interaction may be repulsive or
attractive. Similarly, the rotational interaction may tilt the two
objects so that they glide toward or away from each other. In
addition, we can apply time-dependent forcing, as described in
Sec.~\ref{sec:ProgrammedForcing}. Depending on these conditions the
pair may experience along the trajectory both attraction and
repulsion. This may give rise to bound states, where the separation
between the two objects is kept constant or oscillates about a
constant value.

\begin{figure}
\centerline{\includegraphics[width=0.32\textwidth]{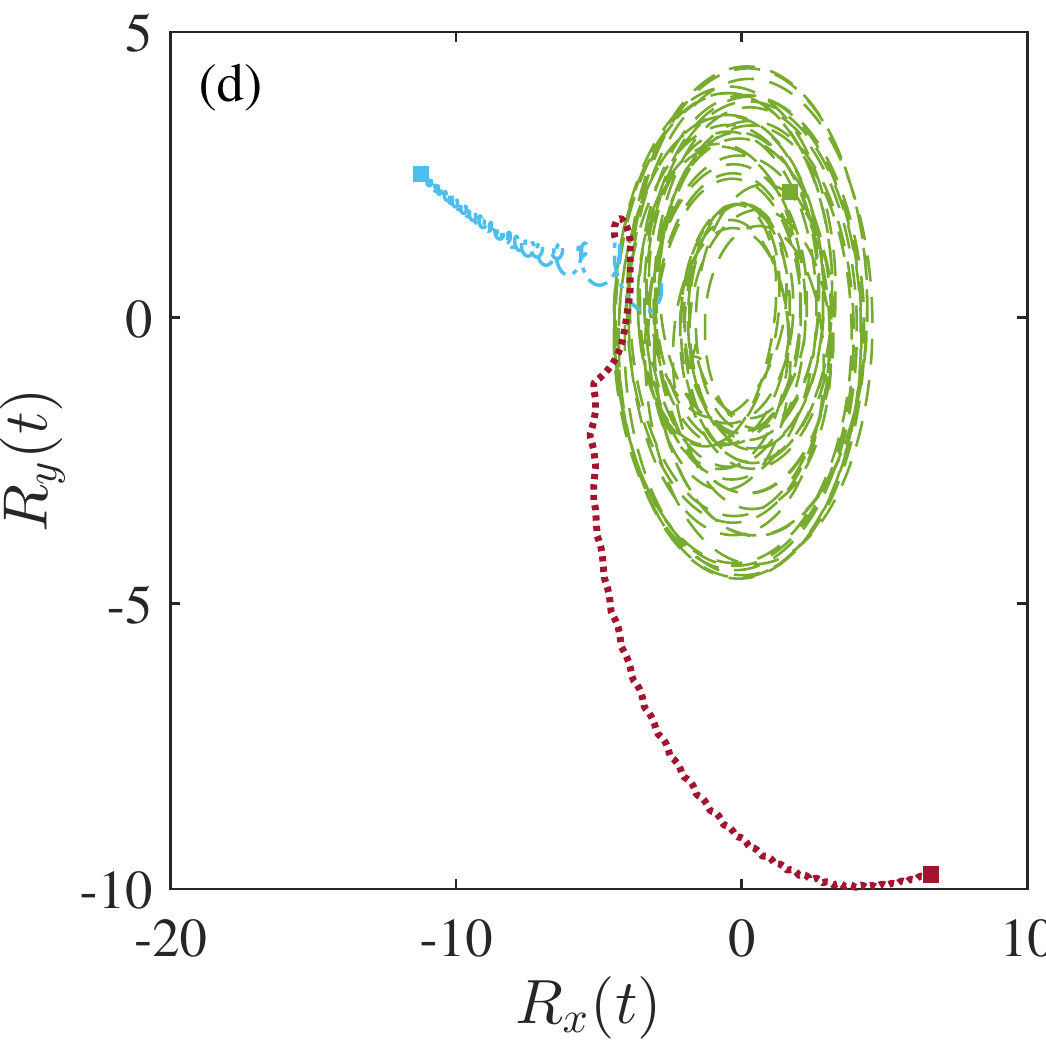}
\hspace{0.15cm}
\includegraphics[width=0.32\textwidth]{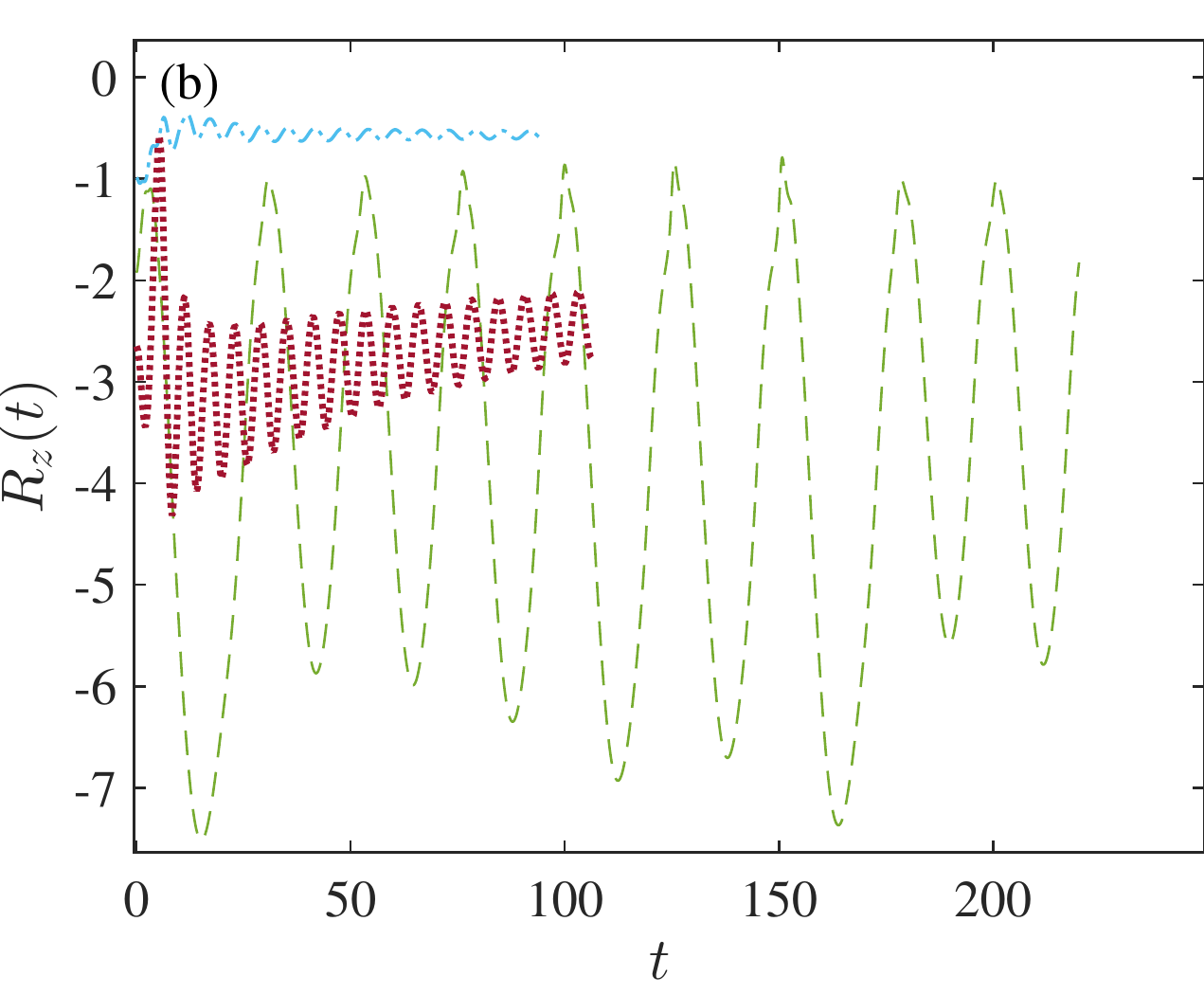}
\hspace{0.1cm}
\includegraphics[width=0.32\textwidth]{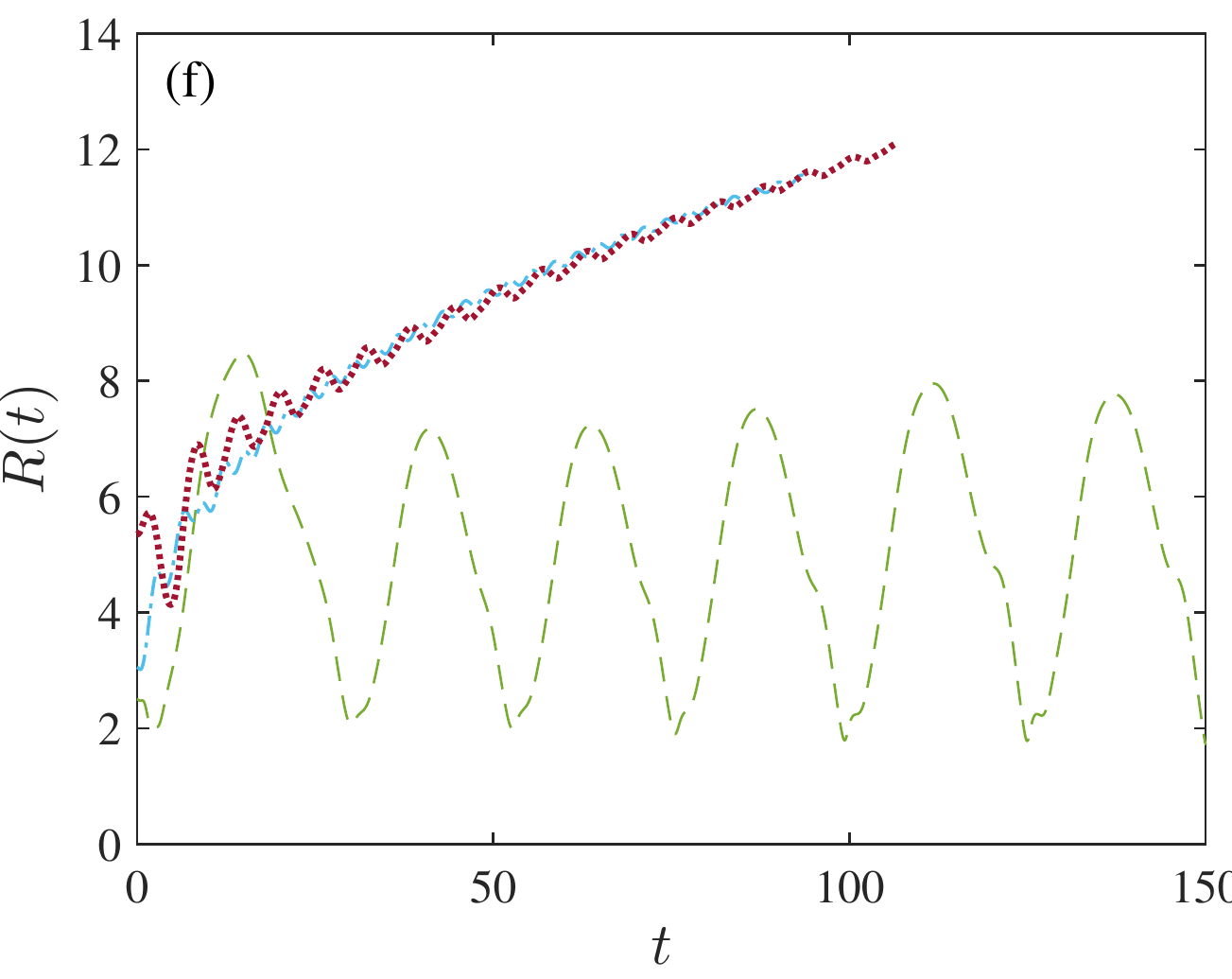}}
\caption{Trajectories of pairs of axially aligning objects. The three
types of curves correspond to three different randomly generated
4-Stokeslet objects. The distances between Stokeslets within an object
are taken randomly between $0$ and $1$. The separation between the
objects is presented in these units. The objects are subject to a
constant force in the $-z$ direction plus a weak rotating force in the
$xy$ plane. The additional force rotates with the same frequency as
the individual object and tilts the total force by an angle of
$0.1\pi$. Left: pair separation projected perpendicular to the
direction of average force. The small squares indicate the locations
at the end of the simulations. Center: pair separation along the
direction of average force. Right: total separation.  Two of the pairs
(black dotted and blue dash-dotted curves) move apart
horizontally. The third pair (green dashed) appears trapped in a
confined region, the separation oscillating about a constant value
(bound state). The elliptic appearance of this orbit is due to the
unequal horizontal and vertical plotting scale. The vertical
separation oscillates for all pairs. Taken from
Ref.~\cite{Goldfriend:2015ek}.}
\label{fig:bound}
\end{figure}

Figure~\ref{fig:bound} demonstrates such a bound state. The figure
shows the trajectories of three pairs, each made of different
axially aligning Stokeslet objects. The objects start from perfect
alignment. They are subject to one of the time-dependent forcing
protocols described in Sec.~\ref{sec:ProgrammedForcing}\,---\,a weak
rotating horizontal force on top of the constant vertical
force. Two of the object pairs show repulsion. The third pair (green
curves) gets trapped in a limit cycle where the two objects orbit
around one another.

Bound states are found also for symmetric objects such as spheroids,
disks, and rods at sufficiently small
separations \cite{Claeys:1993zr,Jung:2006ys,Chajwa:2019yq}. The
mechanism in this case is based on tumbling. The rotational
interaction makes the particles continually rotate in opposite
senses. As a result, they glide away from one another, and then toward
each other, and so forth. The mechanism in the case of asymmetric
objects is different.  Their alignment with the force resists
tumbling, and the bound state requires translational attraction.

\section{Collective dynamics}
\label{sec:collective}
\setcounter{equation}{0}
\setcounter{figure}{0}

The motions of viscous fluids, in general, are characterized by strong
and long-range correlations \cite{HansenMcDonald}. The frequent
collisions of the fluid's molecules allow stresses to diffuse quickly
over large distances. Consequently, as described in
Sec.~\ref{sec:StokesletObjects}, exerting a steady force on a small
region in the fluid creates a flow whose velocity decreases with
distance $r$ only as $1/r$. The same long-range effect is responsible
also for correlations between thermal velocity fluctuations in the
fluid, which decay slowly (as a power law) in both space and time
\cite{HansenMcDonald}.

The motions of particles dispersed in the fluid inherit this
cooperative behavior. It is accentuated further under external
drive. Each of the forced particles exerts on the fluid a localized
force as the one mentioned above. The resulting flows couple the
particles' velocities over large distances. This many-body coupling
leads to complex dynamics. A vivid example is the large chaotic
structures observed in the sedimentation of spherical particles, even
when the Reynolds number, concentration, and thermal motion are all
negligibly small
\cite{SegrePRL1997,LeiPRL2001,Ramaswamy2001,Guazzelli2011}.
Remarkably, the strong coupling leads to the prediction that the
velocity fluctuations of the sedimenting particles at steady state
should increase indefinitely with system
size \cite{CaflischPhysFluids1985}. This prediction is not confirmed
in experiments. It relies on the assumption that the forced particles
are randomly distributed in space\,---\,an issue to which we return toward the
end of this section.

Even richer collective behaviors are found if the suspended particles
have a more complicated response to the driving and to the surrounding
flows.  As we shall see, the responses of a single object
and a pair of objects to the drive, which were outlined in
Secs.~\ref{sec:ConstantForce} and \ref{sec:HydrodynamicInteraction},
leave clear footprints in the continuous description of the
many-particle suspension.

We first discuss in Sec.~\ref{sec:fluidsymmetry} the symmetries obeyed
by continuous viscous fluids close to equilibrium and the ways in
which these symmetries can be broken. In Secs. \ref{sec:chiralfluids_active} and \ref{sec:chiralfluids_passive}
we discuss the unusual behaviors of active and passive chiral
fluids. In Sec.~\ref{sec:asymmetricsuspension}, we present
distinctive phenomena taking place in a sedimenting suspension of
asymmetric particles.

\subsection{Symmetries of a viscous fluid}
\label{sec:fluidsymmetry}

Consider a fluid with a steady flow velocity $\vecv(\vecr)$ and stress
tensor $\tensigma(\vecr)$. If the fluid is invariant to translations,
its total momentum is fixed. Viscous stresses are associated then with
velocity gradients and not with the flow velocity itself. The tensor
of velocity gradients can be written as a sum of a symmetric and an
anti-symmetric strain-rate tensors,
\begin{equation}
  \partial_iv_j = (v_{ij}+\bar{v}_{ij})/2,\ \ \
  v_{ij}\equiv\partial_iv_j+\partial_jv_i,\ \ \
  \bar{v}_{ij} \equiv \partial_iv_j-\partial_jv_i.
\end{equation}
To the leading orders in both the strength and nonuniformity of the
flow, the viscous stress tensor is written as a linear response to the
velocity gradients,
\begin{equation}
  \sigma_{ij} = \eta_{ijkl} v_{kl} + \bar{\eta}_{ijkl} \bar{v}_{kl},
\end{equation}
where $\teneta$ and $\bar{\eta}_{ijkl}$ are tensors of viscosity
coefficients. By construction, the viscosity tensors satisfy
\begin{equation}
  \eta_{ijkl} = \eta_{ijlk},\ \ \ \bar{\eta}_{ijkl} = -\bar{\eta}_{ijlk}.
\label{tenetasymmetry}
\end{equation}
  
If the constitutive behavior of the fluid is invariant to rotations
(as distinguished, for example, from a liquid crystal), its total
angular momentum is fixed. This constraint can be shown to demand a
symmetric stress tensor \cite{LandauLifshitzElasticity},
$\sigma_{ij}=\sigma_{ji}$; hence,
\begin{equation}
  \mbox{symmetry under rotations:}\ \ \ \ \eta_{ijkl}
  = \eta_{jikl},\ \ \ \bar{\eta}_{ijkl} = \bar{\eta}_{jikl}.
\label{tenetarotation}
\end{equation}
If the response is of a fluid at equilibrium, the viscosity tensors
are related to equilibrium stress fluctuations. Equilibrium
fluctuations are time-reversible. From Onsager's relations this
implies \cite{OnsagerPR1931},
\begin{equation}
  \mbox{symmetry under time reversal:}\ \ \ \ \eta_{ijkl} = \eta_{klij},
  \ \ \ \bar{\eta}_{ijkl} = \bar{\eta}_{klij}.
\label{tenetatime}
\end{equation}
If all three equations (\ref{tenetasymmetry})--(\ref{tenetatime}) are
satisfied, we find
$\bar{\eta}_{ijkl}=-\bar{\eta}_{ijlk}=-\bar{\eta}_{lkij}
=-\bar{\eta}_{klij}=-\bar{\eta}_{ijkl}$, \ie
$\bar{\eta}_{ijkl}=0$. The conclusion is that $\bar{\eta}_{ijkl}$ vanishes
unless either the symmetry under rotation or the one under
time-reversal is broken.

The symmetries of the fluid may be broken in various ways. An object
in contact with the fluid, whose velocity is fixed (\eg a stationary
object) can break momentum conservation. Similarly, an object whose
angular velocity is given (\eg an active rotor) can violate
angular-momentum conservation. The latter would lead to asymmetric
contributions to the stress tensor, as mentioned above. Time-reversal
symmetry is broken if the ambient fluid is out of equilibrium. This is
relevant, in particular, to active fluids. Such scenarios allow for
viscosity tensors that do not obey Eqs.~(\ref{tenetarotation}) and
(\ref{tenetatime}). As a result, unusual behaviors emerge, as we will see
below.

\subsection{Active chiral fluids}
\label{sec:chiralfluids_active}

To demonstrate such an unusual behavior let us examine a contribution to
the stress tensor which is proportional to the flow vorticity,
\begin{equation}
  \sigma^{\rm ch}_{ij} = \alpha_{ijk}\omega_k,\ \ \ \
  \vecomega = \nabla\times\vecv.
\label{sigmach1}
\end{equation}
We can rewrite it as $\sigma^{\rm ch}_{ij}=(1/2)\alpha_{ijm}\epsilon_{mkl}
\bar{v}_{kl}$. Thus, the viscosity tensor associated with this
contribution is $\bar{\eta}^{\rm
ch}_{ijkl}=(1/2)\alpha_{ijm}\epsilon_{mkl}$. Since
$\epsilon_{mlk}=-\epsilon_{mkl}$, $\bar{\eta}^{\rm ch}$ satisfies
Eq.~(\ref{tenetasymmetry}). Thus, according to
Sec.~\ref{sec:fluidsymmetry}, a viscous fluid which is both
rotation-invariant and time-reversible cannot have a stress
contribution of the form (\ref{sigmach1}).

However, if either of these two symmetries is broken, such a stress
contribution is allowed. For example, it emerged in a theory for a
two-dimensional active fluid of rotors \cite{BanerjeeNatComm2017},
which has been supported by experiment \cite{SoniNatPhys2019}. There,
$\alpha$ is found proportional to the rotors' intrinsic angular
momentum. Similar contributions have been included in generic
hydrodynamic theories of chiral active
fluids \cite{FurthauerEPJE2012,MarkovichNJP2019}.

\begin{figure}
\centerline{\includegraphics[width=0.5\textwidth]{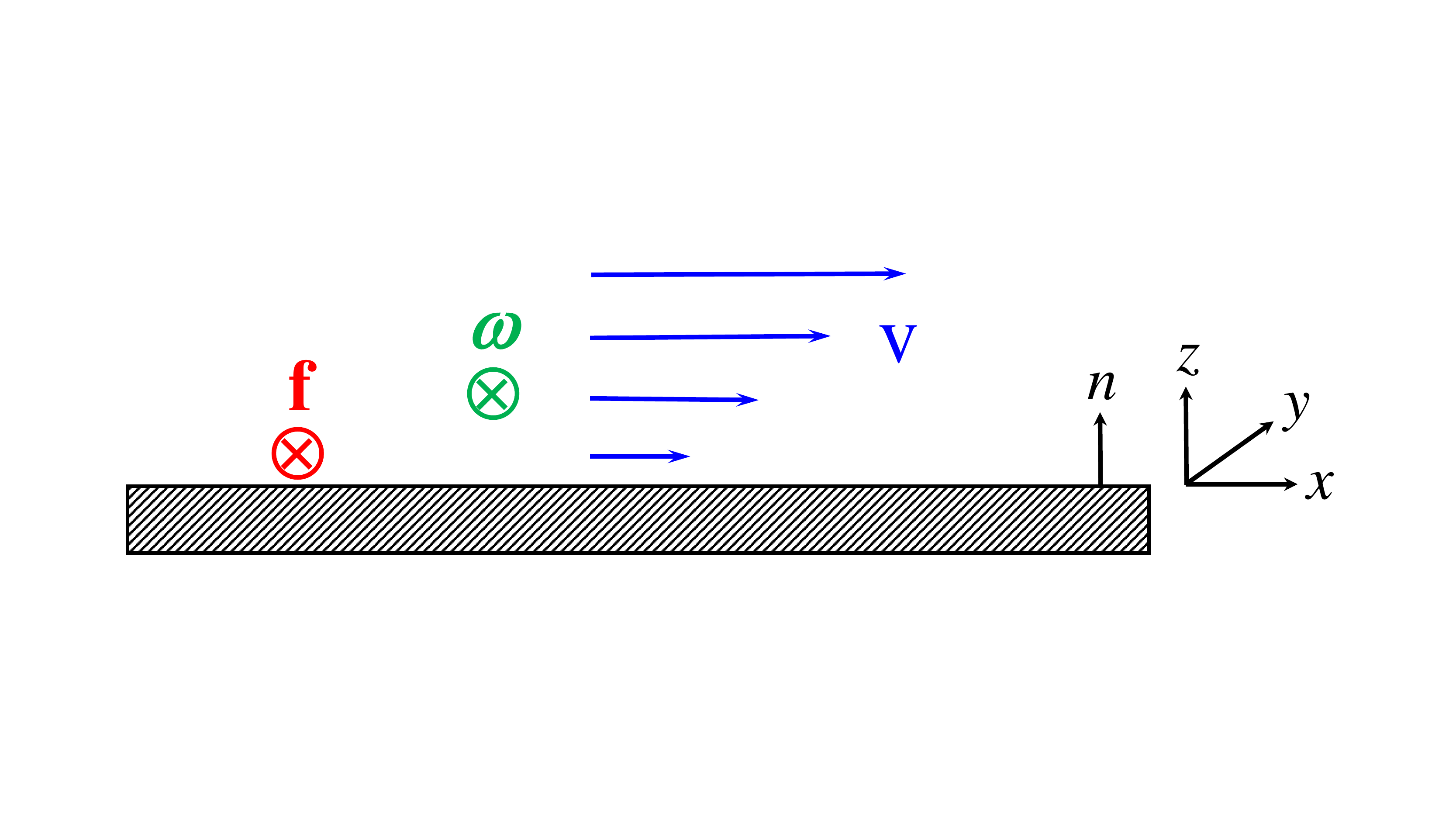}
\includegraphics[width=0.5\textwidth]{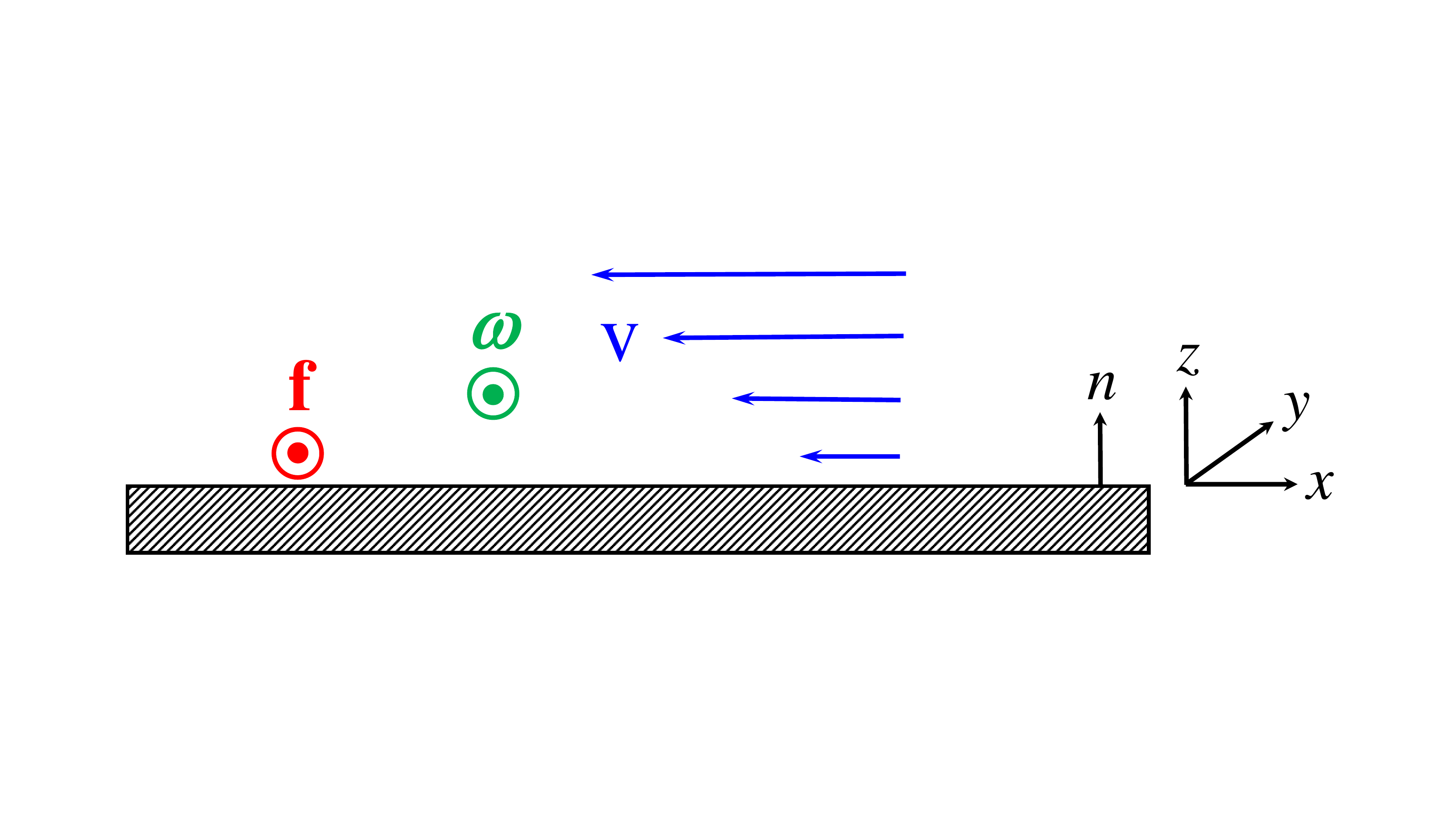}}
\vspace{-1cm}
\caption{Consequence of a chiral stress proportional to the vorticity. Left:
shear flow near a stationary surface leads to a force on the surface
in the normal, neutral direction. Right: reversing the flow direction
reverses the direction of the normal force.}
\label{fig:chiralfluid1}
\end{figure}

Here is a simple consequence of this stress contribution. Consider
a linear flow profile forming at $t=0$ near a stationary surface,
$\vecv(\vecr)=\gamma z \xhat$ (see Fig.~\ref{fig:chiralfluid1}). The
vorticity at this moment is $\vecomega=\gamma\yhat$. In an ordinary
viscous fluid the nonzero stress component is
$\sigma_{xz}=\eta\gamma$, where $\eta=\eta_{xzzx}$. There are no
forces in the neutral ($y$) direction. In the considered
symmetry-broken fluid, however, we have also $\sigma^{\rm ch}_{yz}
= \alpha\omega_y = \alpha\gamma$, where $\alpha=\alpha_{yzy}$. As a
result, the surface is pushed in the $y$ direction by a force per unit
area $f=\sigma^{\rm ch}_{yz}n_z = \alpha\gamma$. If the flow is
reversed, the surface will be pushed in the opposite ($-y$) direction;
see Fig.~\ref{fig:chiralfluid1}.

This is a chiral response. We can imagine that the fluid contains
driven or active chiral particles which in the shear flow move
horizontally (Fig.~\ref{fig:chiralfluid1}). As they do so, they or the
fluid in which they are placed apply on the surface a force component
in the vorticity direction. We note that externally driven or active
components are necessary in this case to supply the $y$-momentum
imparted to the surface. For example, removing momentum in the $x$
direction while adding momentum in the $y$ direction can be achieved
by an active torque. If we took particles of reversed chirality, the
force would be in the opposite direction. Thus, $\alpha$ must change
sign under inversion; hence the term {\it odd
viscosity} \cite{AvronJStatPhys1998}. Another way to express this
property is to consider a binary mixture of particles of both
chiralities and define a ``chiral density'' $\cch$ as the difference
in densities of the two \cite{Andreev:2010fk}. If we write
$\alpha=\alpha_0\cch$, the viscosity becomes manifestly odd and the
whole effect vanishes for a symmetric (racemic) mixture.

The behavior demonstrated here resembles the Hall effect, where the
flow is like the electric current, $\cch$ the magnetic field, and $f$
the resulting Hall potential. The Hall resistance
(relation between Hall potential and electric current) changes sign
when the direction of the magnetic field is reversed. Similarly, the
odd viscosity changes sign when the chiral density is reversed,
$\alpha(-\cch)=-\alpha(\cch)$. As with a magnetic field, a
three-dimensional odd fluid cannot be
isotropic \cite{AvronJStatPhys1998}. Thus, the behavior described here
can be found either in 2D fluids \cite{SoniNatPhys2019} or in 3D
anisotropic ones \cite{FurthauerEPJE2012,MarkovichNJP2019}.

\subsection{Passive chiral fluids}
\label{sec:chiralfluids_passive}

A different hydrodynamic theory for chiral fluids was presented in
Ref.~\cite{Andreev:2010fk}. Here the broken inversion symmetry
produces a chiral response without making the stress tensor
asymmetric. The theory, therefore, is valid for passive fluids of
chiral particles, without activity or intrinsic torques. It can do so
by considering higher-order derivatives of the flow,
\begin{equation}
  \sigma^{\rm ch}_{ij} = \beta_{ijkl}[\partial_k(\cch\omega_l)
  + \partial_l(\cch\omega_k)].
\label{sigmach}
\end{equation}
The linear transport coefficients $\beta_{ijkl}$ are not viscosities
as they multiply second-order derivatives of velocity. By construction
$\beta_{ijkl}=\beta_{ijlk}$. If we demand $\beta_{ijkl}=\beta_{jikl}$,
$\tensigma^{\rm ch}$ is symmetric as required.  In non-chiral fluids a
stress term proportional to $\partial_i\omega_j$ is forbidden since it
is a pseudo-tensor, changing sign under inversion, while the stress
(flux of linear momentum) is even. In a chiral fluid, with a nonzero $c^{\rm ch}$, both sides are even under inversion.

\begin{figure}
\centerline{\includegraphics[width=0.5\textwidth]{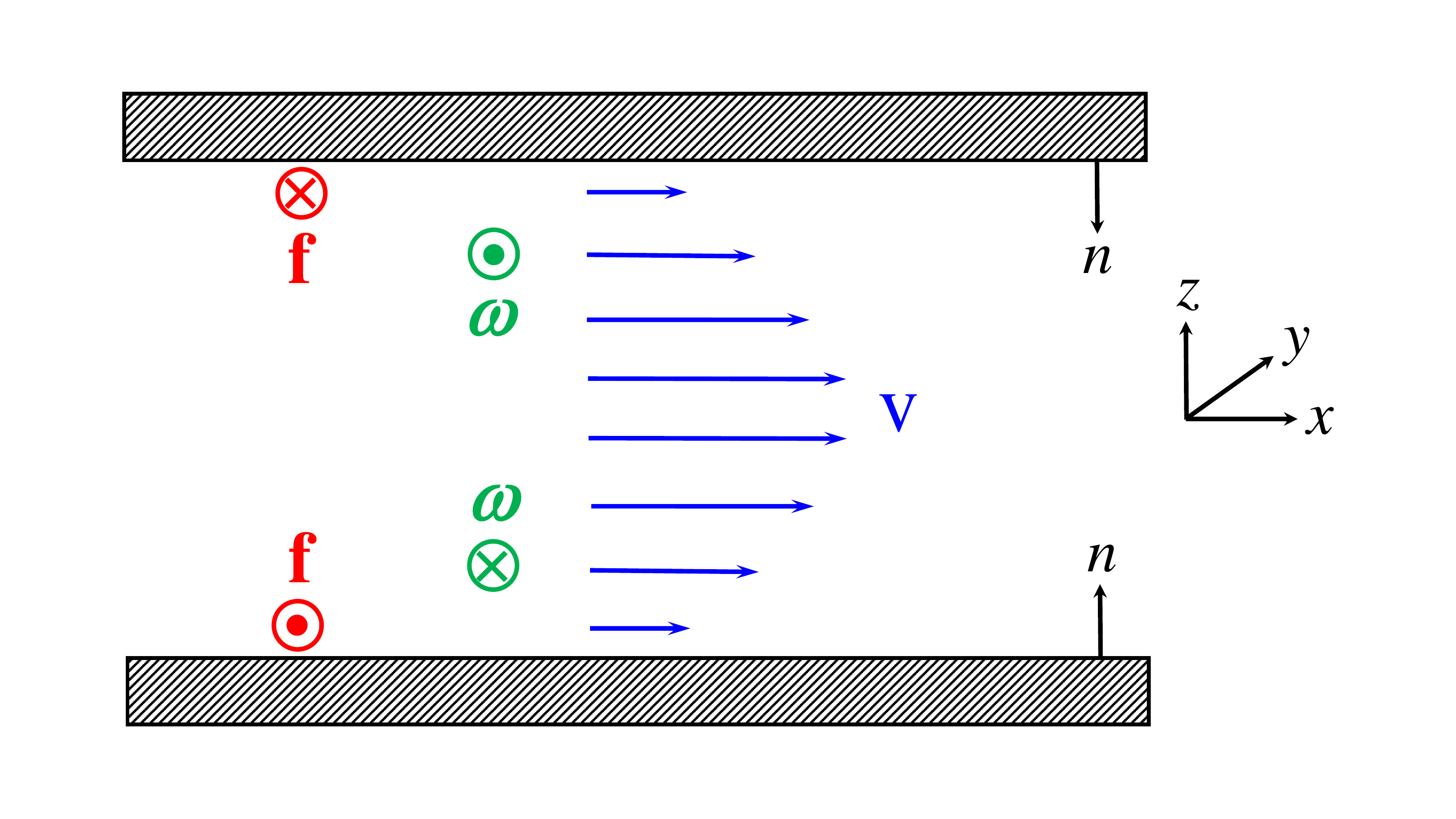}
\includegraphics[width=0.5\textwidth]{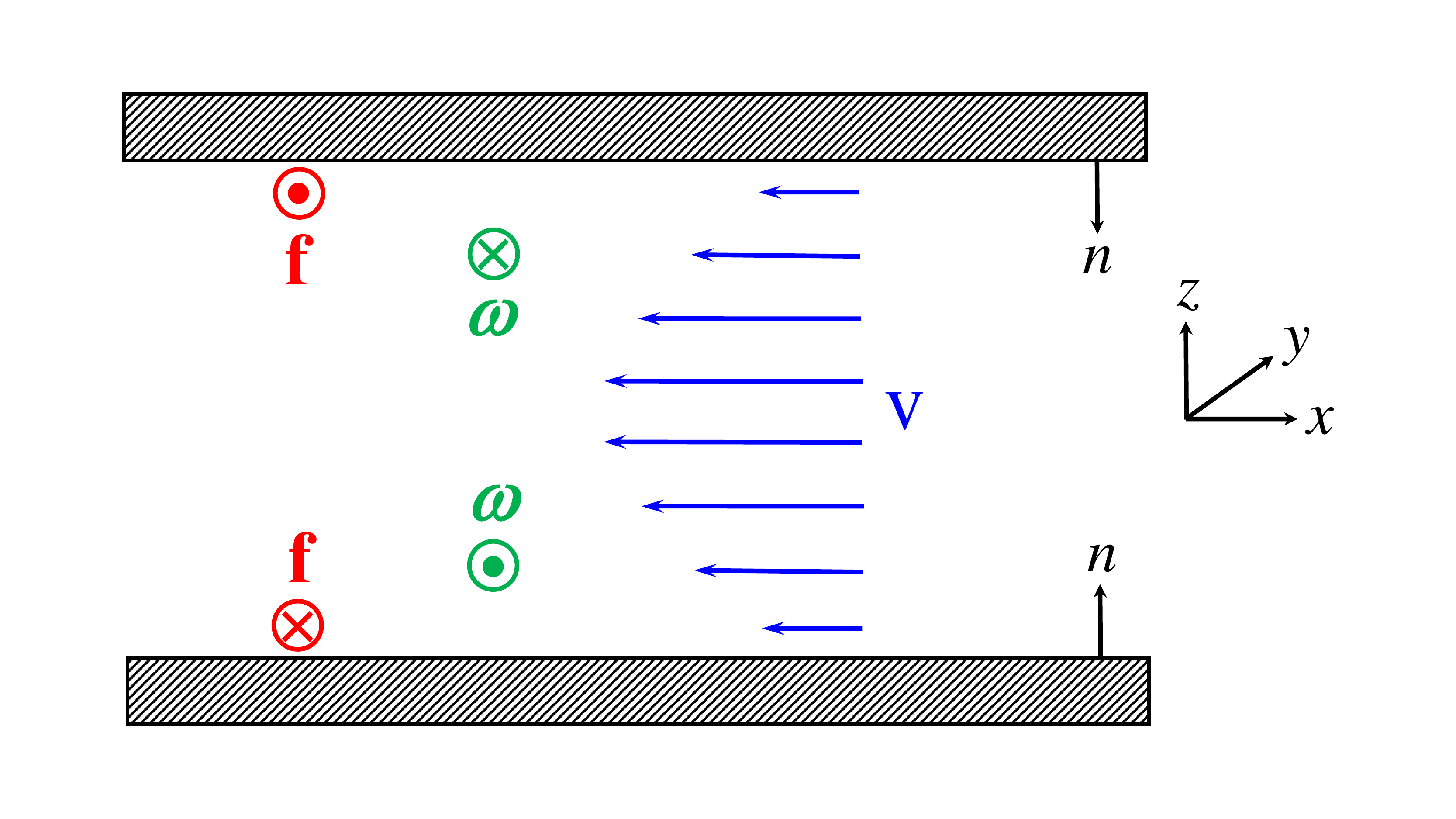}}
\caption{Consequence of a chiral stress proportional to vorticity
gradients \cite{Andreev:2010fk}.  Left: a Poiseuille flow through a
slit leads to forces on the two boundaries of the slit in opposite
directions along the neutral (vorticity) axis. Right: reversing the
flow direction switches the directions of these forces.}
\label{fig:chiralfluid2}
\end{figure}

The chiral stress tensor of Eq.~(\ref{sigmach}) vanishes for a linear
shear profile with uniform $\cch$ as the one used in
Fig.~\ref{fig:chiralfluid1}. To demonstrate a simple implication, we
consider an instantaneous parabolic (Poiseuille) flow profile in a
slit of width $w$, as shown in Fig.~\ref{fig:chiralfluid2}. The flow
velocity at this moment is $\vecv = (3/2)v_0(1 - 4z^2/w^2)\xhat$,
where $v_0$ is the cross-section-averaged flow velocity.  The
vorticity is $\vecomega=\nabla\times\vecv=-(12v_0/w^2) z\yhat$.  For a
non-chiral fluid, by symmetry, no stresses exist along the neutral
direction ($y$). For a chiral fluid, when we substitute $\vecomega$
and a uniform nonzero $\cch$ into Eq.~(\ref{sigmach}), we get a stress
component $\sigma^{\rm ch}_{yz}=-\zeta v_0$,
$\zeta=24\cch\beta_{yzyz}/w^2$. Thus, the two walls of the slit, whose
normals point in the $\pm z$ directions, experience forces of opposite
signs in the $y$ direction\,---\,the slit is sheared along $y$. See
Fig.~\ref{fig:chiralfluid2}. If the flow is reversed, the shearing
switches to the opposite direction.

Also here, the behavior resembles the Hall effect. The ``Hall
resistance'' $\zeta\sim\cch$ determines the force difference along the
neutral axis and is odd under a change of sign of the ``field''
$\cch$. Unlike the active chiral fluid described earlier, here there
is no need for force or torque sources; the $y$-momentum imparted to
one wall is balanced by the $(-y)$-momentum imparted to the other.

It is instructive to see whether the response of passive chiral
objects to shear flow, as discussed in earlier sections, is consistent
with this continuum description. Consider the ``paddles'' described in
Sec.~\ref{sec:NoForce} (see Fig.~\ref{fig:PaddleTwist}), and imagine
putting many such objects in a Poiseuille flow between two parallel
plates as in Fig.~\ref{fig:chiralfluid2}. As described in
Sec.~\ref{sec:NoForce}, the linear shear flow near each surface makes
the chiral objects migrate in the vorticity ($\pm y$)
directions \cite{Makino:2005ty,Makino:2008fj}. The migration velocity
$V_y$ is proportional to the shear rate, $V_y \sim v_0 a/w$. This
would cause opposite backflows of fluid mass, pushing one wall in the
$y$ direction and the other in the $-y$ direction. The corresponding
shear rate in the $yz$ plane, and the resulting stress, are proportional
to $V_y/w \sim v_0 a/w^2$. This is in line with the result obtained
above from the continuum theory, $\sim c^{\rm ch}\beta v_0/w^2$. It is
in contrast with the stress $\sim \alpha v_0/w$ obtained for the
active fluid of Sec.~\ref{sec:chiralfluids_active}. As mentioned in
Sec.~\ref{sec:NoForce}, the appearance of migration as a linear
response to shear flow requires anisotropy of the objects'
orientations, \eg a finite P\'eclet number. Thus the linear continuum
theory presented in this section cannot be constructed from
equilibrium, isotropically oriented chiral objects.

The collective flow of suspensions of paddle objects was considered in
Ref.~\cite{Makino_2004} using analytical theory and simulations. The
authors analyzed the effective viscoelastic modulus of the suspension,
as well as its nonlinear viscoelasticity, arising from the stochastic,
anisotropic orientations of the objects. As far as we could see, they
did not explicitly report the collective chiral effect of
Ref.~\cite{Andreev:2010fk} and Eq.~(\ref{sigmach}).

If we introduce an external force into the chiral fluid discussed
above, it will contribute to the chiral flux through coupling to
gradients of velocity and vorticity \cite{Andreev:2010fk}. Thus, if
the base flow is uniform, the chiral effect of fluctuations in the
drive would be weak, \ie only quadratic in the fluctuation. We present
next a stronger collective effect occurring in suspensions of
asymmetric objects.

\subsection{Suspension of asymmetric particles}
\label{sec:asymmetricsuspension}

\begin{figure}
\centerline{\includegraphics[width=0.4\textwidth]{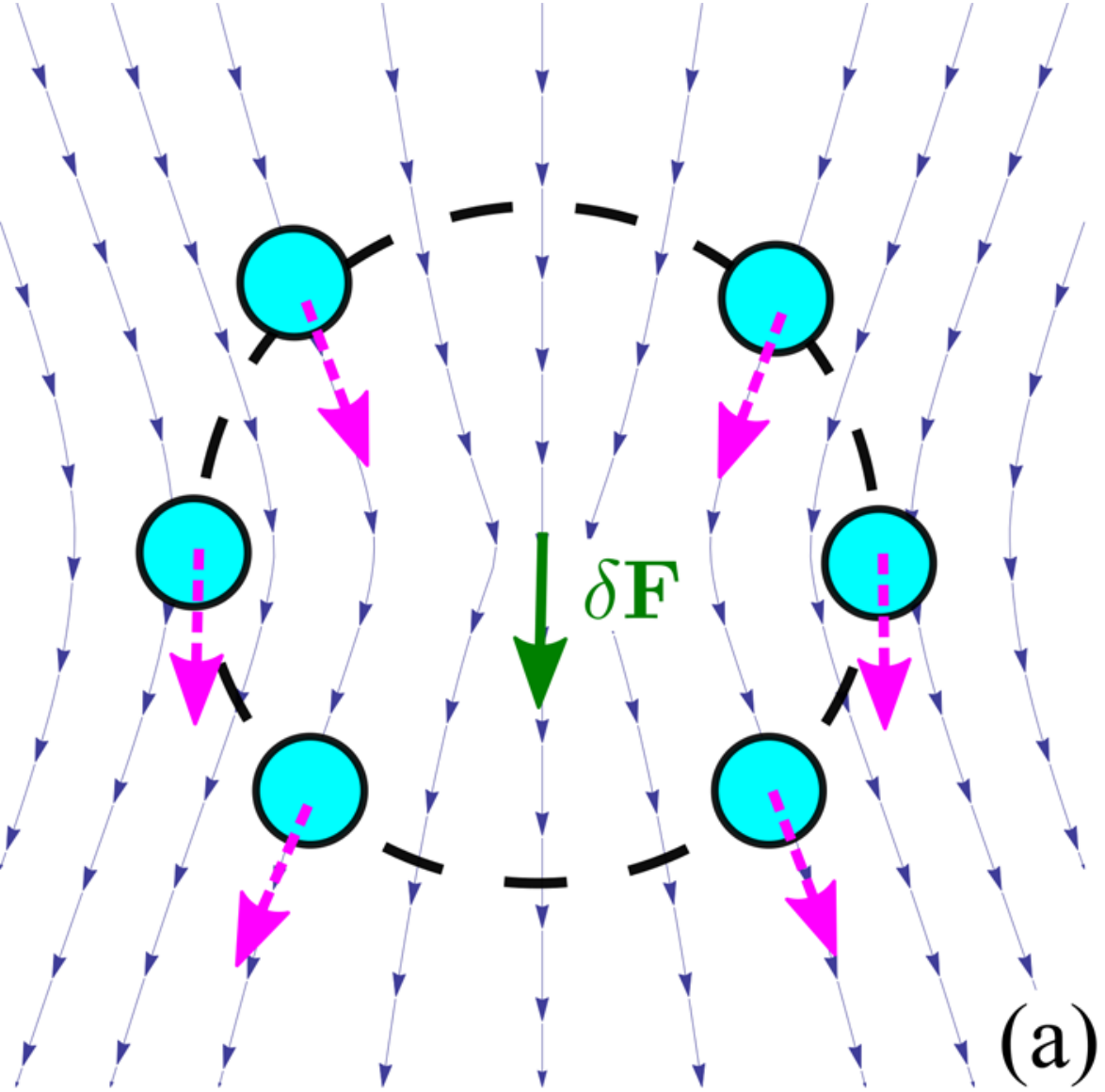}
\hspace{0.5cm} \includegraphics[width=0.4\textwidth]{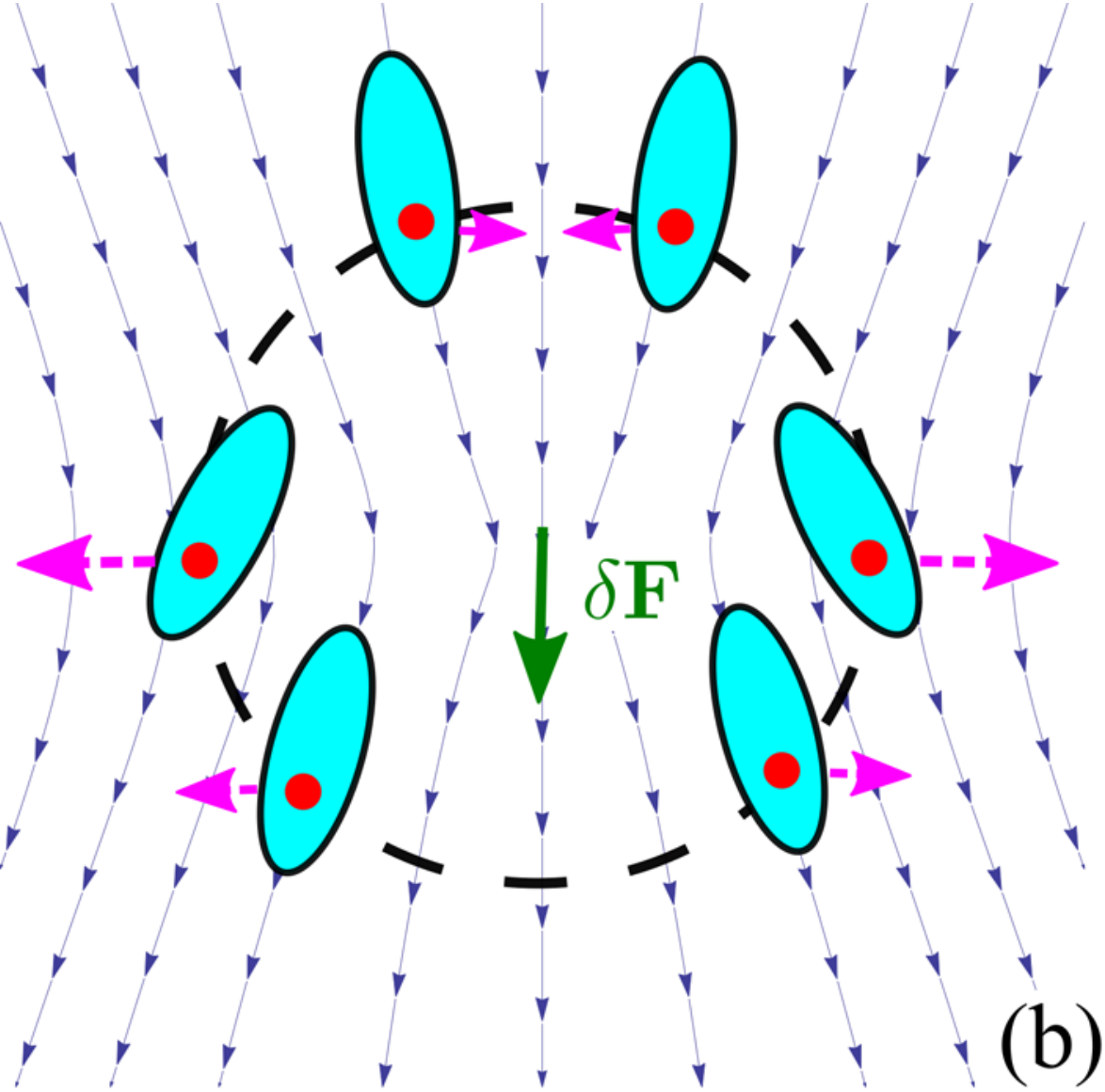}}
\caption{Response of forced colloids to flow-velocity gradients around them.
Here the flow-velocity gradients are caused by an excess force on the
fluid located at the center. (a) The isotropic response of symmetric
objects makes them follow the flow lines. As a result, the objects'
flow, like the fluid's, is volume-conserving. (b) The anisotropic
response of asymmetric objects allows them to deviate from the flow
lines and thus increase or decrease their local concentration. The
drawn objects are ellipsoids with a forcing point shifted along the
large axis. These are asymmetric but non-chiral objects. Taken from
Ref.~\cite{Goldfriend:2017mz}.}
\label{fig:glide}
\end{figure}

Let us return to the object symmetries discussed in
Sec.~\ref{sec:ObjectSymmetries} and compare, for example, three types
of ellipsoid or box: (a) one whose forcing point (\eg center of mass
under gravity) is at its center; (b) another whose forcing point is
shifted from the center along one of the principal axes; and (c) a
third whose forcing point is displaced away from the principal
axes. The first is symmetric, the second is asymmetric but non-chiral,
and the third is chiral. In response to flow around it, the symmetric
ellipsoid, having no intrinsic direction, would move along the flow
lines. The asymmetric non-chiral object, although not rotating in
response to a force, does have an intrinsic direction (from its center
to the forcing point). Thus, in general, its motion would deviate from
the direction of the local flow lines. See Fig.~\ref{fig:glide}. The
same holds for the chiral object. This ``gliding'' property of
asymmetric particles affects their collective dynamics in a
suspension.

Consider a dilute suspension which is isotropic and uniform before a
force is applied. Then a force is turned on. A concrete example is a
dilute sedimenting suspension. Suppose that a local fluctuation in
concentration creates a slightly heavier region, applying a slightly
stronger force density on the host fluid. The additional flow would
advect particles in the vicinity. If the particles are symmetric, they
would follow those flow lines. Since the host fluid can be assumed
incompressible, the flux of the symmetric particles would be
incompressible as well. Thus, to leading order in the fluctuation, the
advection cannot affect the local concentration of symmetric
particles. The effect is at least quadratic in the fluctuation
\cite{LevinePRL1998,Ramaswamy2001}. As mentioned above, the response of the
chiral flux of Ref.~\cite{Andreev:2010fk} shares this weakness.

Asymmetric particles can deviate from the flow lines of the host fluid
and glide more into or more out of the heavier region; they can
deplete the region or fill it further. See Fig.~\ref{fig:glide}
again. This effect is linear in the
fluctuation \cite{Goldfriend:2017mz}. It is an advective mechanism of
either suppression or enhancement of concentration fluctuations, which
is faster than diffusion. As a result, velocity fluctuations are
suppressed or enhanced as well. We note that the effect is not related
to chirality {\it per se} but to the anisotropic response of asymmetric
objects to flow-velocity gradients. Chiral objects have this
anisotropic response as special cases.

This collective behavior of asymmetric suspensions crucially depends
on the existence of two relatively moving substances\,---\,suspended
particles and host fluid. Thus the homogeneously flowing chiral fluids
discussed in the preceding subsections cannot account for it. For
the asymmetric suspensions we present a simplified ``two-fluid''
model, which is a variation on the one presented in
Ref.~\cite{Goldfriend:2017mz}. The particles, each driven by an
external force $\vecF=-F\zhat$, are represented by a continuous fluid
of concentration $c$, velocity $\vecV$, and pressure
$P$. We assume that variations in the orientations of
the asymmetric particles are weak, such that an additional orientation
field is unnecessary. The host fluid is incompressible and has
velocity $\vecv$ and pressure $p$.

The two fluids are coupled by friction. To account for this coupling
we first consider the drag force that an isolated object with velocity
$\vecV$ experiences in a nonuniform flow $\vecv(\vecr)$. We write it
as an expansion in object size $a$,
\begin{equation}
  F^{\rm d}_i = (\tenmobs)^{-1}_{ij} \left( V_j - v_j - a \Phi_{jkl} \partial_l v_k
  - a^2 \Psi_{jklm}\partial_l\partial_m v_k - \ldots \right),
\label{faxen_general}
\end{equation}
where $\tenmobs=\RT_{VF}$ is the object's self-mobility matrix, and
the dimensionless tensors $\tenPhi$ and $\tenPsi$ depend solely on the
object's shape. For a spherical object $(\tenmobs)^{-1}_{ij}=6\pi\eta
a\delta_{ij}$, $\tenPhi=0$, $\Psi_{jklm}=(1/6)\delta_{jk}\delta_{lm}$,
and the higher-order terms vanish (because $\nabla^4\vecv=0$ in Stokes
flow) \cite{Happel-Brenner}. For an arbitrarily shaped object
$\tenPhi\neq 0$, and we can approximate $\vecF^{\rm d}$ by the two
leading terms in Eq.~(\ref{faxen_general}). The tensor $\tenPhi$,
characterizing the velocity response of an object to nonuniform flow,
has already appeared in the discussion of the hydrodynamic interaction
between two objects,
Sec.~\ref{sec:multipole}.\footnote{We note that if
the object is isotropically oriented, the orientation-averaged
$\tenPhi$ vanishes \cite{Makino:2005ty}. Thus, the anisotropy due to
the drive is vital for this theory.}

The overdamped dynamics of the two fluids are described by the
following equations:
\begin{eqnarray}
  0 &=& -\nabla P + c\vecF - c\vecF^{\rm d}, \label{stokesobjects}\\
  0 &=& -\nabla p + \eta\nabla^2\vecv + c\vecF^{\rm d}, \label{stokesfluid}\\
  0 &=& \nabla\cdot\vecv, \label{massflowfluid} \\
 \partial_t c &=& -\nabla\cdot(c\vecV). \label{massflowobjects}
\end{eqnarray}
The first two equations account for the balance of forces within the two fluids, ensuring the conservation of total momentum. The last two equations account for the conservation of mass of the two fluids separately.  Focusing on deviations from uniform sedimentation, we replace $c$ by
$c_0+c$, $\vecV$ by $\vecV_0+\vecV$, etc., where
$(c_0,\vecV_0,\vecv_0,P_0,p_0)$ correspond to uniform sedimentation,
and expand to linear order in
$(c,\vecV,\vecv,P,p)$. Equations~(\ref{stokesobjects})--(\ref{massflowfluid})
yield the usual linear relation between force and flow,
\begin{equation}
  v_i(\vecr) = \int d\vecr\,' \oseen_{iz}(\vecr-\vecr\,') (-F),
\label{twofluidoseen}
\end{equation}
where $\tenoseen$ is the Oseen tensor presented in
Sec.~\ref{sec:StokesletObjects} [Eq.~(\ref{eq:Oseen})]. Substituting
$\vecv(\vecr)$ from Eq.~(\ref{twofluidoseen}) in
Eq.~(\ref{faxen_general}), and the resulting $\vecF^{\rm d}$ in
Eq.~(\ref{stokesobjects}), gives
\begin{equation}
  V_i = -F \int d\vecr\,' \left[ \oseen_{iz}(\vecr-\vecr\,') +
    a\Phi_{ikl}\partial_l\oseen_{kz}(\vecr-\vecr\,') \right]
    c(\vecr\,') - (1/c_0) \mobs_{ij} \partial_jP.
\label{Vcollect}
\end{equation}
In Fourier space ($f(\vecr) \rightarrow \tilde f(\vecq)$),
\begin{equation}
  V_i = -F \left[ \tilde\oseen_{iz}(\vecq) +
    i a\Phi_{ikl} q_l\tilde\oseen_{kz}(\vecq) \right]
    \tilde c(\vecq) - (i/c_0) \mobs_{ij} q_j \tilde P,
\label{Vcollectq}
\end{equation}
where $\tilde\oseen_{ij}(\vecq)=(\eta
q^2)^{-1}(\delta_{ij}-q_iq_j/q^2)$. Note that $\tilde\oseen \sim
q^{-2}$ at small $q$.

Consider for a moment the relation between stochastic fluctuations in
$\vecV$ and in $c$. In the limit of small $q$ Eq.~(\ref{Vcollectq})
gives $\langle |\tilde V_z(\vecq)|^2\rangle = F^2 \tilde\oseen_{zz}^2(\vecq) \langle
|\tilde c(\vecq)|^2\rangle \sim (F/\eta)^2 q^{-4} \langle
|\tilde c(\vecq)|^2\rangle$. For normal concentration fluctuations (as in a
random distribution of particles) $\langle |\tilde c(q\rightarrow
0)|^2\rangle = \mbox{const}$. In this case, therefore, $\langle
|\tilde V_z(\vecq)|^2\rangle$ increases with small $q$ as $1/q^4$. This
implies that the velocity fluctuations of individual particles, $\langle
V_z^2(r\rightarrow 0)\rangle \sim \int d\vecq q^{-4}$, diverge as
$L$ with the system size $L$. As mentioned above, this
prediction \cite{CaflischPhysFluids1985} is not borne out by
experiments. This apparent paradox would evidently disappear if
density fluctuations were suppressed, such that $\langle
|\tilde c(q\rightarrow 0)|^2\rangle \sim q^\alpha$ with $\alpha> 1$. Such a
state is called ``hyperuniform'' \cite{TorquatoPhysRep2018}. We will
see shortly that it does occur in the asymmetric suspension.

Returning to Eq.~(\ref{Vcollect}), we expand the pressure as $P\simeq
(\partial P/\partial c_0)c \equiv \kappa c$. The coefficient $\kappa$
characterizes the resistance of the assembly of particles to
compression, for example, as a result of hard core repulsion. (It need
not be the same as the equilibrated thermal coefficient.) In a dilute
suspension we expect $P\sim c_0$, and hence $\kappa=\mbox{const}$. We
substitute $\vecV$ from Eq.~(\ref{Vcollect}) in
Eq.~(\ref{massflowobjects}) and transform to Fourier
space. Equation~(\ref{massflowobjects}) becomes $\partial_t \tilde
c(\vecq,t) = -\Gamma(\vecq) \tilde c$, with the following decay rate of
concentration fluctuations,
\begin{equation}
  \Gamma(\vecq) = \kappa \mobs_{ij}q_iq_j + ac_0F \Phi_{ikl}q_iq_l
  \tilde\oseen_{kz}(\vecq).
\label{Gamma}
\end{equation}
The first term, quadratic in $q$, represents anisotropic diffusive
relaxation with the diffusion tensor
$\tenD=\kappa\mobs_{ij}\sim\kappa/(\eta a)$. The second term arises from
the anisotropic hydrodynamic interactions. Since $\tilde\oseen \sim
1/q^2$, it is dominant at large length scales. The screening length
beyond which the diffusive relaxation is suppressed is
\begin{equation}
  \xi = \left(\varphi a^2c_0F/\kappa\right)^{-1/2},
\end{equation}
where $\varphi$ is an asymmetry parameter obtained from the components
of $\tenPhi$ \cite{Goldfriend:2017mz}.

Several observations follow from Eq.~(\ref{Gamma}). (a) If the object
responds isotropically, with $\Phi_{ikl}$ of the form
$\lambda_l\delta_{ik}$ (along with its rotations and reflections), the
second, advective term vanishes since $q_k\tilde\oseen_{kz}=0$
(incompressibility of the host fluid). Thus, the screening mechanism
requires object asymmetry. Without it, the divergence of the advective
terms in $\vecV$ [Eq.~(\ref{Vcollect})] vanishes. This reflects the
compliance of symmetric objects, and defiance of asymmetric ones, to
follow the direction of the local flow lines when they are under
force, as shown pictorially in Fig.~\ref{fig:glide}. (b) The
advective term vanishes also for $\vecq=q\zhat$, because
$\tilde\oseen_{kz}(q\zhat)=0$. Thus, in the exact direction of the
force there is only normal diffusive relaxation. (c) The advective
term may be negative, in which case the large-scale collective modes
$q<|\xi|^{-1}$ become unstable. In the case of asymmetrically forced
ellipsoids (Fig.~\ref{fig:glide}(b)), the response tensor $\tenPhi$
and resulting asymmetry parameter $\varphi$ can be calculated
exactly. Depending on whether the forcing point is shifted along the
large or small axis, the object is found to have, respectively,
$\varphi>0$ (advective stabilization) or $\varphi<0$ (advective
destabilization) \cite{Goldfriend:2017mz}.

These observations concerning the collective dynamics are consistent
with the analysis in Sec.~\ref{sec:multipole} concerning pair
interactions at large separation. As we have seen in
Sec.~\ref{sec:multipole}, the tensor $\tenPhi$ determines whether two
objects move together or apart under force.

The suspension's stochastic dynamics can be studied by including
random fluxes $\vecj(\vecr,t)$ in the mass-flow equation
(\ref{massflowobjects}) (in the form of a $-\nabla\cdot\vecj$ term on
its right-hand side). This leads to the following structure factor at
steady state \cite{Goldfriend:2017mz}:
\begin{equation}
  S(\vecq) \equiv c_0^{-1}\langle |\tilde c|^2\rangle_{\vecq,\omega=0}
  = [c_0\Gamma(\vecq)]^{-1}\,q_iq_j\langle \tilde j_i\tilde
  j_j^*\rangle_{\vecq,\omega=0}.
\end{equation}
Since $\Gamma(q\ll\xi^{-1}) \sim q^0$, and assuming $\langle
\tilde j_i\tilde j_j^*\rangle(q\ll\xi^{-1}) \sim \mbox{const}$, we get $S(q)\sim
q^2$ for small $q$\,---\,a signature of hyperuniformity
\cite{TorquatoPhysRep2018}. As we have seen above, such a
hyperuniform state removes the divergence of velocity fluctuations
with system size. As in the case of the advective relaxation rate, in
the exact direction of the force ($\zhat$) the distribution of particles
is not hyperuniform.

\begin{figure}\label{fig:sedimentation}
\centerline{\includegraphics[width=0.33\textwidth]{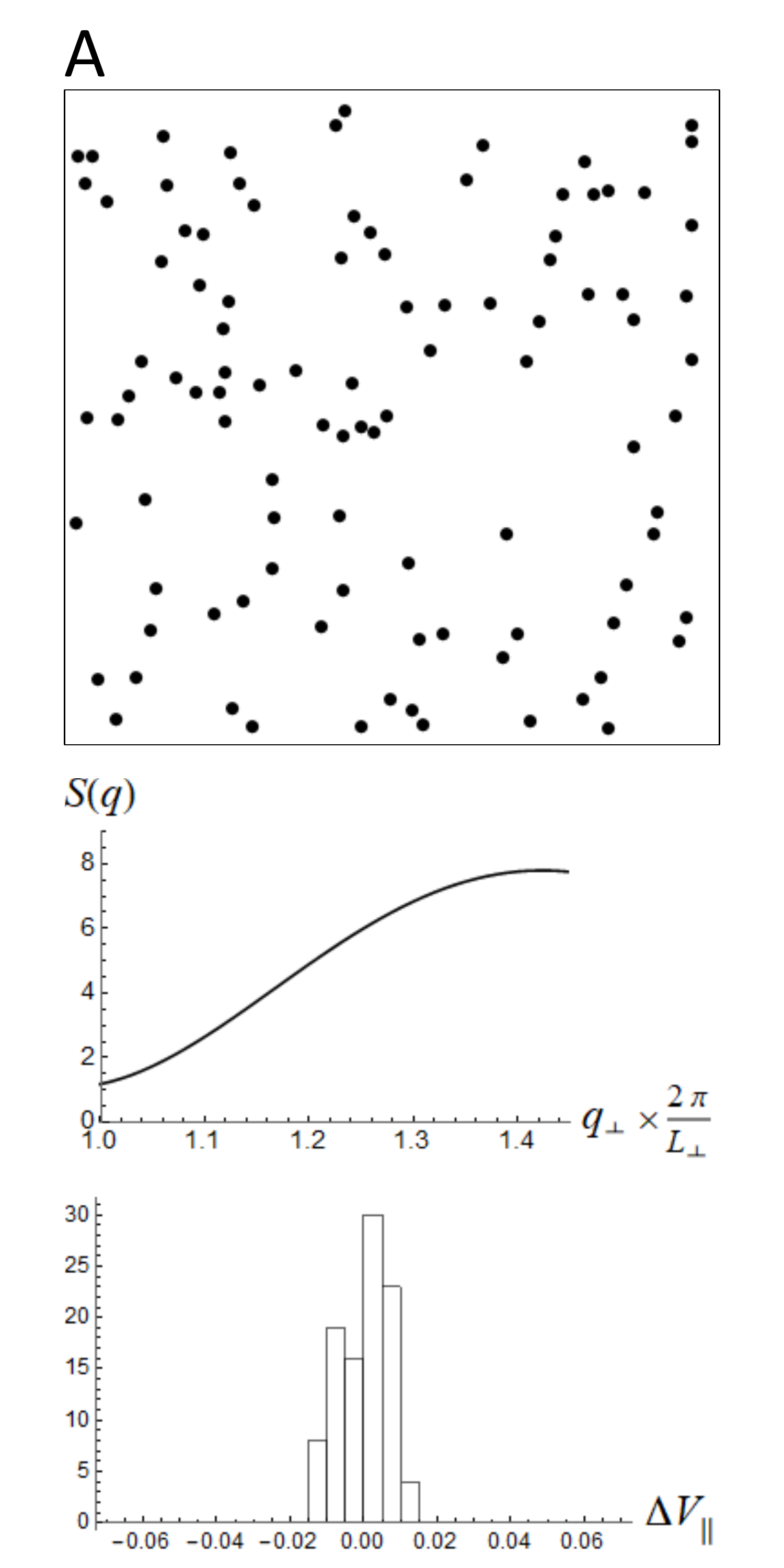}
\includegraphics[width=0.33\textwidth]{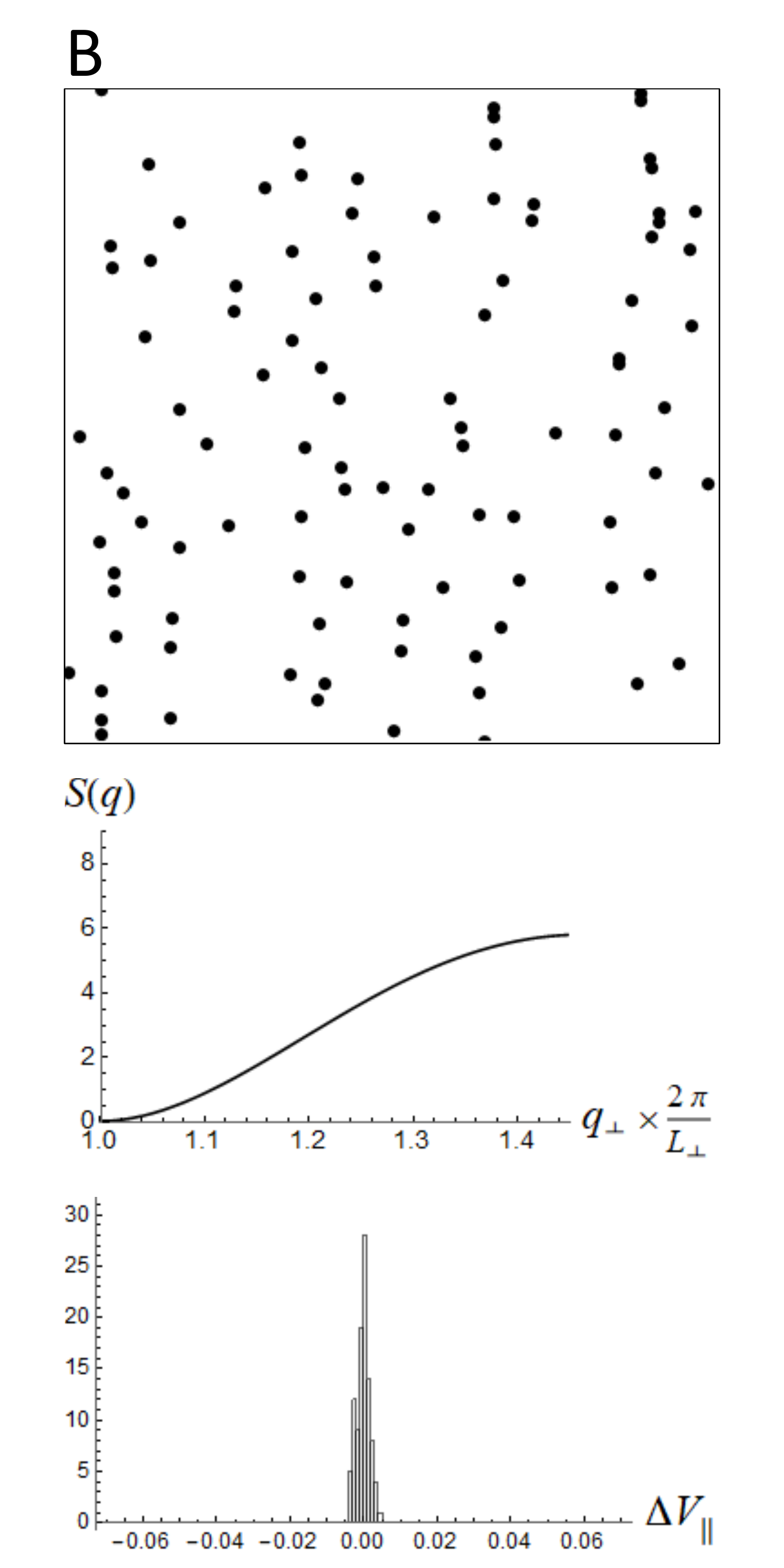}
\includegraphics[width=0.33\textwidth]{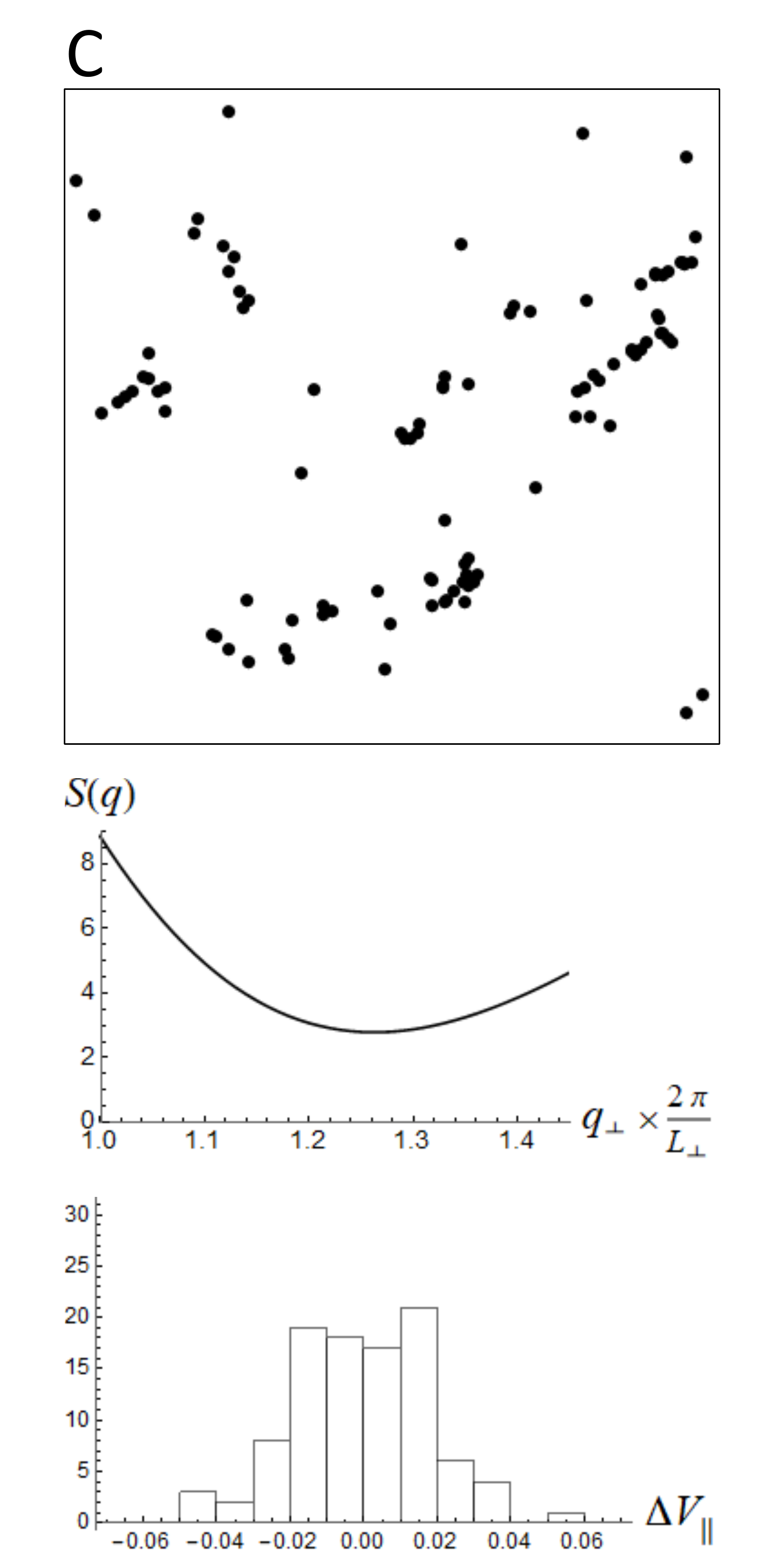}}
\caption{Simulation results of forced particles in two dimensions. The
particles are self-avoiding disks, interacting hydrodynamically as in
a fluid membrane. The response tensor $\tenPhi$ is made asymmetric to
mimic asymmetric particles. The simulation contains 100 disks of
diameter 50 times smaller than the side of the square simulation
box. The force points to the right ($x$). Periodic boundary conditions
are applied. (A) An initial state where the particles are distributed
randomly. The structure factor reaches a finite value at the minimum
$q_y$. Also shown is the distribution of velocities of individual
objects in the direction of the force. (B) A much later snapshot for
the case of a stabilizing $\tenPhi$. The structure factor decays
almost to zero at the minimum $q_y$, reflecting suppressed large-scale
density fluctuations (hyperuniformity) transverse to the force. The
distribution of velocities parallel to the force is much narrower. (C)
A late snapshot for the case of a destabilizing $\tenPhi$, showing
particle clusters. The structure factor increases, reflecting the
enhanced density fluctuations. The distribution of velocities
broadens.}
\end{figure}

Figure \ref{fig:sedimentation} shows results from simulations of
forced disks moving in a fluid membrane. The disks avoid overlap. They
interact via two-dimensional hydrodynamic interactions with an
asymmetric $\tenPhi$, to mimic the hydrodynamic response of asymmetric
particles. We show an initial state where the particles are
distributed randomly, a snapshot at long time for a stabilizing
$\tenPhi$, and a similar snapshot for a destabilizing $\tenPhi$. The
corresponding structure factors reflect normal, suppressed, and
enhanced density fluctuations, respectively, transverse to the
force. The distribution of velocities parallel to the force becomes
narrower or wider correspondingly.

To summarize this part, the anisotropic response of forced asymmetric
objects to the nonuniform flow caused by concentration fluctuations
allows them to glide into or out of regions of different
concentration. This either suppresses or enhances concentration
fluctuations. The mechanism is directly related to the one causing two
forced objects to either repel or attract one another hydrodynamically
(Sec.~\ref{sec:HydrodynamicInteraction}). The former behavior leads to
a nondiffusive fast relaxation and a hyperuniform distribution of
particles. The latter creates unstable clustering. All these striking
phenomena have not yet been verified experimentally. Relevant
experiments are underway \cite{BurtonPrivate}; see the suggestive
results in Fig.~\ref{fig:BurtonDimers}.

\section{Discussion}\label{sec:discussion}
\setcounter{equation}{0}
\setcounter{figure}{0}


				In recent years micron-scale objects dispersed in fluids have shown a wealth of striking and useful behavior based on simple aspects of the individual dispersed object.  This includes exotic positional ordering\cite{Talapin:2009kq, Corte:2008fk, Lin:2000ai}, striking rheological behavior\cite{Liu:2001lq, James:2019db} and strong control over external fields like nematic order parameters\cite{Lapointe:2009qe} or light \cite{Man:2013dp}.  In this article, we have focused on a complementary aspect of these dispersions: they may create controllable self-organized structure by the effect of asymmetric shape on the motion of the surrounding fluid.  This controllability is due to the tensorial character of the objects' linear response to driving.  Here we survey the prospects for organizing colloidal dispersions in the future using these tensorial responses. 
		
				We have noted that asymmetry in colloidal objects provides handles for manipulating them, separating them from other particles or organizing their rotational states or controlling their overall density fluctuations.   In general such handles will be of interest in tandem with specific features of the objects, either to perturb their phase behavior, to construct a multi-object assembly, or to create an organized collective motion.  For all these purposes it is valuable to know in greater depth how tensorial responses can enable desired states.
									
				Our investigation of single particle properties gave a glimpse into creating aligned and synchronized rotation in a dispersion. As we have noted, all types of linear rotational response to a driving {\em vector} are subject to a common type of dynamics governed by a simple 3 $\cross$ 3 tensor of shape-dependent quantities. This includes explicit external forces on the object or phoretic gradients like electric fields.  We gave evidence that these effects are strong enough to be practical in the lab.  But only the simplest cases of strongly chiral ``axially-aligning" objects were treated.  A major neglected aspect is the case of more weakly chiral objects that can reach several alternative final states of rotation.  These objects thus have a richer dynamics than the axially-aligning objects that we understand well.  Even a general understanding of what initial conditions determine the choice of the final state appears to be lacking.  Also we lack a description of the final states in terms of the eigenstates of the matrix.  
				
				An important new aspect of these multiple-alignment objects is their response to programmed forcing, generalizing the programmed forcing effects for axially-aligning objects treated in Sec. \ref{sec:ProgrammedForcing}.  The multiple alignment property means that constant forcing of a dispersion leads to a mixture of at least two different populations with different alignments.  The potential behavior of this mixture is of interest in itself.  The prospect of manipulating this mixed population into one or another of the available alignment states is an intriguing new subject to explore.  
				
				As we have seen, perturbations of vector form are only a subset of the possibilities for manipulating objects.  Another possibility is to control the flow environment of the objects, as seen in Sec. \ref{sec:ChiralShear}.  An object experiences this flow environment via the local five-component shear tensor. Thus the proportionality of angular velocity $\vector \Omega$ to the shear tensor is a 15-component response function.  Section \ref{sec:ChiralShear} demonstrates one useful manipulation: the response can move particles of opposite chirality away from each other, thus effecting separation.  For general objects there is 
an equation of motion giving the time derivative of the response tensor to the tensor itself, in analogy to Sec. \ref{sec:MotionLaws}.  If $\del v$ is constant in the lab frame, one may determine the motion by regarding the motion of $\del v$ in the body frame as done in Sec \ref{sec:MotionLaws}.  Much of this motion amounts to rotation about the vorticity direction, the so-called Jeffery orbits mentioned in Sec. \ref{sec:ChiralShear}.  Doi and Makino\cite{Makino:2005ty} determined the ultimate distribution of orientations for specific shapes, but a more general description including nonperiodic features analogous to the orbits of Sec. \ref{sec:MotionLaws} seems possible. Since Jeffery motion of chiral objects need not be periodic, shear-induced driving need not show the strong convergence to a vanishing fraction of all orientations, as with the vector driving of Secs. \ref{sec:ConstantForce} and \ref{sec:NoForce}.  A further open question is whether a time-varying, programmed shear might increase this self-organization.  
						
				The phenomenon of convergence to an oriented state is remarkable in itself, as noted in Sec. \ref{sec:MotionLaws}.  This convergence is more subtle than the simple alignment of \eg a compass needle in a  magnetic field.  There the alignment results from minimizing an interaction energy.  However, in tensorial alignment, there need not be a unique orientation.  Further, the final state may be an orbit that cycles through many orientations relative to the driving vector. Nevertheless, the final state of orientation is always an arbitrarily small subset of the initial set of possible orientations: the self-organizing tendency is qualitatively as strong as that of a compass needle.  The mutual alignment under programmed forcing shows another case of emergent self-organization.  This mutual alignment is subtle in a further way: the same kind of forcing that results in complete alignment can destroy that alignment if the forcing is simply increased moderately.   To understand the basic principles underlying self organization in these simple, dissipative dynamical systems would be of great value.
				
				A dispersion of synchronized rotating objects can create striking collective responses.  A flash of bleaching light transverse to the rotation axis creates a periodically changing optical absorption from each object.  Two successive flashes give a distinctive label to objects that have rotated an integer number of times between the flashes.  If programmed forcing has created a common orientation of all the objects, they may then be manipulated by further forcing into any chosen orientation---for example the orientation of the unstable fixed point.  Any subsequent perturbation can trigger a large coherent reversal of direction.  Such responses can be used to manipulate the fluid's flow or its optical or acoustic properties. 

				The imperfections of the real world naturally limit the self-organized responses described above.  Above we have mentioned rotational diffusion, which randomizes the orientation and degrades any orientational order achieved by driving.  Further, the heterogeneity of any experimental dispersion of particles means that different particles respond differently to the same driving.  Thus over time any common periodicity is smeared away.   Any given particle-to-particle variability in the $\RT$ tensors imposes a time window within which periodic responses can be seen.  As with magnetic resonance phenomena, there is a potential of compensating for this variability by forms of driving that unwind the unwanted smearing.  Finally interactions among the objects alter their response to driving, as they respond to one another as well as to the the external driving field.  Though such interaction complicates the response, it has the potential to create new collective states, which we now consider.


As we have seen, the hydrodynamic interaction between two forced
objects generally degrades their alignment. This effect should be
dominant in dilute suspensions, jeopardizing the stability of the
synchronized rotating states mentioned above. At higher
concentrations, however, the collective effect of interactions may be
significantly different, or even opposite, due to mutual cancellation
of rotational interactions. Thus the role of hydrodynamic interactions
in the stabilization of synchronized rotating states in concentrated
suspensions remains to be clarified.


We have seen that hydrodynamic repulsion or attraction among many
forced asymmetric colloids give rise to unusual collective structures
(Sec.~\ref{sec:asymmetricsuspension}). Repulsion makes particles
self-organize into statistically hyperuniform arrangements transverse
to the force. Attraction creates unstable dynamic clusters. This
suggests a means to control the stability of forced suspensions and
fluidized beds via the shapes of particles. The conclusions of
Sec.~\ref{sec:collective} were based on linear response. Nonlinear
terms which couple concentration and velocity
fluctuations \cite{LevinePRL1998} may further stabilize or destabilize
the collective dynamics.

Other collective structures might form if the interactions are more
complex, as in the case of a pair-interaction which creates bound
states, or under programmed forcing that strengthens the
alignment. The formation of correlated structures in aligned
suspensions makes one contemplate the possibility of convergence to a
small set of translational, not only orientational, states. The
dynamic hyperuniform arrangement suppresses anomalous velocity
fluctuations and the ensuing increased dissipation. The same would be
achieved, in principle, by a static hyperuniform arrangement or even
an ordered one.

Continuum theory has been very helpful in reaching these insights concerning the collective dynamics of asymmetric particles. In the case of collective chiral response, the present connection between the continuum theory of passive
chiral fluids \cite{Andreev:2010fk} and the behaviors of individual
chiral objects in shear flow is only heuristic
(Sec.~\ref{sec:chiralfluids_passive}). Obtaining the continuum
description from coarse-graining of a suspension of chiral objects
would be very valuable.

\section*{Conclusion}\label{sec:Conclusion}
The asymmetric, driven colloids treated in this review show a range of 
soft matter dynamics that complements the much-studied domain of active 
materials.  These phenomena show a surprising and little-explored richness both in single 
object dynamics and in collective dynamics.  These all arise from the fluid environment in the tractable regime of Stokes flow.  Thus the phenomena are tractable and designable.  Accordingly they give 
promise of enabling materials that can be finely controlled and 
organized by external influences.  Though we have treated only fluid 
dispersions, there are clearly analogous phenomena in viscoelastic 
fluids and liquid crystals.

\section*{Acknowledgments}\label{sec:Acknowledgements}

We are grateful to Lara Braverman, Jonah Eaton, Tomer Goldfriend,
Nathan Krapf, Brian Moths, and Aaron Mowitz for their contributions to
the research reviewed here. We thank Justin Burton and Xiaolei Ma for
sharing their unpublished experimental results. We thank Robert
Deegan, Michelle Driscoll, Oscar Gonzalez, Alex Leshansky, John Maddocks, and Boris Spivak for
helpful discussions. We acknowledge support from the US--Israel
Binational Science Foundation (Grant No.~2012090). HD acknowledges
support from the Israel Science Foundation (Grant No.~986/18).

\section*{References}

\bibliographystyle{iopart-num}
\bibliography{chiral_refs}

\end{document}